\documentclass[aps,prb,twocolumn,amsmath,amssymb,superscriptaddress,floatfix,eqsecnum]{revtex4-1}

\usepackage{amsfonts}
\usepackage{amsthm}
\usepackage{mathrsfs}
\usepackage{cases}
\usepackage{url}
\usepackage{framed}
\usepackage{cancel,soul}
\usepackage[dvipdfmx]{graphicx}
\usepackage{multirow}
\usepackage[pdftex]{color}
\usepackage{bm}
\usepackage[normalem]{ulem}
\usepackage[colorlinks=true,linktoc=page,linkcolor=blue,citecolor=blue]{hyperref}

\newcommand{\bea}{\begin{eqnarray}}
\newcommand{\eea}{\end{eqnarray}}
\def\:={\,\raisebox{0.85pt}{.}\hspace{-2.78pt}\raisebox{2.85pt}{.}\!\!=\,}
\def\=:{\,=\!\!\raisebox{0.85pt}{.}\hspace{-2.78pt}\raisebox{2.85pt}{.}\,}

\begin{document}

\title{
Quantum phase transitions beyond Landau-Ginzburg theory
in one-dimensional space revisited
      }

\author{Christopher Mudry}
\affiliation{Condensed Matter Theory Group, Paul Scherrer Institute, 
CH-5232 Villigen PSI, Switzerland}
\affiliation{Institute of Physics,
\'Ecole Polytechnique F\'ed\'eerale de Lausanne (EPFL), CH-1015
Lausanne, Switzerland}

\author{Akira Furusaki}
\affiliation{Condensed Matter Theory Laboratory, RIKEN, Wako, Saitama,
351-0198, Japan}
\affiliation{RIKEN Center for Emergent Matter Science (CEMS), Wako, Saitama,
351-0198, Japan}

\author{Takahiro Morimoto}
\affiliation{Department of Physics, University of California, Berkeley,
California 94720, USA}
\affiliation{Materials Science Division, Lawrence Berkeley National Laboratory, Berkeley, California 94720, USA}

\author{Toshiya Hikihara}
\affiliation{Faculty of Science and Technology, Gunma University, Kiryu
376-8515, Japan}
\date{\today}

\begin{abstract} 
The phase diagram of the
quantum spin-1/2 antiferromagnetic $J^{\,}_{1}$-$J^{\,}_{2}$ XXZ chain
was obtained by Haldane using bosonization techniques
in Refs.~\onlinecite{Haldane1982,Haldane1982E}.
It supports three distinct phases for $0\leq J^{\,}_{2}/J^{\,}_{1}<1/2$,
i.e.,  a gapless algebraic spin liquid phase,
a gapped long-range ordered Neel phase,
and a gapped long-range ordered dimer phase.
Even though the Neel and dimer phases are not related hierarchically
by a pattern of symmetry breaking, it was shown that
they meet along a line of quantum critical points with
a U(1) symmetry and central charge $c=1$.
Here, we extend the analysis made by Haldane on
the quantum spin-1/2 antiferromagnetic $J^{\,}_{1}$-$J^{\,}_{2}$ XYZ chain
using both bosonization and numerical techniques. 
We show that there are three Neel phases and the dimer phase
that are separated from each other by six planes of phase boundaries
realizing Gaussian criticality when $0\leq J^{\,}_{2}/J^{\,}_{1}<1/2$.
We also show that each long-range ordered phase harbors
topological point defects (domain walls)
that are dual to those across the phase boundary
in that a defect in one ordered phase locally binds
the other type of order around its core.
By using the bosonization approach, 
we identify the critical theory that describes simultaneous proliferation
of these dual point defects, and show that it supports
an emergent U(1) symmetry that originates from
the discrete symmetries of the XYZ model.
To confirm this numerically, we perform exact diagonalization and
density-matrix renormalization-group calculations and show
that the critical theory is characterized by the central charge $c=1$
with critical exponents that are
consistent with those obtained from the bosonization approach.
Furthermore, we generalize the field theoretic description
of direct continuous phase transition to higher dimensions,
especially in $d=3$, by using a non-linear sigma model (NLSM)
with a topological term.
In particular, we propose  the $\pi$-flux phase on the cubic lattice
with local quartic interactions
as a platform for direct continuous phase transition
and deconfined criticality. We discuss possible phase diagrams for the
$\pi$-flux phase on the cubic lattice with these quartic interactions
from the renormalization flow of NLSMs.
\end{abstract}

%75.10.Pq 	Spin chain models
%75.10.Jm 	Quantized spin models, including quantum spin frustration
%64.70.Tg 	Quantum phase transitions
\maketitle
\tableofcontents

\section{Introduction}

A paradigm for a phase transition with
spontaneous symmetry breaking is
the antiferromagnetic phase transition of quantum spin systems
with antiferromagnetic exchange interactions.
The antiferromagnetic phase is characterized
by a local order parameter, the staggered magnetization.
In dimensions above the lower critical dimension, it is sufficient
to account for the smooth fluctuations of the order parameter,
the spin waves, in order to describe the second-order transition from the
antiferromagnetic to the paramagnetic phase.

The discovery of the Berezinskii-Kosterlitz-Thouless transition%
~\cite{Berezinskii71,Berezinskii72,Kosterlitz73,Kosterlitz74}
demonstrated that fluctuations that are point-wise singular can
also drive classical continuous phase transitions,
while the spin waves are only good enough
to downgrade long-range order to quasi-long-range order 
owing to the low dimensionality of space.
In the context of quantum phase transitions, 
Haldane in Ref.\ \onlinecite{Haldane88}
and Read and Sachdev in Ref.\ \onlinecite{Read89}
pointed out that the proliferation of hedgehog defects in
$(2+1)$-dimensional spacetime
have the potential to drive a transition from an antiferromagnetic
ground state to a spin dimerized ground state in 
spin-1/2 two-dimensional (2D) systems. Because both phases
break spontaneously distinct symmetries of the microscopic Hamiltonian
(the spin SU(2) symmetry in the antiferromagnetic phase and
the point-group symmetry of the lattice in the spin dimerized phase),
such a transition was originally considered to be non-generic
and discontinuous.

This interpretation, derived as it is
from the conventional wisdom based on the Landau-Ginzburg
paradigm for symmetry breaking,
was questioned in a series of theoretical papers,%
~\cite{Senthil04,Senthil04PRB,Senthil04Levin,TanakaHuPRL2005,Senthil06Fisher}
where it was  proposed that a direct continuous quantum phase transition
between the antiferromagnetic and spin dimerized (valence bond solid) phases
of 2D spin-1/2 systems is generic when driven by the proliferation of
nontrivial point defects. In this scenario,
the nature of the transition is encoded by the duality relating
the point defects in the two phases. 
Namely, a point defect in one phase
binds the order of the other phase in its core. 
At the transition both types of point defects proliferate.
It was also argued that the critical theory is described by
a doublet of bosonic matter fields that are
coupled to a non-compact U(1) gauge field. 
This emerging scalar quantum electrodynamics is
in its deconfined Coulomb phase,
and hence, was called deconfined quantum criticality.%
~\cite{Senthil04}

A series of numerical studies of 2D lattice models that were designed
with the potential to host deconfined quantum criticality,
have been performed.%
~\cite{Sandvik2007,Melko2008,Jiang2008,Kuklov2008,Lou2009,Sandvik2010,Sen2010,Banerjee2010,Banerjee2011,NahumPRX2015,Pujari2015,Gazit2018,ZhangPRL2018}
However, confirming numerically the existence of
deconfined quantum criticality has been a hard task.
First, one must identify the proper
two-dimensional lattice model that may host a direct continuous quantum phase
transition between two phases that break spontaneously and in distinct ways
the symmetries of the lattice Hamiltonian. Second, the numerics must rule out a
quantum phase transition that is weakly discontinuous.

Whereas the original proposal for deconfined quantum criticality
referred to a direct phase transition between antiferromagnetic order
and valence bond (dimerization) order,
one may consider many-body quantum systems that are not quantum magnets,
one may consider different choices of the ordered phases,
and one may work in spaces with a dimensionality other than two.%
~\cite{TanakaHuPRL2005,Senthil06Fisher,Ryu2009,Hosur10,Gaemi10,Gaemi2012,Sato2017,Liu18assaad,You2018,Metlitski2018,Bi2018}
In particular, if one considers discrete symmetries instead of continuous
ones, one may seek examples of deconfined quantum criticality characterizing
one-dimensional (1D) lattice Hamiltonians.
This approach has recently been  advocated in Ref.\ \onlinecite{Jiang2018} 
in the context of quantum spin-1/2 chains.

In this paper, we report analytical and numerical studies
of quantum spin-1/2 chains supporting the
core idea of quantum criticality beyond the Landau-Ginzburg paradigm,
namely that a direct continuous quantum phase transition
between two ordered phases
can be interpreted as the proliferation of point defects
that nucleate the order across the transition.
Specifically, we identify the critical theory in $(1+1)$-dimensional spacetime
by the bosonization approach and show that the critical theory supports
an emergent U(1) symmetry.
Performing exact diagonalization and density-matrix renormalization-group
(DMRG) calculations, we numerically confirm
that the critical theory is characterized by the central charge $c=1$
with critical exponents that are consistent
with those from the bosonization approach.

We also argue that higher-dimensional analogs of deconfined quantum
criticality can be obtained from fermionic tight-binding Hamiltonians
that support a Dirac semi-metallic phase with certain contact
interactions.  Mean-field decoupling of interactions naturally leads
to a non-linear sigma model (NLSM) augmented by a topological term.  This
topological term describes the mutual relationship between the defects
in the two ordered phases and is responsible for deconfined quantum
criticality.
As an example in three-dimensional (3D) space, we demonstrate that 
the $\pi$-flux phase on the cubic lattice with contact interactions
gives a natural platform for a duality between point defects
that nucleate the antiferromagnetic (dimer) order
at the core in the dimer (antiferromagnetic) phase.

The paper is organized as follows.
The case of 1D space, $d=1$, is treated in
Sec.\ \ref{sec: J1-J2 XYZ model on a linear chain}.
The case of 3D space, $d=3$, is treated in
Sec.\ \ref{sec: Dirac semimetallic phase in d>1-dimensional space ...}.
A summary is given in Sec.\ \ref{sec: Summary}.

\section{$J^{\,}_{1}$-$J^{\,}_{2}$ XYZ model on a (linear) chain}
\label{sec: J1-J2 XYZ model on a linear chain}

\subsection{Symmetries and phases}

The sites of a spin chain 
are denoted by the letter $l=1,\cdots,L$, where
the number of sites $L$ is assumed to be an even integer.
We study the quantum spin-1/2 XYZ chain with the antiferromagnetic
nearest-neighbor $J^{\,}_{1}>0$ and next-nearest-neighbor
$J^{\,}_{2}\geq0$ couplings defined by the Hamiltonian
\begin{subequations}
\label{eq: def HXYZ}
\begin{align}
H\:=&\,
J^{\,}_{1}
\sum_{l=1}^{L}
\left(
S^{x}_{l}\,S^{x}_{l+1}
+
\Delta^{\,}_{y}\,
S^{y}_{l}\,S^{y}_{l+1}
+
\Delta^{\,}_{z}\,
S^{z}_{l}\,S^{z}_{l+1}
\right)
\nonumber\\
&
+
J^{\,}_{2}
\sum_{l=1}^{L}
\left(
S^{x}_{l}\,S^{x}_{l+2}
+
\Delta^{\,}_{y}\,
S^{y}_{l}\,S^{y}_{l+2}
+
\Delta^{\,}_{z}\,
S^{z}_{l}\,S^{z}_{l+2}
\right),
\label{eq: def HXYZ a}
\end{align}
where we have chosen to impose the periodic boundary conditions
$S^{\alpha}_{l+L}\equiv S^{\alpha}_{l}$ ($\alpha=x,y,z$).
The spin operators obey the SU(2) algebra
\begin{equation}
\left[S^{\alpha}_{l},S^{\beta}_{l'}\right]=
\mathrm{i}\,\delta^{\,}_{l,l'}\,\epsilon^{\alpha\beta\gamma}\,S^{\gamma}_{l},
\qquad
\bm{S}^{2}_{l}=\frac{3}{4},
\label{eq: def HXYZ b}
\end{equation}
with $\epsilon^{\alpha\beta\gamma}$ the fully antisymmetric Levi-Civita tensor
where $\alpha,\beta,\gamma=x,y,z$ and $l,l'=1,\cdots,L$.
The exchange-anisotropy parameters
$\Delta^{\,}_{y}$ and $\Delta^{\,}_{z}$ are non-negative numbers
($\Delta^{\,}_{y},\Delta^{\,}_{z}\ge0$).
The Hamiltonian thus depends on three dimensionless positive parameters,
namely
\begin{equation}
\mathcal{J}\:=\frac{J^{\,}_{2}}{J^{\,}_{1}}\geq0,
\qquad
\Delta^{\,}_{y}\geq0,
\qquad
\Delta^{\,}_{z}\geq0.
\label{eq: def HXYZ c}
\end{equation}
We restrict our discussion to the case of weak frustration,
\begin{equation}
\mathcal{J}<\frac{1}{2}
\label{eq: def HXYZ d}
\end{equation}
\end{subequations}
for which the nearest-neighbor exchange coupling $J^{\,}_{1}$
is the dominant interaction and sets the dimension of energy.
In other words, we are not concerned with other phases that can
exist for $\mathcal{J}>\frac12$, such as the UUDD (up-up-down-down)
Neel-ordered phase that appears in the limit $\Delta^{\,}_{z}\gg1$, etc.
When $\Delta^{\,}_{y}=0$, Hamiltonian (\ref{eq: def HXYZ a})
is equivalent to the one recently studied by Jiang and Motrunich
in Ref.\ \onlinecite{Jiang2018}.
Reference \onlinecite{Xu2018arXiv} also reports numerical simulations
of the spin-1/2 XYZ Hamiltonian (\ref{eq: def HXYZ a}).

We first discuss the symmetries of the model defined by Eq.\
(\ref{eq: def HXYZ}). Hamiltonian $H$ is invariant
under the following transformations:\\

\noindent(i)
$\pi$-rotations about the $x$, $y$, and $z$ axes in spin space,
\begin{subequations}
\begin{align}
&R^{x}_{\pi}: (S^{x}_{l}, S^{y}_{l}, S^{z}_{l})\mapsto(S^{x}_{l},-S^{y}_{l},-S^{z}_{l}),
\label{eq: R_pi^x}\\
&R^{y}_{\pi}: (S^{x}_{l}, S^{y}_{l}, S^{z}_{l})\mapsto(-S^{x}_{l},S^{y}_{l},-S^{z}_{l}),
\label{eq: R_pi^y}\\
&R^{z}_{\pi}: (S^{x}_{l}, S^{y}_{l}, S^{z}_{l})\mapsto(-S^{x}_{l},-S^{y}_{l},S^{z}_{l}),
\label{eq: R_pi^z}
\end{align}\\

\noindent(ii)
translation by one lattice site,
\begin{equation}
T: (S^{x}_{l},S^{y}_{l},S^{z}_{l})\mapsto(S^{x}_{l+1},S^{y}_{l+1},S^{z}_{l+1}),
\label{eq: T}
\end{equation}\\

\noindent(iii)
inversion about the site $l=0\, (\equiv L)$,
\begin{equation}
P: (S^{x}_{l},S^{y}_{l},S^{z}_{l})\mapsto
(S^{x}_{L-l},S^{y}_{L-l},S^{z}_{L-l}),
\label{eq: P}
\end{equation}\\

\noindent(iv)
time reversal,
\begin{equation}
\Theta: (S^{x}_{l},S^{y}_{l},S^{z}_{l})\mapsto(-S^{x}_{l},-S^{y}_{l},-S^{z}_{l}).
\label{eq: Theta}
\end{equation}
\end{subequations}
We note that inversion about the center of a nearest-neighbor bond
of the lattice is obtained by combining the site-inversion
$P$ with the lattice translation $T$. Similarly, we note that
$R^{z}_{\pi}=R^{y}_{\pi}\,R^{x}_{\pi}$.
Equations (\ref{eq: R_pi^x})--(\ref{eq: R_pi^z})
thus imply the existence of a global internal 
$\mathbb{Z}^{\,}_{2}\times \mathbb{Z}^{\,}_{2}$ symmetry.  

As we shall show in the following subsections,
the ground-state phase diagram of the
quantum spin-1/2 antiferromagnetic $J^{\,}_{1}$-$J^{\,}_{2}$ XYZ chain
defined by Eq.\ (\ref{eq: def HXYZ})
consists of the following four gapped phases:
Neel${}^{\,}_{x}$, Neel${}^{\,}_{y}$, Neel${}^{\,}_{z}$,
and valence-bond-solid (VBS or dimer) phases:\\

\noindent(1)
\underline{Neel${}^{\,}_{x}$ phase.} 
The symmetries $R^{y}_{\pi}$, $R^{z}_{\pi}$, $T$, $\Theta$
are spontaneously broken.
The order parameter is
\begin{subequations}
\begin{equation}
O^{\,}_{\mathrm{N}^{\,}_{x}}\:=
\frac{1}{L}\sum_{l=1}^{L}
(-1)^{l}\,\langle S^x_l \rangle,
\label{eq: order parameter Neel x}
\end{equation}
where we recall that $L$ is the number of sites
and $\langle A\rangle$ is the expectation-value of the operator $A$
in a ground state. The composition $T\,\Theta$ is a symmetry.\\

\noindent(2)
\underline{Neel${}^{\,}_{y}$ phase.} 
The symmetries $R^{z}_{\pi}$, $R^{x}_{\pi}$, $T$, $\Theta$
are spontaneously broken.
The order parameter is
\begin{equation}
O^{\,}_{\mathrm{N}^{\,}_{y}}\:=
\frac{1}{L}\sum_{l=1}^{L}
(-1)^{l}\,\langle S^y_l \rangle.
\label{eq: order parameter Neel y}
\end{equation}
The composition $T\,\Theta$ is a symmetry.\\

\noindent(3)
\underline{Neel${}^{\,}_{z}$ phase.}
The symmetries $R^{x}_{\pi}$, $R^{y}_{\pi}$, $T$, $\Theta$
are spontaneously broken.
The order parameter is
\begin{equation}
O^{\,}_{\mathrm{N}^{\,}_{z}}\:=
\frac{1}{L}\sum_{l=1}^{L}
(-1)^{l}\,\langle S^z_l \rangle.
\label{eq: order parameter Neel z}
\end{equation}
The composition $T\,\Theta$ is a symmetry.\\

\noindent(4)
\underline{VBS (dimer) phase.}
The symmetries $P$ and $T$ are spontaneously broken.
We can take
\begin{align}
O^{\,}_{\mathrm{VBS}}\:=
\frac{1}{L}\sum_{l=1}^{L}
(-1)^{l}\,
\langle
S^{x}_{l}\,S^{x}_{l+1}
+
S^{y}_{l}\,S^{y}_{l+1}
+
S^{z}_{l}\,S^{z}_{l+1}
\rangle
\label{eq: order parameter VBS}
\end{align}
\end{subequations}
as an order parameter of the VBS phase.

It is important to note that the $\mathbb{Z}^{\,}_{2}$ symmetries
$R^{\alpha}_{\pi}$ are broken in the Neel${}^{\,}_{\beta}$ phase provided
$\alpha\neq\beta$, while the site-inversion symmetry $P$ is broken in the VBS
phase.  Since the $\pi$-rotation $R^{\alpha}_{\pi}$ and the inversion $P$
are symmetries in the spin and real spaces, respectively, a direct
continuous phase transition between a Neel phase and a VBS phase
cannot be described by the Ginzburg-Landau theory.  As we discuss
below, the phase transitions between gapped ordered phases are
continuous transitions described by a Gaussian theory with
no less than a U(1) symmetry.

\subsection{The phase diagram at $\Delta^{\,}_{y}=1$}
\label{sec: Delta_{y}=1}

We first discuss the case $\Delta^{\,}_{y}=1$ for which the Hamiltonian
$H$ describes the
quantum spin-1/2 antiferromagnetic $J^{\,}_{1}$-$J^{\,}_{2}$ XXZ chain
with an enhanced U(1) symmetry under continuous spin rotations
about the $z$ axis compared to the case when  $\Delta^{\,}_{y}\neq1$.
The quantum spin-1/2 antiferromagnetic $J^{\,}_{1}$-$J^{\,}_{2}$ XXZ chain
has been studied in many publications.%
~\cite{Haldane1982,Haldane1982E,Okamoto1992,Nomura1994,WhiteAffleck1996,Eggert1996,FurukawaSatoFurusaki2010,FurukawaSatoOnodaFurusaki2012}
Its ground-state phase diagram and low-energy effective theory are
well understood; see e.g., Refs.~\onlinecite{Haldane1982}
and \onlinecite{Nomura1994}.  We briefly
review the low-energy effective theory, in order to fix notations and prepare
for the full discussion of the ground-state phase diagram of the
quantum spin-1/2 antiferromagnetic $J^{\,}_{1}$-$J^{\,}_{2}$ XYZ chain
($\Delta^{\,}_{y}\neq1$).

We introduce the Jordan-Wigner fermions $c^{\,}_{l}$ through the relations
\begin{subequations}
\label{eq: def XXZ model}
\begin{align}
&
S^{z}_{l}\=:
c^{\dag}_{l}\,c^{\,}_{l}
-
\frac12\equiv
n^{\,}_{l},
\label{eq: def XXZ model a}
\\
&
S^{+}_{l}\equiv
S^{x}_{l}+\mathrm{i}S^{y}_{l}\=:
c^{\dag}_{l}\,
\exp\!\left(\mathrm{i}\pi\sum_{n<l}c_n^{\dag}\,c_n^{}\right),
\label{eq: def XXZ model b}
\end{align}
with which the Hamiltonian $H$ at $\Delta^{\,}_{y}=1$ is rewritten as
\begin{align}
H^{\,}_{\mathrm{XXZ}}\equiv&\,
J^{\,}_{1}
\sum_l
\left[
\frac{1}{2}
\left(
c^{\dag}_{l+1}\,c^{\,}_{l}
+
c^{\dag}_{l}\,c^{\,}_{l+1}
\right)
+
\Delta^{\,}_{z}\,n^{\,}_{l}\,n^{\,}_{l+1}
\right]
\nonumber\\
&
+
J^{\,}_{2}
\sum_l
\left[
\left(
c^{\dag}_{l+2} c^{\,}_{l}
+
c^{\dag}_{l} c^{\,}_{l+2}
\right)
n^{\,}_{l+1}
+
\Delta^{\,}_{z}
n^{\,}_{l}\,
n_{l+2}
\right]. 
\label{eq: def XXZ model c}
\end{align}
\end{subequations}

When both $\Delta^{\,}_{z}$ and $\mathcal{J}\equiv J^{\,}_{2}/J^{\,}_{1}$
are zero, the Jordan-Wigner fermions are non-interacting and
their energy band is half filled.
We introduce left- and right-moving fermions,
$\psi^{\,}_{L}(x)$ and $\psi^{\,}_{R}(x)$, which describe low-energy
excitations near the two Fermi points at momentum
$k=\pm\pi/2\mathfrak{a}$,
\begin{subequations}
\label{eq: def XXZ model bis}
\begin{equation}
c^{\,}_{l}\approx
\sqrt{\mathfrak{a}}
\left[
e^{+\mathrm{i}\pi x/(2\mathfrak{a})}\,
\psi^{\,}_{\mathrm{L}}(x)
+
e^{-\mathrm{i}\pi x/(2\mathfrak{a})}\,
\psi^{\,}_{\mathrm{R}}(x)
\right],
\label{eq: def XXZ model a bis}
\end{equation}
where $x=l\,\mathfrak{a}$ with $\mathfrak{a}$ the lattice spacing.
We take the continuum limit to
approximate Hamiltonian $H^{\,}_{\mathrm{XXZ}}$ by the integral
$\int\mathrm{d}x\,\mathcal{H}^{\,}_{\mathrm{XXZ}}$
over the Hamiltonian density $\mathcal{H}^{\,}_{\mathrm{XXZ}}$, where
\begin{align}
\mathcal{H}^{\,}_{\mathrm{XXZ}}=&\,
\mathrm{i}v
\left(
\psi^{\dag}_{\mathrm{L}}\,\partial^{\,}_{x}\psi^{\,}_{\mathrm{L}}
-
\psi^{\dag}_{\mathrm{R}}\,\partial^{\,}_{x}\psi^{\,}_{\mathrm{R}}
\right)
\nonumber\\
&
+
g^{\,}_{+}\!
\left(
\bm{:}\!\psi^{\dag}_{\mathrm{L}}\,\psi^{\,}_{\mathrm{L}}\!\bm{:}
+
\bm{:}\psi^{\dag}_{\mathrm{R}}\,\psi^{\,}_{\mathrm{R}}\!\bm{:}
\right)^{2}
\nonumber\\
&
+
g^{\,}_{-}\!
\left(
\bm{:}\!\psi^{\dag}_{\mathrm{L}}\,\psi^{\,}_{\mathrm{L}}\!\bm{:}
-
\bm{:}\!\psi^{\dag}_{\mathrm{R}}\,\psi^{\,}_{\mathrm{R}}\!\bm{:}
\right)^{2}
\nonumber\\
&
+
g^{\,}_{\mathrm{u}}
\left(
\bm{:}\!\psi^{\dag}_{\mathrm{L}}\,\psi^{\dag}_{\mathrm{L}}\!\bm{:}\,
\bm{:}\!\psi^{\,}_{\mathrm{R}}\,\psi^{\,}_{\mathrm{R}}\!\bm{:}
+
\bm{:}\!\psi^{\dag}_{\mathrm{R}}\,\psi^{\dag}_{\mathrm{R}}\!\bm{:}\,
\bm{:}\!\psi^{\,}_{\mathrm{L}}\,\psi^{\,}_{\mathrm{L}}\!\bm{:}
\right).
\label{eq: def XXZ model b bis}
\end{align}
\end{subequations}
Here, $v$ is the velocity ($v>0$), the coupling constants $g^{\,}_{\pm}$ and
$g^{\,}_{\mathrm{u}}$ are matrix elements of forward- and umklapp scatterings,
and the normal-ordered operators are defined using point-splitting, i.e.,
$
\bm{:}\!\mathcal{O}^{\,}_{\mathrm{A}}(x)\,\mathcal{O}^{\,}_{\mathrm{B}}(x)\!\bm{:}
$
is the leading term from the Laurent expansion in powers of $\mathfrak{a}$ of
$
\mathcal{O}^{\,}_{\mathrm{A}}(x)\,\mathcal{O}^{\,}_{\mathrm{B}}(x+\mathfrak{a})
-
\langle
\mathcal{O}^{\,}_{\mathrm{A}}(x)\,\mathcal{O}^{\,}_{\mathrm{B}}(x+\mathfrak{a})
\rangle
$,
where
$\mathcal{O}^{\,}_{\mathrm{A}}$
and
$\mathcal{O}^{\,}_{\mathrm{B}}$
are
$\psi^{\,}_{\mathrm{M}}$
or
$\psi^{\dag}_{\mathrm{M}}$ with
$\mathrm{M}=\mathrm{L},\mathrm{R}$.

Next, we bosonize the fermion fields using the formulas
\begin{subequations}
\label{eq: chiral bosonized HXXZ}
\begin{equation}
\psi^{\,}_{\mathrm{L}}(x)=
\frac{e^{-\mathrm{i}\varphi^{\,}_{\mathrm{L}}(x)}}{\sqrt{2\pi\mathsf{a}}},
\qquad
\psi^{\,}_{\mathrm{R}}(x)=
\frac{e^{+\mathrm{i}\varphi^{\,}_{\mathrm{R}}(x)}}{\sqrt{2\pi\mathsf{a}}},
\label{eq: chiral bosonized HXXZ aaa}
\end{equation}
where $\mathsf{a}$ is a short-distance cutoff
of the order of the lattice spacing $\mathfrak{a}$
and the chiral boson fields
$\varphi^{\,}_{\mathrm{M}}(x)$
with $\mathrm{M}=\mathrm{L},\mathrm{R}$
satisfy the equal-time commutation relations
\begin{align}
&
[\varphi^{\,}_{\mathrm{R}}(x),\varphi^{\,}_{\mathrm{R}}(y)]=
-[\varphi^{\,}_{\mathrm{L}}(x),\varphi^{\,}_{\mathrm{L}}(y)]=
\mathrm{i}\pi\,\mathrm{sgn}(x-y),
\\
&
[\varphi^{\,}_{\mathrm{R}}(x),\varphi^{\,}_{\mathrm{L}}(y)]=
\mathrm{i}\pi.
\label{eq: chiral bosonized HXXZ aa}
\end{align}
The Hamiltonian density, when expressed in terms of this pair
of bosonic chiral fields, takes the form 
\begin{align}
\mathcal{H}^{\,}_{\mathrm{XXZ}}=&\,
\frac{\tilde{g}^{\,}_{+}}{8\pi}\,
\left[\partial^{\,}_{x}(\varphi^{\,}_{\mathrm{L}}+\varphi^{\,}_{\mathrm{R}})\right]^{2}
+
\frac{\tilde{g}^{\,}_{-}}{8\pi}\,
\left[\partial^{\,}_{x}(\varphi^{\,}_{\mathrm{L}}-\varphi^{\,}_{\mathrm{R}})\right]^{2}
\nonumber\\
&
+
\tilde{g}^{\,}_{\mathrm{u}}\,
\cos\!\left[2(\varphi^{\,}_{\mathrm{L}}+\varphi^{\,}_{\mathrm{R}})\right].
\label{eq: chiral bosonized HXXZ a}
\end{align}
\end{subequations}
Here, the parameters $\tilde{g}^{\,}_{\pm}$ and $\tilde{g}^{\,}_{\mathrm{u}}$
are given by
\begin{subequations}
\begin{align}
&
\tilde{g}^{\,}_{+}=
\mathfrak{a}\,J^{\,}_{1}\!
\left[1+\frac{4}{\pi}(\Delta^{\,}_{z}+\mathcal{J})\right],
\label{eq: chiral bosonized HXXZ b}
\\
&
\tilde{g}^{\,}_{-}=
\mathfrak{a}\,J^{\,}_{1}\!\left(1-\frac{4}{\pi}\mathcal{J}\right),
\label{eq: chiral bosonized HXXZ c}
\\
&
\tilde{g}^{\,}_{\mathrm{u}}=
\frac{\mathfrak{a}\,J^{\,}_{1}}{2\pi^{2}\mathsf{a}^{2}}
\left[\Delta^{\,}_{z}-\mathcal{J}(2+\Delta^{\,}_{z})\right],
\label{eq: chiral bosonized HXXZ d}
\end{align}
in the weak-coupling limit
\begin{equation}
0\leq\Delta^{\,}_{z}\ll1,
\qquad
0\leq\mathcal{J}\ll1/2.
\label{eq: chiral bosonized HXXZ e}
\end{equation}
\end{subequations}
However, the chiral representation
(\ref{eq: chiral bosonized HXXZ a})
of $\mathcal{H}^{\,}_{\mathrm{XXZ}}$
holds beyond the perturbative regime
(\ref{eq: chiral bosonized HXXZ e}).

Instead of the chiral representation
(\ref{eq: chiral bosonized HXXZ a})
of $\mathcal{H}^{\,}_{\mathrm{XXZ}}$,
we shall use the sine-Gordon representation
\begin{subequations}
\label{eq: sine-Gordon rep HXXZ}
\begin{align}
&
\mathcal{H}^{\,}_{\mathrm{XXZ}}=
\frac{v}{2}
\left[
\frac{1}{\eta}
\left(\partial^{\,}_{x}\theta\right)^{2}
+
\eta
\left(\partial^{\,}_{x}\phi\right)^{2}
+
\lambda^{\,}_{\phi}\,
\cos(\sqrt{8\pi}\,\phi)
\right],
\label{eq: sine-Gordon rep HXXZ a}
\\
&
\frac{v}{2\eta}\:=
\tilde{g}^{\,}_{-},
\qquad
2v\eta\:=
\tilde{g}^{\,}_{+},
\qquad
\frac{v\lambda^{\,}_{\phi}}{2}\:=
\tilde{g}^{\,}_{\mathrm{u}}.
\label{eq: sine-Gordon rep HXXZ b}
\end{align}
The sine-Gordon Hamiltonian density $\mathcal{H}^{\,}_{\mathrm{XXZ}}$
is invariant under any constant shift of $\theta$.
In other words, $\mathcal{H}^{\,}_{\mathrm{XXZ}}$
is invariant under the  global U(1) transformation
\begin{equation}
\theta
\mapsto \theta+\hbox{constant (mod $\sqrt{2\pi}$\,)}.
\qquad
\phi\mapsto\phi,  
\label{eq:U(1) symmetry XXZ}
\end{equation}
The freedom in shifting $\theta$ by an arbitrary constant
originates from the U(1) symmetry of the
quantum spin-1/2 antiferromagnetic $J^{\,}_{1}$-$J^{\,}_{2}$ XXZ Hamiltonian,
whereby the compactifiaction radius $\sqrt{2\pi}$ stems from
the relation
(\ref{eq: relation lattice quantum spins and bosonic quantum fields})
between the quantum spin-1/2 degrees of freedom on the lattice
and the quantum fields $\phi$ and $\theta$.
The advantage of the sine-Gordon representation
(\ref{eq: sine-Gordon rep HXXZ a})
of $\mathcal{H}^{\,}_{\mathrm{XXZ}}$
is that the parameter $\eta$  has a simple interpretation.
It controls the exponents of algebraic correlation
functions when the cosine interaction is irrelevant.
Here, we have adopted the conventions from Ref.~\onlinecite{Hikihara2017},
see also references therein.  The nonchiral bosonic fields $\phi$ and
$\theta$ are dual to each other and defined by
\begin{align}
&
\phi(x)\:=
\frac{1}{\sqrt{2\pi}}
\left[\varphi^{\,}_{\mathrm{L}}(x)+\varphi^{\,}_{\mathrm{R}}(x)\right],
\label{eq: sine-Gordon rep HXXZ c}
\\
&
\theta(x)\:=
\frac{1}{\sqrt{8\pi}}
\left[\varphi^{\,}_{\mathrm{L}}(x)-\varphi^{\,}_{\mathrm{R}}(x)\right].
\label{eq: sine-Gordon rep HXXZ d}
\end{align}
They inherit the commutation relation
\begin{equation}
[\phi(x),\theta(y)]=\mathrm{i}\Theta(y-x),
\label{eq: sine-Gordon rep HXXZ e}
\end{equation}
\end{subequations}
where $\Theta(x)$ is the Heaviside function
taking the value $1/2$ at the origin.
We note that the dimensionful coupling constant
$\lambda^{\,}_{\phi}=2\tilde{g}^{\,}_{\mathrm{u}}/v$ is a function of
$\Delta^{\,}_{z}$ and $\mathcal{J}$.
In the weak-coupling limit
(\ref{eq: chiral bosonized HXXZ e}),
it is seen that $\lambda^{\,}_{\phi}$ changes
its sign from positive to negative as $\mathcal{J}$ is increased holding
$\Delta^{\,}_{z}$ fixed.

When the coupling $\lambda^{\,}_{\phi}$
of the sine-Gordon interaction in the effective
Hamiltonian density (\ref{eq: sine-Gordon rep HXXZ a})
flows to zero in the low-energy limit, the ground state is a critical phase,
the Tomonaga-Luttinger liquid (TLL) phase,
described by the Gaussian Hamiltonian
\begin{equation}
H^{\,}_{\eta}\:=
\frac{v}{2}\,
\int\mathrm{d}x\,
\left[
\frac{1}{\eta}\,
(\partial^{\,}_{x}\theta)^{2}
+
\eta\,
(\partial^{\,}_{x}\phi)^{2}
\right].
\label{eq: Gaussian fixed point}
\end{equation}
The TLL phase realizes a $c=1$ conformal field theory
in $(1+1)$-dimensional spacetime.
It is invariant under the global U(1) transformation
(\ref{eq:U(1) symmetry XXZ}).
The Gaussian Hamiltonian (\ref{eq: Gaussian fixed point}) inherits
this U(1) symmetry from that of the sine-Gordon
Hamiltonian density (\ref{eq: sine-Gordon rep HXXZ a}).
Moreover, the Gaussian Hamiltonian (\ref{eq: Gaussian fixed point})
is also invariant under the global U(1) transformation
\begin{equation}
\phi
\mapsto \phi+\hbox{constant  (mod }\sqrt{2\pi}\,),
\qquad
\theta\mapsto\theta.
\label{eq: 2nd U(1) symmetry}
\end{equation}
Therefore, the Gaussian Hamiltonian (\ref{eq: Gaussian fixed point})
has the
U(1)${}^{\,}_{\theta}\times$U(1)${}^{\,}_{\phi}$ symmetry,
or, equivalently,
the U(1)${}^{\,}_\mathrm{L}\times$U(1)${}^{\,}_\mathrm{R}$ symmetry.
This global U(1)$\times$U(1)
symmetry is enhanced to a global SU(2) symmetry when
$\Delta^{\,}_{y}=\Delta^{\,}_{z}=1$ 
that originates from the SU(2) symmetry of the
quantum spin-1/2 antiferromagnetic $J^{\,}_{1}$-$J^{\,}_{2}$ XXX Hamiltonian.
At the SU(2) symmetric point $\Delta^{\,}_{y}=\Delta^{\,}_{z}=1$,
$\eta=1$ must necessarily hold in the effective Hamiltonian
(\ref{eq: Gaussian fixed point}).
Conversely, if $\eta=1$ holds in the effective theory,
then $\mathcal{H}^{\,}_{\eta=1}$ supports a global SU(2) symmetry,
as we now explain.

The spin operators are related to the bosonic fields by 
\begin{subequations}
\label{eq: relation lattice quantum spins and bosonic quantum fields} 
\begin{align}
&
S^{z}_{l}\approx
\frac{\mathfrak{a}}{\sqrt{2\pi}}\partial^{\,}_{x}\phi(x)
+
a^{\,}_{1}(-1)^{l}\,
\sin\!\left(\sqrt{2\pi}\phi(x)\right),
\label{eq: relation lattice quantum spins and bosonic quantum fields a}
\\
&
S^{+}_{l}\approx
e^{+\mathrm{i}\sqrt{2\pi}\theta(x)}\,
\left[
a^{\,}_{2}(-1)^{l}
+
a^{\,}_{3}
\sin\!\left(\sqrt{2\pi}\phi(x)\right)
\right],
\label{eq: relation lattice quantum spins and bosonic quantum fields b}
\end{align}
\end{subequations}
where $a^{\,}_{1}$, $a^{\,}_{2}$, and $a^{\,}_{3}$
are real numbers that are functions of $\mathcal{J}$
and $\Delta^{\,}_{z}$.  From Eq.\
(\ref{eq: relation lattice quantum spins and bosonic quantum fields b})
the $S^{x}_{l}$ and
$S^{y}_{l}$ operators are written as
\begin{subequations}
\begin{align}
&
\begin{split}  
S^{x}_{l}=&\,
a^{\,}_{2}\,
(-1)^{l}\,
\cos\!\left(\sqrt{2\pi}\,\theta(x)\right)
\\
&\,\qquad
+
\mathrm{i}a^{\,}_{3}\,\sin\left(\sqrt{2\pi}\,\theta(x)\right)
\sin\!\left(\sqrt{2\pi}\,\phi(x)\right),
\end{split}
\label{S^x_l}
\\
&
\begin{split}
S^{y}_{l}=&\,
a^{\,}_{2}\,
(-1)^{l}\,
\sin\!\left(\sqrt{2\pi}\,\theta(x)\right)
\\
&\,\qquad
-
\mathrm{i}a^{\,}_{3}\,\cos\!\left(\sqrt{2\pi}\,\theta(x)\right)
\sin\!\left(\sqrt{2\pi}\,\phi(x)\right).
\end{split}
\label{S^y_l}
\end{align}
\end{subequations}
Here, the commutator $[\phi(x),\theta(x)]=\mathrm{i}/2$ was used.

Motivated by the four order parameters
(\ref{eq: order parameter Neel x})--(\ref{eq: order parameter VBS})
that were defined on the lattice,
we define in the field theory the following four
fields whose non-vanishing ground-state expectation value
signal long-range order.
From Eqs.\ (\ref{eq: relation lattice quantum spins and bosonic quantum fields a}), (\ref{S^x_l}), and (\ref{S^y_l}),
the triplet of fields whose non-vanishing ground-state expectation value
signal Neel order are
\begin{subequations}
\begin{align}
&
N^{\,}_{x}(x)\:=\cos\!\left(\sqrt{2\pi}\,\theta(x)\right),
\label{eq: order parameter Neel x bis}
\\
&
N^{\,}_{y}(x)\:=\sin\!\left(\sqrt{2\pi}\,\theta(x)\right),
\label{eq: order parameter Neel y bis}
\\
&
N^{\,}_{z}(x)\:=\sin\!\left(\sqrt{2\pi}\,\phi(x)\right).
\label{eq: order parameter Neel z bis}
\end{align}
The field whose non-vanishing ground-state expectation value
signal VBS (dimer) long-range order is%
~\cite{Hikihara2017}
\begin{equation}
D(x)\:=\cos\!\left(\sqrt{2\pi}\,\phi(x)\right).
\label{eq: order parameter VBS bis}
\end{equation}
\end{subequations}

The $\pi$-rotations $R^{x}_{\pi}$, $R^{y}_{\pi}$, and $R^{z}_{\pi}$
in spin space,
the lattice translation $T$, the site inversion $P$,
and the (anti-unitary) time-reversal $\Theta$ act on the bosonic fields as 
\begin{subequations}
\label{TPThetaR}
\begin{align}
R^{x}_{\pi}&: (\phi,\theta)\mapsto(-\phi,-\theta),
\label{Rxpi on (phi,theta)}\\
R^{y}_{\pi}&: (\phi,\theta)\mapsto(-\phi,\sqrt{\pi/2}-\theta),
\label{Rypi on (phi,theta)}\\
R^{z}_{\pi}&: (\phi,\theta)\mapsto(\phi,\theta+\sqrt{\pi/2}\,),
\label{Rzpi on (phi,theta)}\\
T&: (\phi,\theta)\mapsto(\phi+\sqrt{\pi/2},\theta+\sqrt{\pi/2}\,),
\label{T on (phi,theta)}\\
P&: (\phi,\theta)\mapsto(-\phi+\sqrt{\pi/2},\theta),
\label{P on (phi,theta)}\\
\Theta&: (\phi,\theta)\mapsto(-\phi,\theta+\sqrt{\pi/2}),
\label{Theta on (phi,theta)}
\end{align}
\end{subequations}
respectively.  The U(1) spin rotation symmetry about the $z$ axis of
the quantum spin-1/2 antiferromagnetic $J^{\,}_{1}$-$J^{\,}_{2}$ XXZ chain
is generated by the infinitesimal transformation
\begin{equation}
(\phi,\theta)\mapsto(\phi,\theta+\delta\theta).
\label{eq: infinitesimal U(1) symmetry}
\end{equation}

\begin{figure}[t!]
\begin{center}
\includegraphics[width=0.4\textwidth]{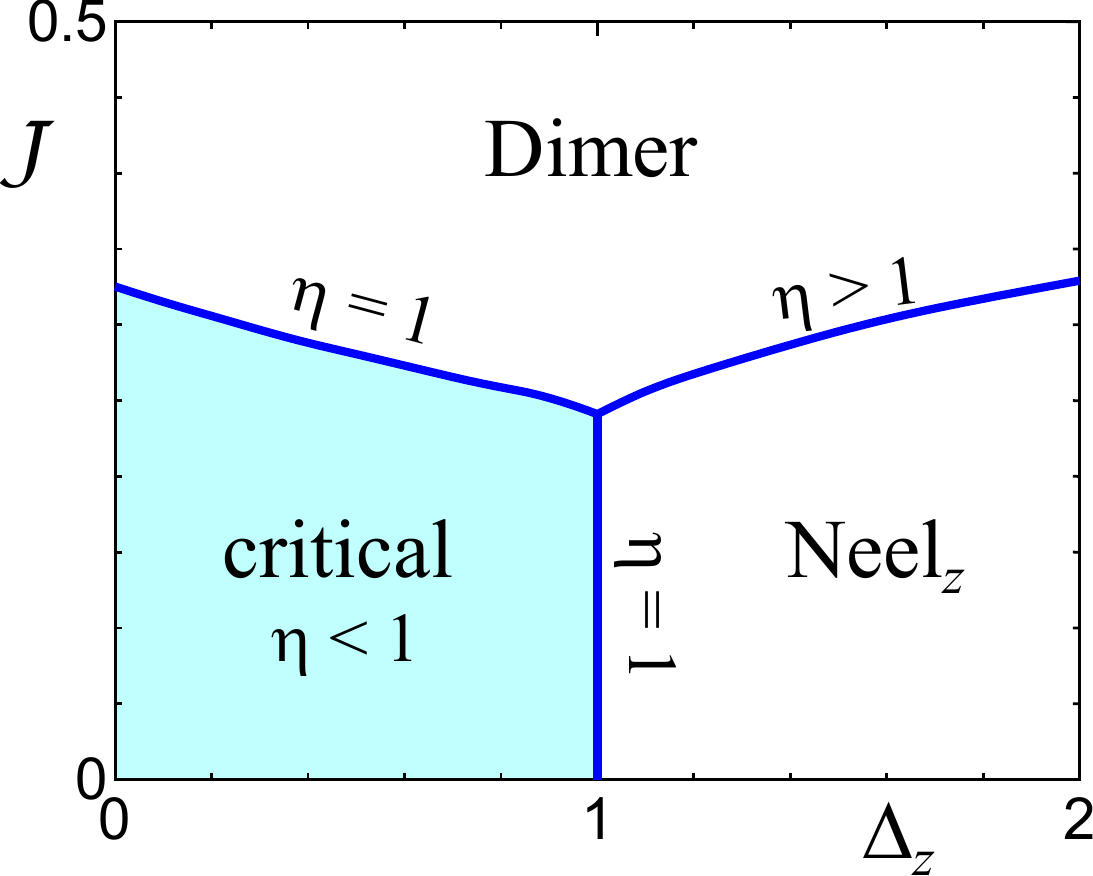}
\caption{(Color online)
Ground-state phase diagram of the
quantum spin-1/2 antiferromagnetic $J^{\,}_{1}$-$J^{\,}_{2}$ XXZ Hamiltonian
(\ref{eq: def XXZ model c})
with the Ising exchange anisotropy $\Delta^{\,}_{z}$
and the exchange ratio
$\mathcal{J}=J^{\,}_{2}/J^{\,}_{1}$.
The phase boundaries are taken from Ref.\ \onlinecite{Nomura1994}.
When $0\le\Delta^{\,}_{z}<1$ and
$0\le\mathcal{J}<\mathcal{J}^{\,}_{\mathrm{c}}(\Delta^{\,}_{z})$,
the system is in the critical phase with in-plane spin-correlation exponent
$\eta<1$ ($\eta=1$ on the boundary of the critical phase).  The Umklapp
coupling vanishes along the phase boundary with $\eta>1$
separating the gapped dimer (VBS) and Neel${}^{\,}_{z}$ phases that break
different $\mathbb{Z}^{\,}_{2}$ symmetries. 
\label{Fig: haldane PRB 82 fig2}
         }
\end{center}
\end{figure}

Figure \ref{Fig: haldane PRB 82 fig2} is a schematic picture of the
ground-state phase diagram of the
quantum spin-1/2 antiferromagnetic $J^{\,}_{1}$-$J^{\,}_{2}$ XXZ chain
obtained by Haldane. The phase diagram supports a critical phase
(extended over a finite region of parameter space)
and two gapped phases. These long-range ordered phases
are separated by a phase boundary that realizes
a line of quantum critical points, each of which realizes a
$c=1$ conformal field theory with U(1) symmetry.

The critical phase defined by
\begin{equation}
0\leq\Delta^{\,}_{z}<1,
\qquad
0\leq\mathcal{J}<\mathcal{J}^{\,}_{\mathrm{c}}(\Delta^{\,}_{z}),
\label{eq: def TL phase XXZ j1-J2}
\end{equation}
is governed by the Gaussian Hamiltonian density
$\mathcal{H}^{\,}_{\eta}$ defined by Eq.\
(\ref{eq: Gaussian fixed point}) with $\eta<1$.
The parameter $\eta$ takes the value $\eta=1/2$ at the
free-fermion point $\mathcal{J}=\Delta^{\,}_{z}=0$ and continuously
increases until $\eta=1$, the value of which defines
the phase boundaries to the gapped phases.
The cosine term $\lambda^{\,}_{\phi}\,\cos(\sqrt{8\pi}\,\phi)$ in the
effective theory (\ref{eq: sine-Gordon rep HXXZ}) has scaling
dimension $2/\eta>2$. It is thus
an irrelevant perturbation to the Gaussian Hamiltonian
(\ref{eq: Gaussian fixed point}).
This critical phase is characterized by quasi-long-range order
for all correlation functions of local operators. In particular,
the two-point functions
$\langle N^{\,}_{x}(x)\,N^{\,}_{x}(0)\rangle$
and
$\langle N^{\,}_{y}(x)\,N^{\,}_{y}(0)\rangle$
show the slowest decay proportional to $|x|^{-\eta}$,
whereas
$\langle N^{\,}_{z}(x)\,N^{\,}_{z}(0)\rangle$
and
$\langle D(x)\,D(0)\rangle$
decay like $|x|^{-1/\eta}$.
The isotropy in the decay of the correlation functions
$\langle N^{\,}_{x}(x)\,N^{\,}_{x}(0)\rangle$
and
$\langle N^{\,}_{y}(x)\,N^{\,}_{y}(0)\rangle$
is a consequence of the global U(1) symmetry
(\ref{eq: infinitesimal U(1) symmetry})
that the Gaussian Hamiltonian (\ref{eq: Gaussian fixed point}) enjoys.
At $\eta=1$, all four two-point functions decay like $|x|^{-1}$.
The isotropy in the decay of the correlation functions
$\langle N^{\,}_{x}(x)\,N^{\,}_{x}(0)\rangle$,
$\langle N^{\,}_{y}(x)\,N^{\,}_{y}(0)\rangle$,
$\langle N^{\,}_{z}(x)\,N^{\,}_{z}(0)\rangle$
is a consequence of the hidden (non-manifest) global SU(2) symmetry
of $\mathcal{H}^{\,}_{\eta}$ when $\eta=1$.
It follows that the critical points on the upper edge
of the critical phase (\ref{eq: def TL phase XXZ j1-J2})
shown in Fig.~\ref{Fig: haldane PRB 82 fig2}
has a global SU(2) symmetry.

Outside the critical phase (\ref{eq: def TL phase XXZ j1-J2}),
the cosine term
$\lambda^{\,}_{\phi}\cos(\sqrt{8\pi}\,\phi)$ is relevant and opens an
energy gap. The resulting ground state is either the Neel${}^{\,}_{z}$
phase or the dimer phase, depending on the sign of the coupling
constant $\lambda^{\,}_{\phi}$.  If $\lambda^{\,}_{\phi}<0$, then the
cosine term pins the $\phi$ field at $\phi=0$ or $\sqrt{\pi/2}$ (mod
$\sqrt{2\pi}$), and the dual field $\theta$ is disordered.  This leads
to $D=+1$ or $-1$ (if we ignore quantum fluctuations for simplicity),
and $N^{\,}_{x}=N^{\,}_{y}=N^{\,}_{z}=0$; the ground state is in the dimer phase.
If $\lambda^{\,}_{\phi}>0$, then the cosine term pins the $\phi$ field at
$\phi=\sqrt{\pi/8}$ or $3\sqrt{\pi/8}$ (mod $\sqrt{2\pi}$), and the
dual field $\theta$ is disordered.  This leads to $N^{\,}_{z}=+1$ or $-1$
(if we ignore quantum fluctuations for simplicity), and
$D=N^{\,}_{x}=N^{\,}_{y}=0$; the
ground state is in the Neel${}^{\,}_{z}$ phase.

The phase transition between the dimer phase and the Neel${}^{\,}_{z}$
phase for $\Delta^{\,}_{z}>1$ is determined by the condition
$\lambda^{\,}_{\phi}=0$; see Fig.~\ref{Fig: haldane PRB 82 fig2}.
The critical theory at the phase transition is the $c=1$
Gaussian Hamiltonian (\ref{eq: Gaussian fixed point})
with $\eta>1$.%
~\cite{Haldane1982}
At any dimer-Neel${}^{\,}_{z}$ critical point, the two-point functions
$\langle N^{\,}_{z}(x)\,N^{\,}_{z}(0)\rangle$
and
$\langle D(x)\,D(0)\rangle$
show the slowest algebraic decay $\sim|x|^{-1/\eta}$,
whereas
$\langle N^{\,}_{x}(x)\,N^{\,}_{x}(0)\rangle$
and
$\langle N^{\,}_{y}(x)\,N^{\,}_{y}(0)\rangle$
decay like $\sim|x|^{-\eta}$.

Along the SU(2) invariant line $\Delta^{\,}_{z}=1$
the phase transition from the critical Tomonaga-Luttinger liquid phase
to the dimer phase is known to occur at%
~\cite{Okamoto1992,Eggert1996}
\begin{equation}
\mathcal{J}=\mathcal{J}^{\star}_{\mathrm{c}}\equiv
\mathcal{J}^{\,}_{\mathrm{c}}(\Delta^{\,}_{z}=1)=
0.2411\ldots.
\label{eq: def cal J star c}
\end{equation}

We now examine the effects of breaking the U(1) symmetry
(\ref{eq:U(1) symmetry XXZ}) with small
$|\Delta^{\,}_{y}-1|>0$. Following the steps of the Jordan-Wigner
transformation and bosonization, we find that the deviation of
$\Delta^{\,}_{y}$ from unity yields the perturbation
\begin{align}
(1-\Delta^{\,}_{y})\big(S^{x}_{l}\,S^{x}_{l+1}-S^{y}_{l}\,S^{y}_{l+1}\big)
&
=
\frac{1-\Delta^{\,}_{y}}{2}\big(S^{+}_{l}\,S^{+}_{l+1}+\mathrm{H.c.}\big)
\nonumber\\
&
\approx
a^{2}_{2}\left(\Delta^{\,}_{y}-1\right)\cos(\sqrt{8\pi}\,\theta),
\label{eq: effective pertubation XXZ to XYZ}
\end{align}
which should be added to the effective theory
(\ref{eq: sine-Gordon rep HXXZ}).
The operator $\cos(\sqrt{8\pi}\,\theta)$
is invariant under the transformations (\ref{TPThetaR})
and has the scaling dimension $2\eta$.
It is thus a relevant perturbation to the
Gaussian Hamiltonian (\ref{eq: Gaussian fixed point})
with $\eta<1$ in the critical phase
\begin{equation}
0\le\Delta^{\,}_{z}<1,
\qquad  
0\leq\mathcal{J}<\mathcal{J}^{\,}_{\mathrm{c}}(\Delta^{\,}_{z}),
\qquad
\Delta^{\,}_{y}=1.
\label{eq: def critical plane in R3}
\end{equation}
As such, the potential $(\Delta^{\,}_{y}-1)\cos(\sqrt{8\pi}\,\theta)$ pins
the $\theta$ field at $\theta=0$ or $\sqrt{\pi/2}$ (mod $\sqrt{2\pi}$)
for $\Delta^{\,}_{y}<1$ and at $\theta=\sqrt{\pi/8}$ or
$3\sqrt{\pi/8}$ (mod $\sqrt{2\pi}$) for $\Delta^{\,}_{y}>1$.  This
means that the critical phase (\ref{eq: def critical plane in R3})
is located exactly on the boundary between the
Neel${}^{\,}_{x}$ phase at $\Delta^{\,}_{y}<1$ and the Neel${}^{\,}_{y}$
phase at $\Delta^{\,}_{y}>1$.  On the other hand, the
$\cos(\sqrt{8\pi}\,\theta)$ operator is an irrelevant perturbation in
both the dimer phase and the Neel${}^{\,}_{z}$ phase where the dual
$\phi$ field is pinned by the $\cos(\sqrt{8\pi}\,\phi)$ perturbation.
More importantly, the $\cos(\sqrt{8\pi}\,\theta)$ perturbation is an
irrelevant perturbation to the Gaussian Hamiltonian
(\ref{eq: Gaussian fixed point}) with
$\eta>1$ on the phase boundary between the Neel${}^{\,}_{z}$ and the dimer
phase. Hence, the Neel${}^{\,}_{z}$-dimer phase boundary is a
two-dimensional surface
of $c=1$ Gaussian criticality that extends out of the plane
$\Delta^{\,}_{y}=1$.

It turns out that the criticality on the phase boundary between the
dimer and Neel${}^{\,}_{x}$ phases or between the dimer and
Neel${}^{\,}_{y}$ phases for $\Delta^{\,}_{y}\ne1$ is also described by
the Gaussian Hamiltonian (\ref{eq: Gaussian fixed point}).
On the upper edge of the critical phase (\ref{eq: def critical plane in R3}),
whose low-energy theory is the Gaussian model
(\ref{eq: Gaussian fixed point})
with $\eta=1$, both $\cos(\sqrt{8\pi}\,\phi)$ and
$\cos(\sqrt{8\pi}\,\theta)$ are marginal operators with scaling
dimension 2.
The competition between these two dual operators is
known \cite{Lecheminant2002} to yield a line of $c=1$ fixed points,
whose basin of attraction forms a critical plane of the
Neel${}^{\,}_{x}$-dimer and Neel${}^{\,}_{y}$-dimer phase boundaries.
We will discuss the criticality between gapped phases in more detail
below.

\subsection{Global phase diagram}
\label{subsec: global phase diagram}

We deduce the global phase diagram and criticality of the
quantum spin-1/2 antiferromagnetic $J^{\,}_{1}$-$J^{\,}_{2}$ XYZ chain
(\ref{eq: def HXYZ})
from the analysis of the effective sine-Gordon Hamiltonian density 
(\ref{eq: sine-Gordon rep HXXZ})
perturbed by (\ref{eq: effective pertubation XXZ to XYZ}).
This prediction is validated numerically
in Sec.\ \ref{subsec: Numerical results}.
Parameter space for this phase diagram is the three-dimensional slab
\begin{equation}
\Delta^{\,}_{y}\geq0,
\qquad
\Delta^{\,}_{z}\geq0,
\qquad
0\leq\mathcal{J}\leq1/2,
\label{eq: 3D parameter space for global phase diagram}
\end{equation}
of $\mathbb{R}^{3}$.
The global phase diagram for the ground states 
of Hamiltonian (\ref{eq: def HXYZ})
is symmetric:\\

\noindent
(1) about the plane defined by $\Delta^{\,}_{y}=\Delta^{\,}_{z}$
in the three-dimensional parameter space
(\ref{eq: 3D parameter space for global phase diagram}),\\

\noindent
(2)
under cyclic permutations of the
indices $x$, $y$, and $z$ entering either the anisotropies
$\Delta^{\,}_{x}$,
$\Delta^{\,}_{y}$,
and
$\Delta^{\,}_{z}$,
where we have fixed $\Delta^{\,}_{x}\equiv1$,
or the Neel phases
Neel${}^{\,}_{x}$,
Neel${}^{\,}_{y}$,
and Neel${}^{\,}_{z}$.\\

We define
$\mathcal{J}^{\,}_{\mathrm{c}}(\Delta^{\,}_{y},\Delta^{\,}_{z})$ 
to be the critical value of $\mathcal{J}\equiv J^{\,}_{2}/J^{\,}_{1}$
above which the ground state is in the dimer phase.
By symmetry
\begin{equation}
\mathcal{J}^{\,}_{\mathrm{c}}(\Delta^{\,}_{y},\Delta^{\,}_{z})=
\mathcal{J}^{\,}_{\mathrm{c}}(\Delta^{\,}_{z},\Delta^{\,}_{y}).
\end{equation}

The phase diagram on the $\Delta^{\,}_{z}=1$ plane in the parameter space
(\ref{eq: 3D parameter space for global phase diagram})
follows from the phase diagram on the
$\Delta^{\,}_{y}=1$ plane in the parameter space
(\ref{eq: 3D parameter space for global phase diagram})
that we derived in Sec.\ \ref{sec: Delta_{y}=1} by interchanging
the Neel${}^{\,}_{y}$ and Neel${}^{\,}_{z}$  phases.
The phase diagram on the $\Delta^{\,}_{z}=1$ plane has thus three phases:\\

\noindent(i)
\underline{the $c=1$ critical phase}
with quasi-long-range order of the easy-plane Neel correlations
$(N^{\,}_{x}, N^{\,}_{z})$ for
\begin{subequations}
\label{eq: phase diagram Deltaz=1 plane}
\begin{equation}
0\leq\mathcal{J}<\mathcal{J}^{\,}_{\mathrm{c}}(\Delta^{\,}_{y},1),
\qquad
0\le\Delta^{\,}_{y}\le1,
\end{equation}\\

\noindent(ii)
\underline{the Neel${}^{\,}_{y}$ phase} for
\begin{equation}
0\leq\mathcal{J}<\mathcal{J}^{\,}_{\mathrm{c}}(\Delta^{\,}_{y},1),
\qquad
\Delta^{\,}_{y}>1,
\end{equation}\\

\noindent (iii)
\underline{the dimer phase} for
\begin{equation}
\mathcal{J}>\mathcal{J}^{\,}_{\mathrm{c}}(\Delta^{\,}_{y},1).
\end{equation}
\end{subequations}
The phase boundary between
the Neel${}^{\,}_{y}$ phase and the dimer phase is a line of $c=1$
critical points with $\eta>1$.

Similarly, the phase diagram on the plane defined by
$\Delta^{\,}_{y}=\Delta^{\,}_{z}$ is also obtained from the phase
diagram on the $\Delta^{\,}_{y}=1$ plane by replacing
$\Delta^{\,}_{z}$ with $1/\Delta^{\,}_{z}$ in the horizontal axis in
Fig.~\ref{Fig: haldane PRB 82 fig2} and by exchanging the
Neel${}^{\,}_{z}$ and Neel${}^{\,}_{x}$ phases. By this logic,
there are three phases on
the $\Delta^{\,}_{z}=\Delta^{\,}_{y}$ plane:\\

\noindent
(i) \underline{the $c=1$ critical phase}
with quasi-long-range order of the easy-plane Neel correlations
$(N^{\,}_{z},N^{\,}_{y})$ for
\begin{subequations}
\label{eq: phase diagram Deltaz=Deltay plane}
\begin{equation}
0\leq\mathcal{J}<\mathcal{J}^{\,}_{\mathrm{c}}(\Delta^{\,}_{z},\Delta^{\,}_{z}),
\qquad
\Delta^{\,}_{z}>1,
\end{equation}\\

\noindent
(ii) \underline{the Neel${}^{\,}_{x}$ phase} for
\begin{equation}
0\leq\mathcal{J}<\mathcal{J}^{\,}_{\mathrm{c}}(\Delta^{\,}_{z},\Delta^{\,}_{z})
\qquad
0\leq\Delta^{\,}_{z}<1,
\end{equation}\\

\noindent
(iii) \underline{the dimer phase} for
\begin{equation}
\mathcal{J}>\mathcal{J}^{\,}_{\mathrm{c}}(\Delta^{\,}_{z},\Delta^{\,}_{z}).
\end{equation}
\end{subequations}

Examples
(\ref{eq: phase diagram Deltaz=1 plane})
and
(\ref{eq: phase diagram Deltaz=Deltay plane})
illustrate that,
whereas the condition
\begin{equation}
\mathcal{J}>\mathcal{J}^{\,}_{\mathrm{c}}(\Delta^{\,}_{y},\Delta^{\,}_{z})
\end{equation}
always selects the dimer phase in the parameter space
(\ref{eq: 3D parameter space for global phase diagram}),
the condition
\begin{equation}
\mathcal{J}<\mathcal{J}^{\,}_{\mathrm{c}}(\Delta^{\,}_{y},\Delta^{\,}_{z})
\end{equation}
selects either one of the three Neel phases
or the critical manifold separating them.
Which one of the Neel phase is selected depends
on which of the anisotropies $\Delta^{\,}_{y}$ or $\Delta^{\,}_{z}$
is the largest:\\

\begin{figure}[t]
\includegraphics[width=0.4\textwidth]{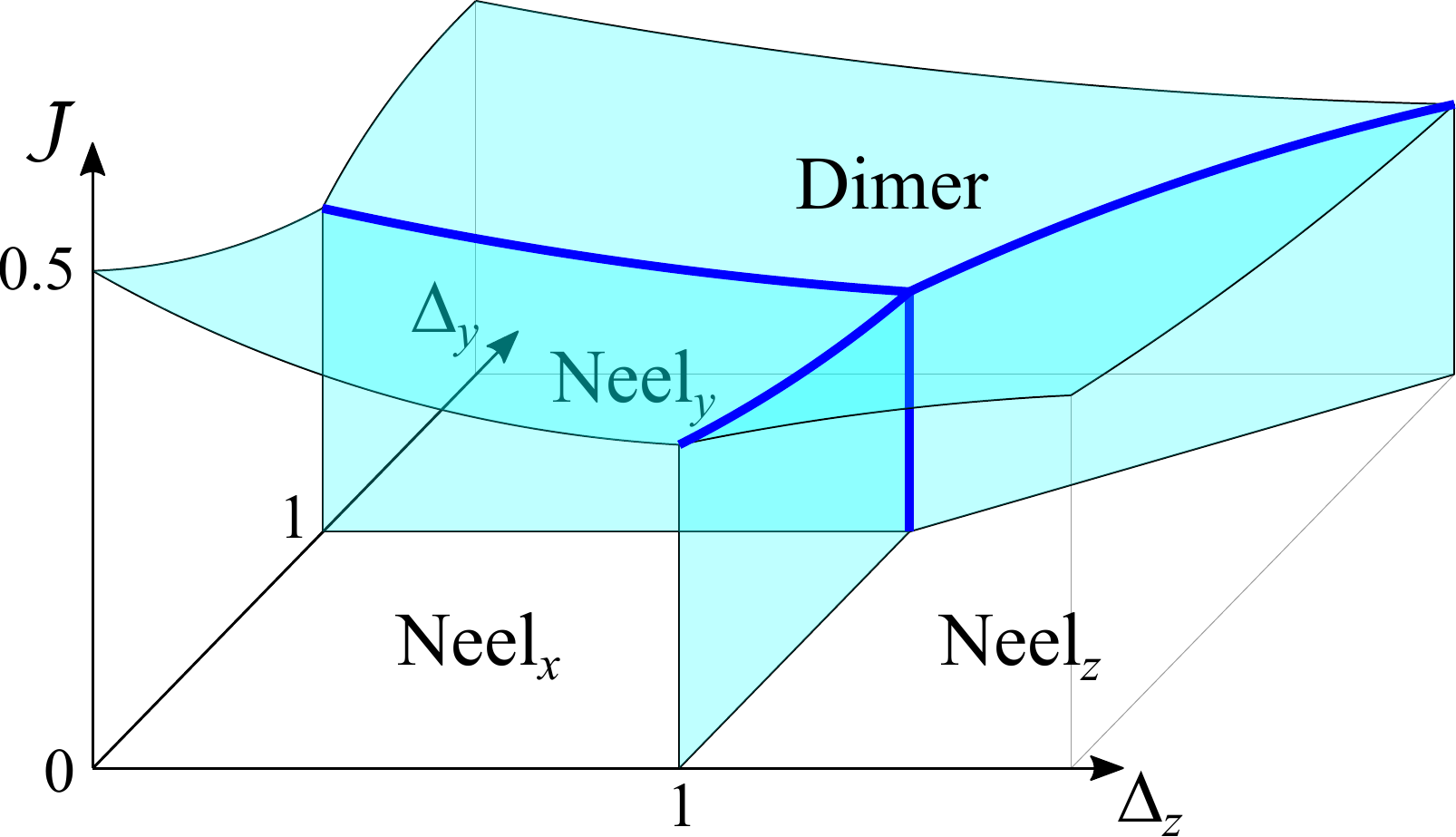}
\caption{(Color online)
Qualitative phase diagram of the
quantum spin-1/2 antiferromagnetic $J^{\,}_{1}$-$J^{\,}_{2}$ XYZ model
in the three-dimensional
parameter space $(\Delta^{\,}_{z},\Delta^{\,}_{y},\mathcal{J})$.
The three Neel phases and the dimer phase are separated by six
planes of phase boundaries, each of which realizes
Gaussian criticality with U(1)$\times$U(1) symmetry.
These six planes join at the four solid blue lines, along 
which the fixed-point theory is the Gaussian model with SU(2) symmetry.
The dimer phase exists for
$\mathcal{J}>\mathcal{J}_c^{\,}(\Delta^{\,}_{y},\Delta^{\,}_{z})$.
The three Neel${}^{\,}_{\alpha}$ phases with $\alpha=x,y,z$
are located below the phase boundaries of the Gaussian criticality,
$\mathcal{J}<\mathcal{J}_c^{\,}(\Delta^{\,}_{y},\Delta^{\,}_{z})$.
The pair of Neel${}^{\,}_{\alpha}$ and Neel${}^{\,}_{\beta}$ phases
with $\alpha<\beta=x,y,z$ are separated by the phase boundary at 
$\Delta^{\,}_{\alpha}=\Delta^{\,}_{\beta}$ of the Gaussian criticality
($\Delta^{\,}_{x}\equiv1$).
We have omitted other phases (such as the UUDD Ising ordered phase)
that can exist for $\mathcal{J}>1/2$.
\label{fig: 3D phase diagram}
        }
\end{figure}

\noindent
(a) \underline{the Neel${}^{\,}_{x}$ phase} is selected when
\begin{subequations}
\begin{equation}
\Delta^{\,}_{y}<1,
\qquad
\Delta^{\,}_{z}<1,
\end{equation}\\

\noindent
(b) \underline{the Neel${}^{\,}_{y}$ phase} is selected when
\begin{equation}
\Delta^{\,}_{y}>1,
\qquad
\Delta^{\,}_{y}>\Delta^{\,}_{z},
\end{equation}\\

\noindent
(c) \underline{the Neel${}^{\,}_{z}$ phase} is selected when
\begin{equation}
\Delta^{\,}_{z}>1,
\qquad
\Delta^{\,}_{z}>\Delta^{\,}_{y}.
\end{equation}
\end{subequations}
The phase boundaries between these three
Neel phases are the $c=1$ critical phases located on the
$\Delta^{\,}_{y}=1$ plane, $\Delta^{\,}_{z}=1$ plane, or
$\Delta^{\,}_{y}=\Delta^{\,}_{z}$ plane, which cross at the SU(2)
symmetric line $\Delta^{\,}_{y}=\Delta^{\,}_{z}=1$.
A schematic picture of the phase diagram in the three-dimensional
parameter space $(\Delta^{\,}_{y},\Delta^{\,}_{z},\mathcal{J})$
is shown in Fig.~\ref{fig: 3D phase diagram}.

The criticality at the Neel-dimer and Neel-Neel phase transitions can
be studied perturbatively near a special point along the SU(2) symmetric
segment [recall Eq.\ (\ref{eq: def cal J star c})]
\begin{equation}
\Delta^{\,}_{y}=\Delta^{\,}_{z}=1,
\qquad
0\leq\mathcal{J}\leq\mathcal{J}^{\star}_{\mathrm{c}},
\qquad
\mathcal{J}^{\star}_{\mathrm{c}}\:=\mathcal{J}^{\,}_{\mathrm{c}}(1,1).
\end{equation}
At the special point
\begin{equation}
\Delta^{\,}_{y}=\Delta^{\,}_{z}=1,
\qquad
\mathcal{J}=\mathcal{J}^{\star}_{\mathrm{c}},
\label{eq: def special point}
\end{equation}
the low-energy theory is the
Gaussian Hamiltonian density with $\eta=1$, i.e.,
\begin{equation}
\mathcal{H}^{\,}_{0}\equiv
\frac{1}{2}
\left[(\partial^{\,}_{x}\theta)^{2}+(\partial^{\,}_{x}\phi)^{2}\right],
\label{eq: def H0}
\end{equation}
where we have set $v=1$ for simplicity.
Away from this special point,
the effective Hamiltonian density
is perturbed by local operators that are invariant under the
symmetry transformations (\ref{TPThetaR}). Among all such local operators,
the less irrelevant ones at the special point
(\ref{eq: def special point})
in parameter space
(\ref{eq: 3D parameter space for global phase diagram})
are the three marginal operators
\begin{equation}
\cos(\sqrt{8\pi}\,\phi),
\quad
\cos(\sqrt{8\pi}\,\theta),
\quad
(\partial^{\,}_{x}\phi)^{2}-(\partial^{\,}_{x}\theta)^{2},
\end{equation}
for they all share the scaling dimension 2 when $\eta=1$.
They are related to the chiral generators of the su(2)${}^{\,}_{1}$ affine
Lie algebra by
\begin{subequations}
\begin{align}
&
J^{\pm}_{\mathrm{L}}\:=
\frac{1}{\mathsf{a}}\,  
e^{\pm\mathrm{i}\sqrt{2}\phi^{\,}_{\mathrm{L}}},
\qquad
J^{z}_{\mathrm{L}}\:=
\frac{1}{\sqrt2}\,\partial^{\,}_{x}\phi^{\,}_{\mathrm{L}},
\\
&
J^{\pm}_{\mathrm{R}}\:=
\frac{1}{\mathsf{a}}\,  
e^{\mp\mathrm{i}\sqrt{2}\phi^{\,}_{\mathrm{R}}},
\qquad
J^{z}_{\mathrm{R}}\:=
\frac{1}{\sqrt2}\,\partial^{\,}_{x}\phi^{\,}_{\mathrm{R}},
\end{align}
where we have introduced yet a second pair of
left- and right-moving chiral bosonic fields, namely
\begin{equation}
\phi^{\,}_{\mathrm{L}}(x)\:=\sqrt{\pi}[\phi(x)+\theta(x)],
\qquad
\phi^{\,}_{\mathrm{R}}(x)\:=\sqrt{\pi}[\phi(x)-\theta(x)].
\end{equation}
\end{subequations}
Here, we chose the normalization convention
for the left-and right-moving currents $J^{\pm}_{\mathrm{M}}$ with
$\mathrm{M}=\mathrm{L},\mathrm{R}$
such that $\langle J^{+}_{\mathrm{M}}(x)\,J^{-}_{\mathrm{M}}(0)\rangle=-1/x^{2}$
for the SU(2) symmetric Gaussian Hamiltonian (\ref{eq: def H0}).
The microscopic origin
of the components $J^{\pm}_{\mathrm{M}}(x)$ and $J^{z}_{\mathrm{M}}(x)$
is the following.
The non-oscillating components of the spin operators $S^{z}_{l}$ and
$S^{+}_{l}$ in Eqs.\
(\ref{eq: relation lattice quantum spins and bosonic quantum fields a})
and
(\ref{eq: relation lattice quantum spins and bosonic quantum fields b})
are equal to $\mathfrak{a}/\sqrt{\pi}$ times
$J^{z}_{\mathrm{L}}+J^{z}_{\mathrm{R}}$
and
$J^{+}_{\mathrm{L}}+J^{+}_{\mathrm{R}}$, respectively, at $\eta=1$.
If we define the linear combinations
\begin{subequations}
\label{eq: def H_JJ}
\begin{equation}
J^{x}_{\mathrm{M}}\:=\frac{J^{+}_{\mathrm{M}}+J^{-}_{\mathrm{M}}}{2},
\qquad
J^{y}_{\mathrm{M}}\:=\frac{J^{+}_{\mathrm{M}}-J^{-}_{\mathrm{M}}}{2\mathrm{i}},
\qquad
\mathrm{M}=\mathrm{L},\mathrm{R},
\label{eq: def H_JJ a}
\end{equation}
we may then define the current-current interaction density\cite{Gogolin}
\begin{align}
\mathcal{H}^{\,}_{JJ}\:=&\,
\lambda^{\,}_{x}\,
J^{x}_{\mathrm{L}}\,
J^{x}_{\mathrm{R}}
+
\lambda^{\,}_{y}\,
J^{y}_{\mathrm{L}}\,
J^{y}_{\mathrm{R}}
+
\lambda^{\,}_{z}\,
J^{z}_{\mathrm{L}}\,
J^{z}_{\mathrm{R}}
\nonumber\\
=&\,
-
\frac{1}{\mathsf{a}^{2}}
\left(\lambda^{\,}_{x}-\lambda^{\,}_{y}\right)
\cos(\sqrt{8\pi}\,\theta)
\nonumber\\
&
-
\frac{1}{\mathsf{a}^{2}}
\left(\lambda^{\,}_{x}+\lambda^{\,}_{y}\right)
\cos(\sqrt{8\pi}\,\phi)
\nonumber\\
&
-
\frac{\pi\lambda^{\,}_{z}}{2}
\left[(\partial^{\,}_{x}\theta)^{2}-(\partial^{\,}_{x}\phi)^{2}\right],
\label{eq: def H_JJ b}
\end{align}
\end{subequations}
where the real-valued couplings
$\lambda^{\,}_{x}$, $\lambda^{\,}_{y}$, and $\lambda^{\,}_{z}$
are dimensionless.

The perturbed Hamiltonian density
$
\mathcal{H}^{\,}_{0}+\mathcal{H}^{\,}_{JJ}
$
should be compared with the perturbed Hamiltonian density
$
\mathcal{H}^{\,}_{\mathrm{XXZ}}+
\frac{A}{\mathsf{a}^{2}}
(\Delta^{\,}_{y}-1)\cos(\sqrt{8\pi}\,\theta),
$
where $A$ is a non-universal positive number of order one,
by demanding that
\begin{equation}
\mathcal{H}^{\,}_{0}+\mathcal{H}^{\,}_{JJ}=
\mathcal{H}^{\,}_{\mathrm{XXZ}}
+
\frac{A}{\mathsf{a}^{2}}\,
(\Delta^{\,}_{y}-1)\cos(\sqrt{8\pi}\,\theta).  
\end{equation}
By matching the couplings of
$\cos(\sqrt{8\pi}\,\theta)$
on both side of this equation and using
the symmetry under cyclic permutations of the indices $x$, $y$, and $z$
with $\Delta^{\,}_{x}\equiv1$, we infer that
\begin{subequations}
\label{eq: relating lambdas to Deltas}
\begin{align}
&
\lambda^{\,}_{x}-\lambda^{\,}_{y}=A\,(1-\Delta^{\,}_{y}),
\label{eq: relating lambdas to Deltas a}\\
&
\lambda^{\,}_{y}-\lambda^{\,}_{z}=A\,(\Delta^{\,}_{y}-\Delta^{\,}_{z}),
\label{eq: relating lambdas to Deltas b}\\
&
\lambda^{\,}_{z}-\lambda^{\,}_{x}=A\,(\Delta^{\,}_{z}-1).
\label{eq: relating lambdas to Deltas c}
\end{align}
\end{subequations}
Furthermore, by matching the coefficients of
$(\partial^{\,}_{x}\theta)^{2}$
and
$(\partial^{\,}_{x}\phi)^{2}$
on both sides of this equation, we deduce that
\begin{subequations}
\begin{equation}
\eta=\sqrt{\frac{1+\pi\lambda^{\,}_{z}}{1-\pi\lambda^{\,}_{z}}}.
\end{equation}
In particular,
\begin{equation}
\eta\approx
1+\pi\lambda^{\,}_{z}
\end{equation}
\end{subequations}
when $|\lambda^{\,}_{z}|\ll1$. Finally,
remembering that the scaling dimensions of
$\cos(\sqrt{8\pi}\,\theta)$
and
$\cos(\sqrt{8\pi}\,\phi)$
in the Gaussian theory (\ref{eq: Gaussian fixed point})
are
\begin{subequations}
\begin{equation}
2\eta\approx 2+2\pi\,\lambda^{\,}_{z},
\qquad |\lambda^{\,}_{z}|\ll1,
\end{equation}
and 
\begin{equation}
\frac{2}{\eta}\approx 2-2\pi\,\lambda^{\,}_{z},
\qquad |\lambda^{\,}_{z}|\ll1,
\end{equation}
\end{subequations}
respectively, we deduce from
the renormalization-group (RG) equation
$\mathrm{d}O/\mathrm{d}\ell=(2-d^{\,}_{O})\,O$
in (1+1)-dimensional spacetime,
where $\mathrm{d}\ell=\mathrm{d}\ln\mathfrak{a}$
and $O$ is an operator whose exact scaling dimension is $d^{\,}_{O}$,
the one-loop RG flows
\begin{subequations}
\begin{align}
&
\frac{\mathrm{d}(\lambda^{\,}_{x}-\lambda^{\,}_{y})}{\mathrm{d}\ell}=
-2\pi\lambda^{\,}_{z}(\lambda^{\,}_{x}-\lambda^{\,}_{y}),
\\
&
\frac{\mathrm{d}(\lambda^{\,}_{x}+\lambda^{\,}_{y})}{\mathrm{d}\ell}=
+2\pi\lambda^{\,}_{z}(\lambda^{\,}_{x}+\lambda^{\,}_{y}),
\end{align}
\end{subequations}
for
$|\lambda^{\,}_{x}-\lambda^{\,}_{y}|,
|\lambda^{\,}_{x}+\lambda^{\,}_{y}|,
|\lambda^{\,}_{z}|\ll1$.
With the help of cyclic permutations of the indices $x$, $y$, and $z$,
we thus obtain the one-loop RG equations
\begin{equation}
\frac{\mathrm{d}\lambda^{\,}_{x}}{\mathrm{d}\ell}=
2\pi\,\lambda^{\,}_{y}\,\lambda^{\,}_{z},
\quad
\frac{\mathrm{d}\lambda^{\,}_{y}}{\mathrm{d}\ell}=
2\pi\,\lambda^{\,}_{z}\,\lambda^{\,}_{x},
\quad
\frac{\mathrm{d}\lambda^{\,}_{z}}{\mathrm{d}\ell}=
2\pi\,\lambda^{\,}_{x}\,\lambda^{\,}_{y},
\label{eq: RG flows anisotropic JJ couplings}
\end{equation}
for the couplings $\lambda^{\,}_{x}$, $\lambda^{\,}_{y}$, $\lambda^{\,}_{z}$
of the current-current interaction density (\ref{eq: def H_JJ b})
at the Gaussian fixed point (\ref{eq: def H0}).

\begin{figure}
\includegraphics[width=0.35\textwidth]{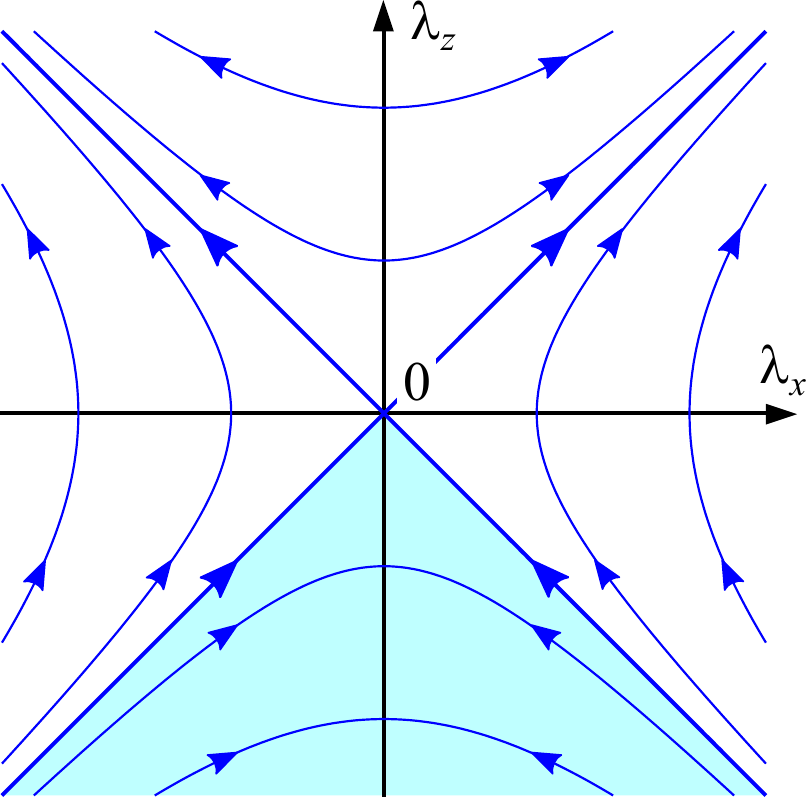}
\caption{(Color online)
Renormalization-group flow diagram of
Eq.\ (\ref{eq: RG flows anisotropic JJ couplings})
on the $\lambda^{\,}_{x}=\lambda^{\,}_{y}$ plane.
The shaded region corresponds to the critical phase in
Fig.~\ref{Fig: haldane PRB 82 fig2}.
\label{fig: RG flow lambda_x=lambda_y}
        }
\end{figure}

\begin{figure}
\includegraphics[width=0.5\textwidth]{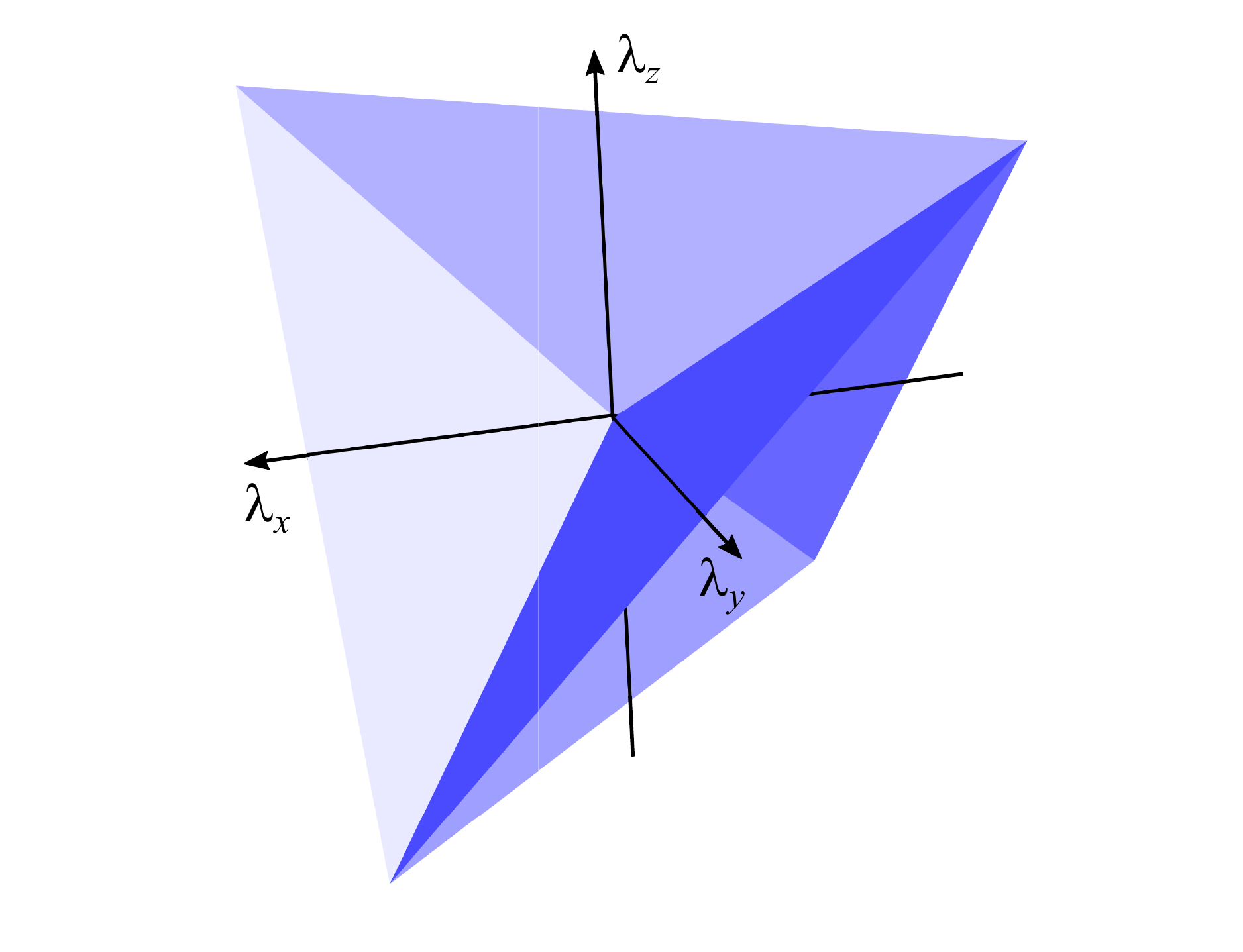}
\caption{(Color online)
Six critical planes separating four gapped phases in the three-dimensional
parameter space $(\lambda^{\,}_{x},\lambda^{\,}_{y},\lambda^{\,}_{z})$.
\label{fig: RG 3D}
        }
\end{figure}

The coupled flow equations
(\ref{eq: RG flows anisotropic JJ couplings})
have three lines of fixed points,
(i)
$\lambda^{\,}_{x}=\lambda^{\,}_{y}=0$,
(ii)
$\lambda^{\,}_{y}=\lambda^{\,}_{z}=0$,
and (iii)
$\lambda^{\,}_{z}=\lambda^{\,}_{x}=0$, where the fixed-point
Hamiltonian is given by Eq.\ (\ref{eq: Gaussian fixed point}).
Let's consider RG flows on the $\lambda^{\,}_{x}=\lambda^{\,}_{y}$
plane, for example; see Fig.~\ref{fig: RG flow lambda_x=lambda_y}.
On this plane the RG
flows in the region defined by
$\lambda^{\,}_{z}\le-|\lambda^{\,}_{x}|$ end up on $\lambda^{\,}_{z}$
axis with $\lambda^{\,}_{z}<0$.  The boundary of this critical region
is the diagonal $\lambda^{\,}_{x}=\pm\lambda^{\,}_{z}$ for
$\lambda^{\,}_{z}<0$ flowing to the origin
$\lambda^{\,}_{x}=\lambda^{\,}_{y}=\lambda^{\,}_{z}=0$.  The RG flows
on the $\lambda^{\,}_{x}=-\lambda^{\,}_{y}$ plane are the reverse of
those on the $\lambda^{\,}_{x}=\lambda^{\,}_{y}$ plane, and the
critical region on the $\lambda^{\,}_{x}=-\lambda^{\,}_{y}$ plane is
given by $\lambda^{\,}_{z}>|\lambda^{\,}_{x}|$.  Similar RG flows can
be obtained for the $\lambda^{\,}_{y}=\pm\lambda^{\,}_{z}$
plane and the $\lambda^{\,}_{z}=\pm\lambda^{\,}_{x}$ plane.  From
these considerations we find that the three-dimensional parameter
space $(\lambda^{\,}_{x},\lambda^{\,}_{y},\lambda^{\,}_{z})$ has six
critical planes on which the low-energy theory is the Gaussian Hamiltonian
(\ref{eq: Gaussian fixed point});
see Fig.\ \ref{fig: RG 3D}.  The
six critical planes form the boundaries of four gapped phases
corresponding to the Neel${}^{\,}_{x}$, Neel${}^{\,}_{y}$,
Neel${}^{\,}_{z}$, and dimer phases.

We note that the critical phase at $\Delta^{\,}_{z}<1$ in
Fig.~\ref{Fig: haldane PRB 82 fig2} corresponds to the critical region
$\lambda^{\,}_{z}<-|\lambda^{\,}_{x}|$ on the
$\lambda^{\,}_{x}=\lambda^{\,}_{y}$ plane.  The phase transition
between the Neel${}^{\,}_{z}$ and dimer phases at $\Delta^{\,}_{z}>1$
corresponds to the positive half of the $\lambda^{\,}_{z}$ axis. We
thus deduce that
\begin{subequations}
\begin{equation}
\lambda^{\,}_{z}=
b\,(\mathcal{J}-\mathcal{J}^{\star}_{\mathrm{c}})
+
c\,\left(\Delta^{\,}_{z}-\frac{1+\Delta^{\,}_{y}}{2}\right),
\end{equation}
where $\Delta^{\,}_{y}=1$ in Fig.~\ref{Fig: haldane PRB 82 fig2}
and $b$ and $c$ are positive constants.
By cyclic permutations of the indices
$x$, $y$, and $z$ with $\Delta^{\,}_{x}\equiv1$, we obtain
\begin{align}
&
\lambda^{\,}_{x}=
b\,(\mathcal{J}-\mathcal{J}^{\star}_{\mathrm{c}})
+
c\,\left(1-\frac{\Delta^{\,}_{y}+\Delta^{\,}_{z}}{2}\right),
\\ &
\lambda^{\,}_{y}=
b\,(\mathcal{J}-\mathcal{J}^{\star}_{\mathrm{c}})
+
c\,\left(\Delta^{\,}_{y}-\frac{\Delta^{\,}_{z}+1}{2}\right).
\end{align}
\end{subequations}
Consistency with Eq.\ (\ref{eq: relating lambdas to Deltas})
demands here that $c=2A/3$.

This perturbative RG analysis is justified for
$|\Delta^{\,}_{y}-1|, |\Delta^{\,}_{z}-1|,
|\mathcal{J}-\mathcal{J}^{\star}_{\mathrm{c}}|\ll1$.  However, we
expect that the global picture of the phase diagram and the $c=1$
Gaussian criticality (\ref{eq: Gaussian fixed point}) at the phase
boundaries should generally be valid beyond this perturbative regime.
Furthermore, the phase transitions between a Neel phase and a dimer
phase are the Gaussian criticality (\ref{eq: Gaussian fixed point})
with $\eta>1$ while the phase transitions between Neel phases are that
with $\eta<1$.
At a Neel${}^{\,}_{\alpha}$-dimer transition with $\alpha=x,y,z$
defined by the condition
\begin{subequations}
\label{scaling Neel-dimer}
\begin{equation}
\mathcal{J}=\mathcal{J}^{\,}_{\mathrm{c}}
\end{equation}
(the dependence of $\mathcal{J}^{\,}_{\mathrm{c}}$ on
$\Delta^{\,}_{y}$ and $\Delta^{\,}_{z}$ is implicit),
the dependencies on the length $L$ of the chain
for the Neel and dimer order parameters are power laws
with the same scaling exponent,
\begin{equation}
\langle N^{\,}_{\alpha}\rangle\sim\langle D\rangle\sim L^{-1/(2\eta)}.
\end{equation}
Moreover, in the thermodynamic limit $L\to\infty$,
their dependencies on $\mathcal{J}-\mathcal{J}^{\,}_{\mathrm{c}}$
are the power laws
\begin{align}
&
\langle N^{\,}_{\alpha}\rangle\sim
\left(\mathcal{J}^{\,}_{\mathrm{c}}-\mathcal{J}\right)^{1/[4(\eta-1)]}\,
\Theta\!\left(\mathcal{J}^{\,}_{\mathrm{c}}-\mathcal{J}\right),
\\
&
\langle D\rangle\sim
\left(\mathcal{J}-\mathcal{J}^{\,}_{\mathrm{c}}\right)^{1/[4(\eta-1)]}\,
\Theta\!\left(\mathcal{J}-\mathcal{J}^{\,}_{\mathrm{c}}\right),
\end{align}
\end{subequations}
where $\Theta(x)$ is the Heaviside function, respectively.
Similarly, at the Neel${}^{\,}_{\alpha}$-Neel${}^{\,}_{\beta}$ transition
with $\alpha<\beta=x,y,z$ defined by the condition
\begin{subequations}
\label{scaling Neel-Neel}
\begin{equation}
\Delta^{\,}_{\alpha}=\Delta^{\,}_{\beta},
\end{equation}
the Neel${}^{\,}_{\alpha}$ and the Neel${}^{\,}_{\beta}$ order parameter
also vanish as power laws as a function of the length $L$ of the chain
with the same scaling exponent,
\begin{equation}
\langle N^{\,}_{\alpha}\rangle\sim\langle N^{\,}_{\beta}\rangle\sim
L^{-\eta/2}.
\end{equation}
Hereto, in the thermodynamic limit $L\to\infty$,
their dependencies on $\Delta^{\,}_{\alpha}-\Delta^{\,}_{\beta}$
are the power laws
\begin{align}
&
\langle N^{\,}_{\alpha}\rangle\sim
\left(\Delta^{\,}_{\alpha}-\Delta^{\,}_{\beta}\right)^{\eta/[4(1-\eta)]}\,
\Theta\!\left(\Delta^{\,}_{\alpha}-\Delta^{\,}_{\beta}\right),
\\
&
\langle N^{\,}_{\beta}\rangle\sim
\left(\Delta^{\,}_{\beta}-\Delta^{\,}_{\alpha}\right)^{\eta/[4(1-\eta)]}\,
\Theta\!\left(\Delta^{\,}_{\beta}-\Delta^{\,}_{\alpha}\right),
\end{align}
\end{subequations}
respectively.
Equations (\ref{scaling Neel-dimer}) and (\ref{scaling Neel-Neel})
express the duality
\begin{equation}
\mathcal{J}-\mathcal{J}^{\,}_{\mathrm{c}},
\
\eta
\qquad
\longleftrightarrow
\qquad
\Delta^{\,}_{\beta}-\Delta^{\,}_{\alpha},
\
1/\eta
\end{equation}
at the level of the scaling variables and
the scaling exponents between
the Neel-dimer and the Neel${}^{\,}_{\alpha}$-Neel${}^{\,}_{\beta}$ transition.
Thereto, the Neel${}^{\,}_{\alpha}$-Neel${}^{\,}_{\beta}$ transition
is an example of a phase transition
beyond the Landau-Ginzburg paradigm, as it separates two
gapped phases breaking spontaneously distinct $\mathbb{Z}^{\,}_{2}$
sectors of the $\mathbb{Z}^{\,}_{2}\times\mathbb{Z}^{\,}_{2}$
symmetry in spin space of the
quantum spin-1/2 antiferromagnetic $J^{\,}_{1}$-$J^{\,}_{2}$ XYZ Hamiltonian
(\ref{eq: def HXYZ}).

Finally, we stress that the critical theory has an emergent
U(1)$\times$U(1) symmetry
that is enhanced relative to the discrete symmetries
($\mathbb{Z}^{\,}_{2}\times\mathbb{Z}^{\,}_{2}$
in spin space,
$\mathbb{Z}\times\mathbb{Z}^{\,}_{2}$ of 1D lattice,
and $\mathbb{Z}^{\,}_{2}$ in time) of the
quantum spin-1/2 antiferromagnetic $J^{\,}_{1}$-$J^{\,}_{2}$ XYZ Hamiltonian
(\ref{eq: def HXYZ}).
We have performed numerical studies to confirm this conjecture.

\subsection{Numerical results}
\label{subsec: Numerical results}

We are going to study the phase diagram of the
quantum spin-1/2 antiferromagnetic $J^{\,}_{1}$-$J^{\,}_{2}$ XYZ Hamiltonian
(\ref{eq: def HXYZ a}) numerically using two complementary methods,
namely exact diagonalization and DMRG.
We will confirm numerically that, by varying the dimensionless coupling
$\mathcal{J}\equiv J^{\,}_{2}/J^{\,}_{1}$ 
while holding $\Delta^{\,}_{y}$ and $\Delta^{\,}_{z}$
fixed to suitable values,
Hamiltonian (\ref{eq: def HXYZ a}) undergoes
a quantum phase transition
between a Neel${}^{\,}_{\alpha}$ 
with $\alpha=x,y,z$ phase and a dimer phase
within the $c=1$ Gaussian universality class
in $(1+1)$-dimensional spacetime.
To this end, we shall study exclusively
the phase transition between the Neel${}^{\,}_{z}$ and dimer phases.
Indeed, the relations between the phase transitions
separating the Neel${}^{\,}_{\alpha}$ phase
with $\alpha=x,y$ from the dimer phase and
the phase transition between the Neel${}^{\,}_{z}$ and dimer phases
were given in Sec.\ \ref{subsec: global phase diagram}.
As a typical example, we focus on the two-dimensional cut
\begin{equation}
\Delta^{\,}_{z}=2.0,
\qquad
0\leq\Delta^{\,}_{y}<\Delta^{\,}_{z},
\qquad
0\leq\mathcal{J}< 0.5,
\label{eq: two-dimensional cut}
\end{equation}
from the three-dimensional parameter space
(\ref{eq: 3D parameter space for global phase diagram})
in which the Neel${}^{\,}_{z}$ to dimer quantum transition
is expected to take place.

\begin{figure*}[t]
\begin{center}
\includegraphics[width=0.4\textwidth]{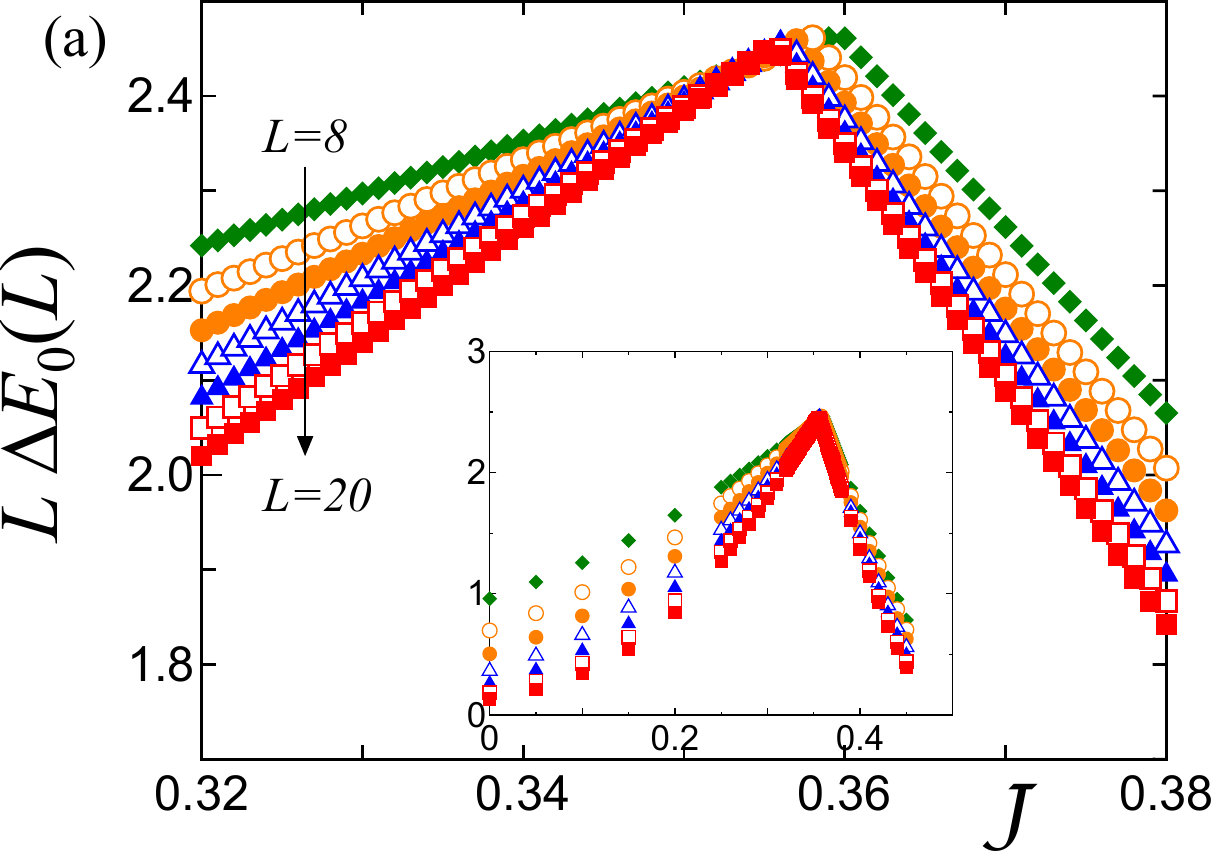}
~~~~~~~~~~~~~~~~~~~~~~~~
\includegraphics[width=0.37\textwidth]{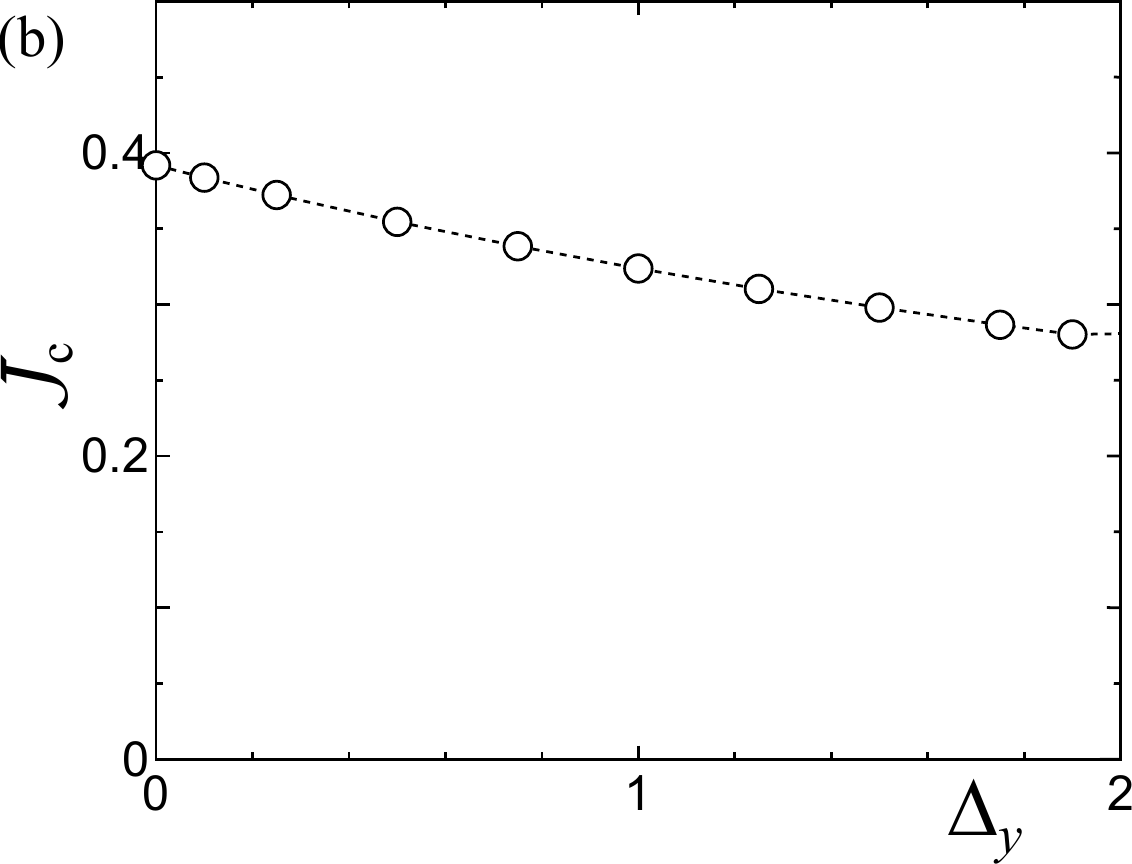}
\caption{(Color online)
(a)
Dependence on $\mathcal{J}\equiv J^{\,}_{2}/J^{\,}_{1}$
of the scaled finite-size gap
$L\,\Delta E^{\,}_{0}(L)$
with
$\Delta E^{\,}_{0}(L)$
defined in Eq.\ (\ref{eq:excitation_gap})
for $\Delta^{\,}_{y}=0.5$, $\Delta^{\,}_{z}=2.0$
and $L=8$ to $L=20$.
The position of the cusp defines
$\mathcal{J}^{\,}_{\mathrm{c}}(L)$.
The inset shows the data for the range $0\leq\mathcal{J}<1/2$,
while the main panel limits the range of data to
the vicinity of the cusp singularity.
(b)
Critical coupling
$\mathcal{J}^{\,}_{\mathrm{c}}=\lim_{L\to\infty}\mathcal{J}^{\,}_{\mathrm{c}}(L)$
at which a continuous quantum phase transition separates
the Neel${}^{\,}_{z}$ from the dimer phase for $\Delta^{\,}_{z}=2.0$
as a function of $0\leq\Delta^{\,}_{y}<2.0$.
        } 
\label{fig: scaled gap and critical cal J}
\end{center}
\end{figure*}

We first analyze the eigenenergy spectrum of
the quantum spin-1/2 antiferromagnetic $J^{\,}_{1}$-$J^{\,}_{2}$ XYZ Hamiltonian
(\ref{eq: def HXYZ a}) 
for a chain hosting $L$ spins obeying periodic boundary conditions.
We assume that $L$ is an even integer.
Eigenstates with ascending eigenenergies are denoted
$\Psi^{\,}_{0}(L)$,
$\Psi^{\,}_{1}(L)$,
$\Psi^{\,}_{2}(L)$,
etc. Their eigenenergies are denoted
$E^{\,}_{0}(L)$,
$E^{\,}_{1}(L)$,
$E^{\,}_{2}(L)$,
etc. The finite-size excitation gap above the ground state is defined by
\begin{equation}
\Delta E^{\,}_{0}(L)\:=
E^{\,}_{1}(L)
-
E^{\,}_{0}(L).
\label{eq:excitation_gap}
\end{equation}
The finite-size excitation gap above the
first excited state is defined by
\begin{equation}
\Delta E^{\,}_{1}(L)\:=
E^{\,}_{2}(L)
-
E^{\,}_{1}(L),
\label{eq: TD gap}
\end{equation}
and so on.
The dependence of $\Delta E^{\,}_{0}(L)$ on $\mathcal{J}$ should
be qualitatively different depending on whether the system is at
or away from a critical point, as we explain below.
On the one hand,
deep either in the Neel${}^{\,}_{z}$ or dimer phases,
$\Delta E^{\,}_{0}(L)$
is expected to decay exponentially fast to zero
with increasing $L$, while
$\Delta E^{\,}_{1}(L)$ remains non-vanishing in the thermodynamic limit
$L\to\infty$. Hence, the finite-size ground and first-excited states
become degenerate while remaining linearly independent 
in the thermodynamic limit $L\to\infty$,
for which a continuum of excitations is separated from the two-fold degenerate
ground states by a gap.
On the other hand, at a putative continuous quantum critical point separating
the Neel${}^{\,}_{z}$ phase from the dimer phase,
the finite-size gap $\Delta E^{\,}_{0}(L)$
between ground and first-excited states
is expected to decay algebraically to zero with increasing $L$,
with a level crossing of the first-excited state $\Psi^{\,}_{1}(L)$ whose
inversion quantum number differs between the Neel${}^{\,}_{z}$ side
and the dimer side. According to this scenario,
we may identify a putative continuous quantum critical point separating
the Neel${}^{\,}_{z}$ phase from the dimer phase
by a cusp singularity in the dependence of $\Delta E^{\,}_{0}(L)$
on $\mathcal{J}$ for fixed $\Delta^{\,}_{y}$, $\Delta^{\,}_{z}$, and $L$.
Indeed, this scenario was verified for the 
quantum spin-1/2 antiferromagnetic $J^{\,}_{1}$-$J^{\,}_{2}$ XXZ chain
in Ref.\ \onlinecite{Nomura1994}.

\begin{figure*}[t]
\begin{center}
\includegraphics[width=0.4\textwidth]{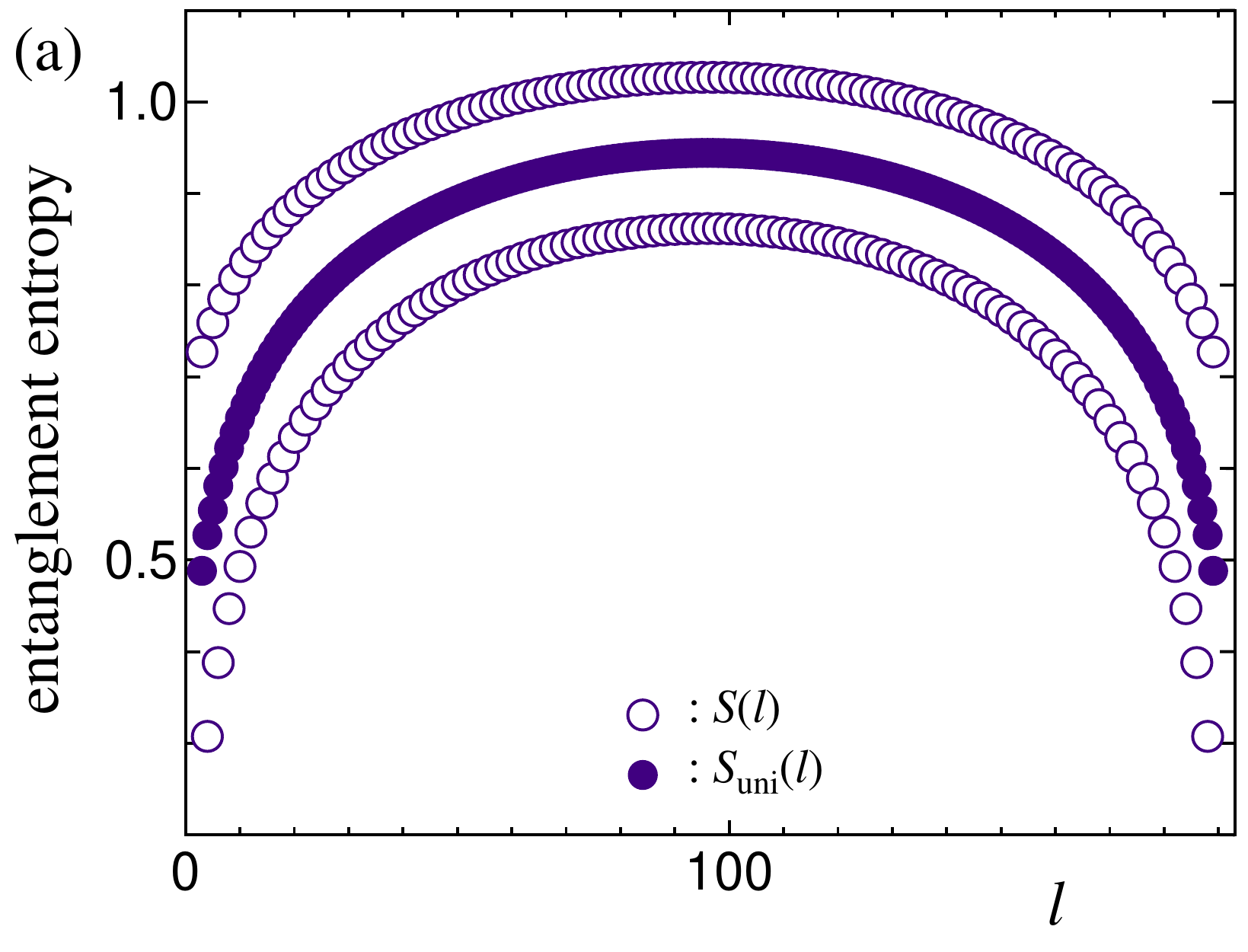}
~~~~~~~~~~~~~~~~~~~~~~~~
\includegraphics[width=0.4\textwidth]{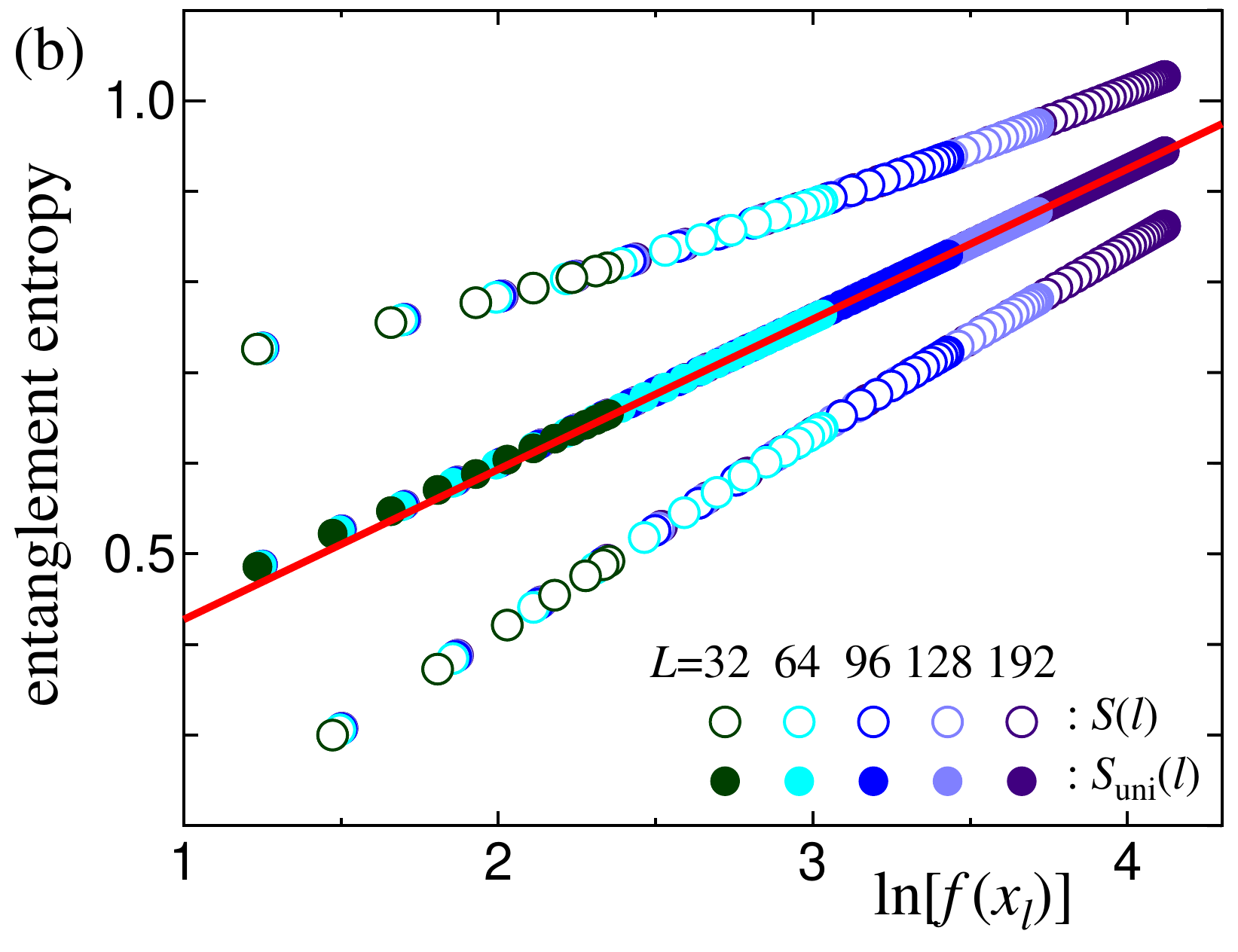}
\caption{(Color online)
Entanglement entropy $\mathcal{S}({{l}})$ (open circles)
and its uniform part $\mathcal{S}^{\,}_{\mathrm{uni}}({{l}})$
(filled circles)
at the quantum critical point $\mathcal{J}=\mathcal{J}^{\,}_{\mathrm{c}}$ 
for $\Delta^{\,}_{z}=2.0$ and $\Delta^{\,}_{y}=0.5$.
(a) $\mathcal{S}({{l}})$ and $\mathcal{S}^{\,}_{\mathrm{uni}}({{l}})$ 
for $L=192$ are plotted as a function of ${{l}}$.
Successive open circles are either above or below the curve defined
by the filled circles. This effect originates from choosing open boundary
conditions.
(b) $\mathcal{S}({{l}})$ and $\mathcal{S}^{\,}_{\mathrm{uni}}({{l}})$ 
for $L=32,64,96,128,192$ are plotted as a function of
$\ln[f(x^{\,}_{l})]$ with $x^{\,}_{l} := {{l}}+\frac{1}{2}$.
      }
\label{fig: entanglement entropy}
\end{center}
\end{figure*}

For several values of $\Delta^{\,}_{y}$,
we have computed the finite-size excitation gap (\ref{eq:excitation_gap})
along the one-dimensional cuts
\begin{equation}
\Delta^{\,}_{z}=2.0,
\qquad
0\leq\mathcal{J}< 0.5,
\label{eq: one-dimensional cut}
\end{equation}
of the two-dimensional cut (\ref{eq: two-dimensional cut})
for $L$ increasing from $L=8$ to $L=20$
using the exact diagonalization (Lanczos) method.
Our results for the one-dimensional cut with $\Delta^{\,}_{y}=0.5$
are presented in
Fig.\ \ref{fig: scaled gap and critical cal J}(a),
by plotting the dependence of $L\,\Delta E^{\,}_{0}(L)$ 
on $\mathcal{J}$. The cusp singularity of
the dependence on $\mathcal{J}$
of the finite-size excitation gap (\ref{eq:excitation_gap})
signals the anticipated level crossing of the first-excited states
$\Psi^{\,}_{1}(L)$ on the Neel${}^{\,}_{z}$ side
crossing energetically with $\Psi^{\,}_{1}(L)$ on the dimer side
of the critical point $\mathcal{J}^{\,}_{\mathrm{c}}$.
On both sides of the cusp, the scaled finite-size gap
$L\,\Delta E^{\,}_{0}(L)$ 
decays with $L$, suggesting the rapid decay (faster than $1/L$)
of $\Delta E^{\,}_{0}(L)$ in the gapped phases.
We also find that $L\,\Delta E^{\,}_{0}(L)$ 
at the cusp is almost independent of $L$,
indicating the critical scaling of the gap,
$\Delta E^{\,}_{0}(L)\sim 1/L$.
These numerical results support our conjecture that the
quantum spin-1/2 antiferromagnetic $J^{\,}_{1}$-$J^{\,}_{2}$ XYZ Hamiltonian
(\ref{eq: def HXYZ a}) with $\Delta^{\,}_{y}$ suitably chosen 
undergoes a continuous quantum phase transition
between the Neel${}^{\,}_{z}$ and dimer phases
with a dynamical scaling exponent $z=1$
upon varying $\mathcal{J}$.

We have extrapolated the critical value $\mathcal{J}^{\,}_{\mathrm{c}}(L)$
determined above by fitting the data to a second-order polynomial of $1/L$
(see Appendix \ref{sec:Appendix}).
The extrapolated value of the critical coupling, $\mathcal{J}^{\,}_{\mathrm{c}}$,
is shown in 
Fig.\ \ref{fig: scaled gap and critical cal J}(b)
as a function of $\Delta^{\,}_{y}$.
With increasing $\Delta^{\,}_{y}$, $\mathcal{J}^{\,}_{\mathrm{c}}$ runs
from $\mathcal{J}^{\,}_{\mathrm{c}}=0.3918(9)$ at $\Delta^{\,}_{y}=0$
towards $\mathcal{J}^{\,}_{\mathrm{c}}=0.276$
at $\Delta^{\,}_{y}=\Delta^{\,}_{z}=2.0$,
the latter was obtained in Ref.\ \onlinecite{Nomura1994}.

Second, we determine the central charge at the
continuous quantum critical point separating
the Neel${}^{\,}_{z}$ and dimer phases.
For this purpose, we analyze the entanglement entropy
$\mathcal{S}({{l}})$
between the left ${{l}}$-site block and the right
$(L-{{l}})$-site block
in the ground state of 
the quantum spin-1/2 antiferromagnetic $J^{\,}_{1}$-$J^{\,}_{2}$ XYZ Hamiltonian
(\ref{eq: def HXYZ a})
when open boundary conditions are imposed.
It is known that the entanglement entropy at criticality
in $(1+1)$-dimensional spacetime scales with $l$ as%
~\cite{HolzheyLW1994,VidalLRK2003,CalabreseC2004,LaflorencieSCA2006,AffleckLS2009}
\begin{subequations}
\begin{equation}
\mathcal{S}({{l}})=
\frac{c}{6}\ln[f(x^{\,}_{l})]
+
\alpha^{\,}_{\mathrm{osc}}\,
E^{\,}_{\mathrm{osc}}({{l}})
+
\mathcal{S}^{\,}_{0},
\label{eq:EEnt_1D_open}
\end{equation}
where $c$ is the central charge,
the function
\begin{equation}
f(x^{\,}_{l})\:=
\frac{L+1}{\pi}\,\sin\left(\frac{\pi\,x^{\,}_{l}}{L+1}\right),
\qquad
x^{\,}_{l}\:=l+\frac{1}{2},
\label{eq:effective_block_size}
\end{equation}
is the effective size of the left ${{l}}$-site block,
$\alpha^{\,}_{\mathrm{osc}}$ is a constant,
the oscillating component of the local bond-energy expectation value,
$E^{\,}_{\mathrm{osc}}({{l}})$,
is defined in Eq.\ (\ref{eq: def Esc}),
and
$\mathcal{S}^{\,}_{0}$ is a constant.
We can therefore estimate the central charge $c$
from the slope of the uniform contribution
\begin{equation}
\mathcal{S}^{\,}_{\mathrm{uni}}({{l}})\:=
\mathcal{S}({{l}})
-
\alpha^{\,}_{\mathrm{osc}}\,
E^{\,}_{\mathrm{osc}}({{l}})
\end{equation}
\end{subequations}
to the entanglement entropy $\mathcal{S}({{l}})$
plotted as a function of
$\ln[f(l+\frac{1}{2})]$.

Using the DMRG method, we have computed the entanglement entropy
$\mathcal{S}({{l}})$ and the oscillating part $E^{\,}_{\mathrm{osc}}({{l}})$
of the local bond energy expectation value for
the quantum spin-1/2 antiferromagnetic $J^{\,}_{1}$-$J^{\,}_{2}$ XYZ Hamiltonian
(\ref{eq: def HXYZ a})
for an open chain hosting up to $L=192$ spins 
at all the continuous quantum critical points
from Fig.\ \ref{fig: scaled gap and critical cal J}(b)
separating the Neel${}^{\,}_{z}$ phase from the dimer phase.
Results for $\Delta^{\,}_{y}=0.5$
are summarized in Fig.\ \ref{fig: entanglement entropy}.
For all the critical values of $\mathcal{J}^{\,}_{\mathrm{c}}$
obtained in Fig.\ \ref{fig: scaled gap and critical cal J}(b),
we find the central charge $c$ obtained for $L=192$ to be
in the range $0.993<c<1.000$,
in agreement with the analytical arguments supporting the claim
that all Neel${}^{\,}_{\alpha=x,y,z}$-dimer quantum critical points
realize a $c=1$ Gaussian conformal-field theory
(CFT) in $(1+1)$-dimensional spacetime.

Finally, we estimate the critical exponent $\eta$ at the transition.
To this end, we have calculated the expectation value of
the local dimer-order operator
\begin{equation}
\begin{split}
O^{\,}_{\mathrm{VBS}}(L)\:=&\,
\langle
S^{x}_{\frac{L}{2}}\,S^{x}_{\frac{L}{2}+1}
+
S^{y}_{\frac{L}{2}}\,S^{y}_{\frac{L}{2}+1}
+
S^{z}_{\frac{L}{2}}\,S^{z}_{\frac{L}{2}+1}
\rangle^{\,}_{L}
\\
&\,
-
\langle
S^{x}_{\frac{L}{2}-1}\,S^{x}_{\frac{L}{2}}
+
S^{y}_{\frac{L}{2}-1}\,S^{y}_{\frac{L}{2}}
+
S^{z}_{\frac{L}{2}-1}\,S^{z}_{\frac{L}{2}}
\rangle^{\,}_{L}
\end{split}
\label{eq:O_dimer_L}
\end{equation}
at the center of an open chain of length $L$
($L$ is chosen a multiple of four).
Here, $\langle A\rangle^{\,}_{L}$
denotes the ground-state expectation value
of the operator $A$ for an open chain of length $L$.
At a continuous quantum critical point,
$O^{\,}_{\mathrm{VBS}}(L)$ is expected to behave as
\begin{equation}
O^{\,}_{\mathrm{VBS}}(L)\sim L^{-\frac{1}{2\eta}}.
\label{eq:scaling_O_dim_L}
\end{equation}

\begin{figure*}[t]
\begin{center}
\includegraphics[width=0.4\textwidth]{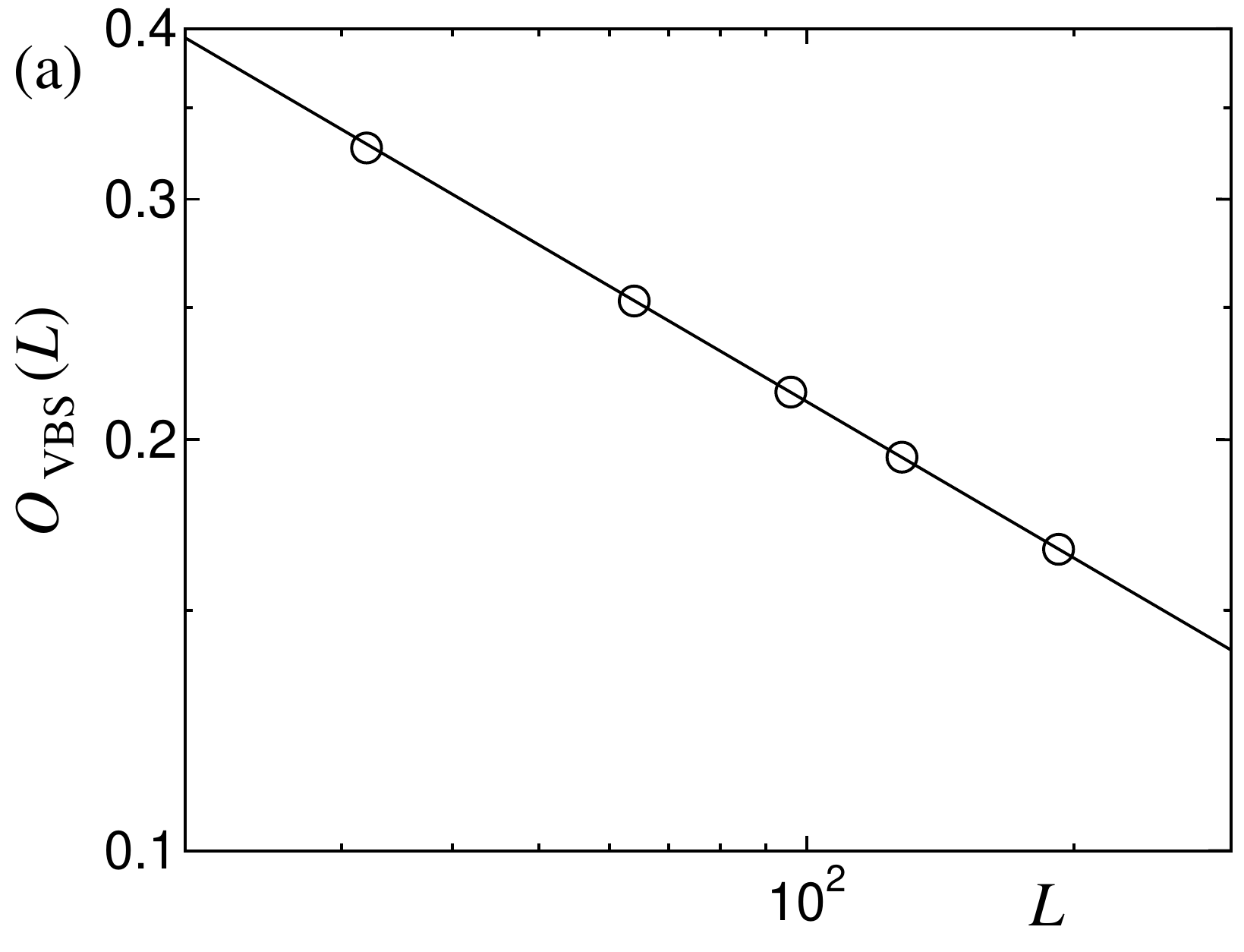}
~~~~~~~~~~~~~~~~~~~~~~~~
\includegraphics[width=0.4\textwidth]{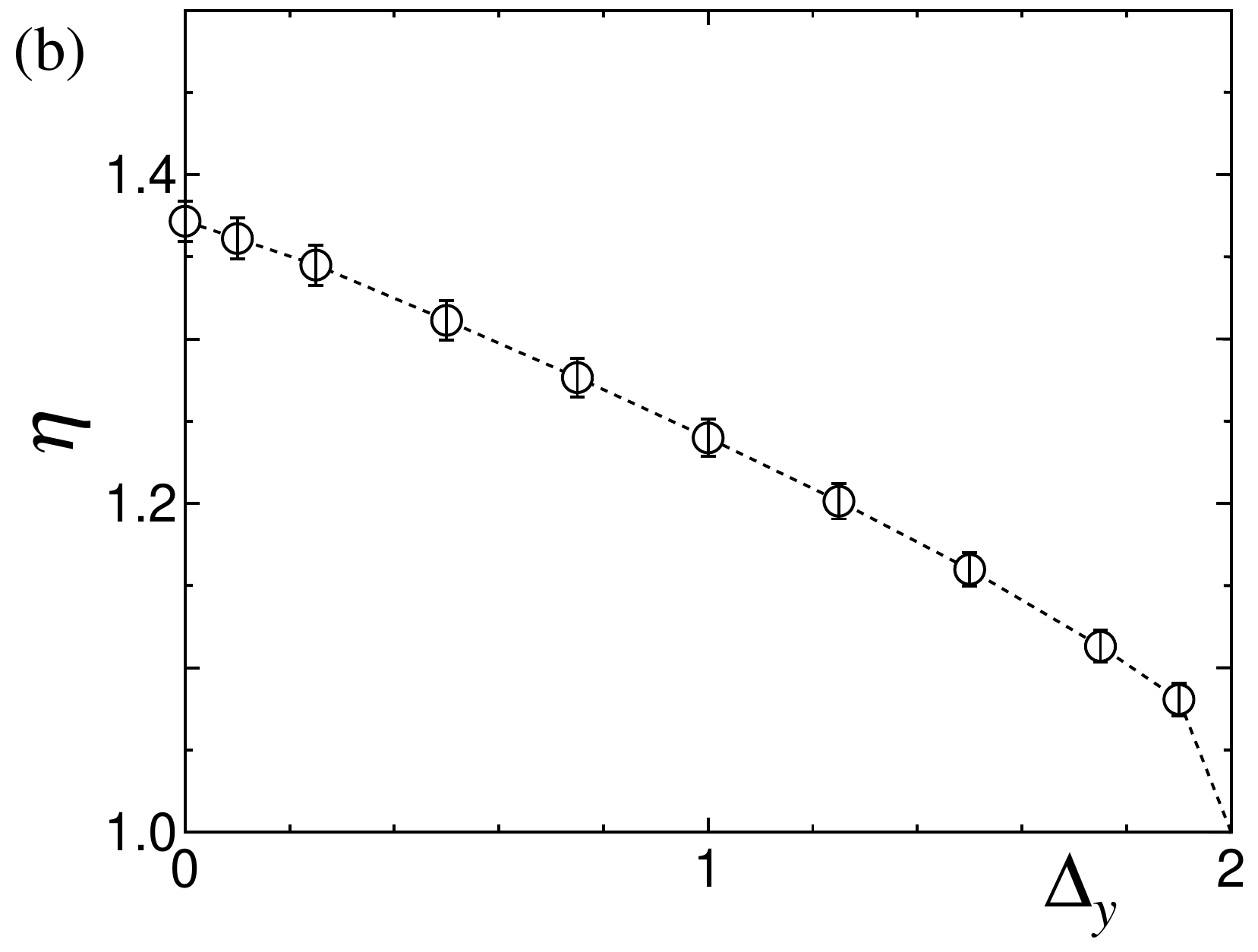}
\caption{
(a) Log-log plot of the dependence
of $O^{\,}_{\mathrm{VBS}}(L)$
defined in Eq.\ (\ref{eq:O_dimer_L})
on $L$ at the critical point $\mathcal{J}=\mathcal{J}^{\,}_{\mathrm{c}}$ 
for $\Delta^{\,}_{z}=2.0$ and $\Delta^{\,}_{y}=0.5$.
(b)
Exponent $\eta$ at the Neel${}^{\,}_{z}$-dimer transition 
for $\Delta^{\,}_{z}=2.0$ as a function of $\Delta^{\,}_{y}$.
        }
\label{fig: eta from DMRG}
\end{center}
\end{figure*}

Figure\ \ref{fig: eta from DMRG}(a) presents our DMRG data of
$O^{\,}_{\mathrm{VBS}}(L)$
for $\Delta^{\,}_{z}=2.0$, $\Delta^{\,}_{y}=0.5$,
and $\mathcal{J}=\mathcal{J}^{\,}_{\mathrm{c}}$.
They show the expected scaling behavior
(\ref{eq:scaling_O_dim_L}).
The estimates of $\eta$ obtained from the fitting
outlined in Appendix \ref{sec:Appendix}
are shown in Fig.\ \ref{fig: eta from DMRG}(b).
It is found that, as anticipated,
$\eta>1$ for $0\leq\Delta^{\,}_{y}<\Delta^{\,}_{z}$
and $\eta$ approaches unity as $\Delta^{\,}_{y}$ approaches $\Delta^{\,}_{z}$,
$\eta-1\propto\sqrt{2-\Delta^{\,}_{y}}$.

\subsection{Revisiting the Neel${}^{\,}_{z}$-VBS transition}
\label{subsec: Revisiting the Neelz-VBS transition}

Having obtained the numerical evidence that the
dimer-Neel${}^{\,}_{\alpha}$ phase boundaries
[i.e., the critical values
$\mathcal{J}^{\,}_{\mathrm{c}}(\Delta^{\,}_{y},\Delta^{\,}_{z})$]
realize a $c=1$ conformal field theory in (1+1)-dimensional spacetime
for a wide range in the two-dimensional parameter space
$(\Delta^{\,}_{y},\Delta^{\,}_{z})$, we return to the
dimer-Neel${}^{\,}_{z}$ phase transition at $\Delta^{\,}_{z}>1$ for
$|\Delta^{\,}_{y}-1|\ll1$. Our aim is to give a complementary
description of this transition.

As a warm up exercise, we consider first
the bosonized theory (\ref{eq: sine-Gordon rep HXXZ}).
Because $\eta>1$ in the Hamiltonian density
(\ref{eq: sine-Gordon rep HXXZ a}), the operator
$\cos(\sqrt{8\pi}\,\theta)$ is irrelevant and thus
plays only a minor role close to the Neel${}^{\,}_{z}$-dimer transition,
a role that we shall ignore.
By inspection of Eq.\ (\ref{eq: order parameter Neel z bis})
we deduce that the $\phi$ field is pinned
at either $\phi=\sqrt{\pi/8}$ or $3\sqrt{\pi/8}$
(mod $\sqrt{2\pi}$)
in the Neel${}^{\,}_{z}$ phase. These two values correspond to two
degenerate Ising-ordered states in the Neel${}^{\,}_{z}$ phase.
Suppose that the ground state is in the Neel${}^{\,}_{z}$ phase and
that there is a domain wall between the two Ising-ordered states
[e.g., $\phi=\sqrt{\pi/8}$ for $x<0$ and $\phi=3\sqrt{\pi/8}$ for
$x>0$ as shown in Fig.\ \ref{fig: defects}(a)].
At the domain wall the $\phi$ field
displays a kink structure that
crosses $\phi=0$ or $\sqrt{\pi/2}$ (mod $\sqrt{2\pi}$),
where the dimer order parameter
(\ref{eq: order parameter VBS bis})
takes a nonvanishing expectation value,
$D=\cos(\sqrt{2\pi}\phi)\ne0$.
Conversely, suppose that the ground state is in the dimer phase in
which there is a domain wall between the two degenerate dimer-ordered
states. By Eq.\ (\ref{eq: order parameter VBS bis}),
the $\phi$ field must then be pinned at either $\phi=0$ or
$\sqrt{\pi/2}$ (mod $\sqrt{2\pi}$).  At the domain wall the $\phi$
field displays a kink structure that crosses $\phi=\sqrt{\pi/8}$ or
$3\sqrt{\pi/8}$ (mod $\sqrt{2\pi}$)
as shown in Fig.~\ref{fig: defects}(b),
where the Neel${}^{\,}_{z}$ order parameter
(\ref{eq: order parameter Neel z bis})
takes a nonvanishing expectation value,
$N^{\,}_{z}=\sin(\sqrt{2\pi}\phi)\ne0$.
Therefore, the center of a
domain wall in the Neel${}^{\,}_{z}$ phase supports a local dimer order
and, conversely, a domain wall in the dimer phase supports a local
Neel${}^{\,}_{z}$ order.

We may draw a parallel to the scenario of
deconfined quantum criticality in two spatial
dimensions that separates easy-plane Neel order from dimer order.
On the one hand, $U(1)$ vortices in the Neel-ordered phase nucleate
local dimer order. On the other hand, $\mathbb{Z}^{\,}_{4}$ vortices
in the dimer ordered phase  nucleate local easy-plane Neel order.
The proliferation of these point defects in one of the ordered phase
destroys this phase in favor of long-range order in the competing phase.%
~\cite{Senthil04,Senthil04PRB,Senthil04Levin}

Let us return to the fermionic theory
with the Hamiltonian (\ref{eq: def XXZ model c})
and (\ref{eq: def XXZ model b bis}).  We first note that the local
Neel${}^{\,}_{z}$ operator can be written in the fermion
representation as
\begin{subequations}
\begin{equation}
n^{\,}_{z}(x)=
\psi^{\dag}_{\mathrm{L}}(x)\,
\psi^{\,}_{\mathrm{R}}(x)
+
\psi^{\dag}_{\mathrm{R}}(x)\,
\psi^{\,}_{\mathrm{L}}(x)=
\Psi^{\dag}(x)\,\sigma^{\,}_{1}\,\Psi(x),
\label{n_{z}(x)}
\end{equation}
while the local dimer operator can be written as
\begin{equation}
d(x)=
-\mathrm{i}\psi^{\dag}_{\mathrm{L}}(x)\,\psi^{\,}_{\mathrm{R}}(x)
+\mathrm{i}\psi^{\dag}_{\mathrm{R}}(x)\,\psi^{\,}_{\mathrm{L}}(x)=
\Psi^{\dag}(x)\,\sigma^{\,}_{2}\,\Psi(x),
\label{d(x)}
\end{equation}
where $\sigma^{\,}_{1}$, $\sigma^{\,}_{2}$, and $\sigma^{\,}_{3}$
are Pauli matrices, and
\begin{equation}
\Psi(x)\equiv
\begin{pmatrix}
\psi^{\,}_{\mathrm{L}}(x)
\\
\psi^{\,}_{\mathrm{R}}(x)
\end{pmatrix}.
\end{equation}
\end{subequations}
Equation (\ref{d(x)}) is obtained from the oscillating contributions in
$S^{+}_{l}\,S^{-}_{l+1}+ S^{-}_{l}\,S^{+}_{l+1}
=c^{\dag}_{l+1}\,c^{\,}_{l}+c^{\dag}_{l}\,c^{\,}_{l+1}$
and $S^{z}_{l}\,S^{z}_{l+1}=n^{\,}_{l}\,n^{\,}_{l+1}$.
Equations (\ref{n_{z}(x)}) and (\ref{d(x)})
encode the correspondence
$(n^{\,}_{z},d)\leftrightarrow(\sigma^{\,}_{1},\sigma^{\,}_{2})$
between the local pair of order parameters $(n^{\,}_{z},d)$
and the pair of Pauli matrices, i.e., two elements of a Clifford algebra.

Taking the two local order operators $n^{\,}_{z}$ and $d$ as mean fields
(or Hubbard-Stratonovich fields), we replace the XXZ Hamiltonian
density (\ref{eq: def XXZ model b bis}) with the mean-field Dirac Hamiltonian
density
\begin{align}
\mathcal{H}^{\,}_\mathrm{MF}(x)\:=&\,
\mathrm{i}v\left(\Psi^{\dag}\sigma^{\,}_{3}\partial^{\,}_{x}\Psi\right)(x)
-g^{\,}_{n}\,
n^{\,}_{z}(x)\,
\left(\Psi^{\dag}\,\sigma^{\,}_{1}\,\Psi\right)(x)
\nonumber\\
&\,
-
g^{\,}_{d}\,
d(x)\,
\left(\Psi^{\dag}\,\sigma^{\,}_{2}\,\Psi\right)(x),
\label{meanfield H_{x}XZ}
\end{align}
which is to be supplemented by the
additive Hubbard-Stratonovich (HS) contributions
\begin{equation}
\mathcal{H}^{\,}_{\mathrm{HS}}(x)\:=
g^{\,}_{n}\,n^{2}_{z}(x)
+
g^{\,}_{d}\,d^{2}(x),
\end{equation}
{where the couplings $g^{\,}_{n}$ and
$g^{\,}_{d}$ are related to the couplings $g^{\,}_{\mathrm{u}}$ and $g^{\,}_{\pm}$
entering the XXZ Hamiltonian density (\ref{eq: def XXZ model b bis}).
Integrating out the Hubbard-Stratonovich fields $n^{\,}_{z}$ and $d$
reproduces $\mathcal{H}^{\,}_{\mathrm{XXZ}}$ approximately.  Therefore,
$\mathcal{H}^{\,}_{\mathrm{MF}}$ can be used as a starting point for the
discussion of the Neel${}^{\,}_{z}$-dimer phase transition.  It is
important to point out that the two order parameters
$n^{\,}_{z}$ and $d$
are Dirac mass terms when they are constants in
(1+1)-dimensional spacetime.

Suppose that we are in the Neel${}^{\,}_{z}$ phase, in which there is
a domain wall.  We thus assume $n^{\,}_{z}(x)=n^{0}_{z}\,\tanh(x/\xi)$ and
$d(x)=0$, where $\xi$ is a width of the domain wall.  We then find
that there is a zero mode localized at $x=0$, which is an eigenstate
of $\sigma^{\,}_{2}$.%
~\cite{Jackiw76}
This implies that the dimer order is locally
generated at the center of a domain wall in the Neel${}^{\,}_{z}$
phase.  Conversely, if we are in the dimer phase with a domain wall,
where $d(x)=d^{\,}_{0}\,\tanh(x/\xi)$ and $n^{\,}_{z}(x)=0$.
Again we obtain a zero mode localized at $x=0$, which is an eigenstate of
$\sigma^{\,}_{1}$.%
~\cite{Jackiw76}
This implies that the Neel${}^{\,}_{z}$ order is
locally generated at the center of a domain wall in the dimer phase.
These considerations run parallel to the discussion on domain walls
in the bosonized theory.

\begin{figure}
\includegraphics[width=0.4\textwidth]{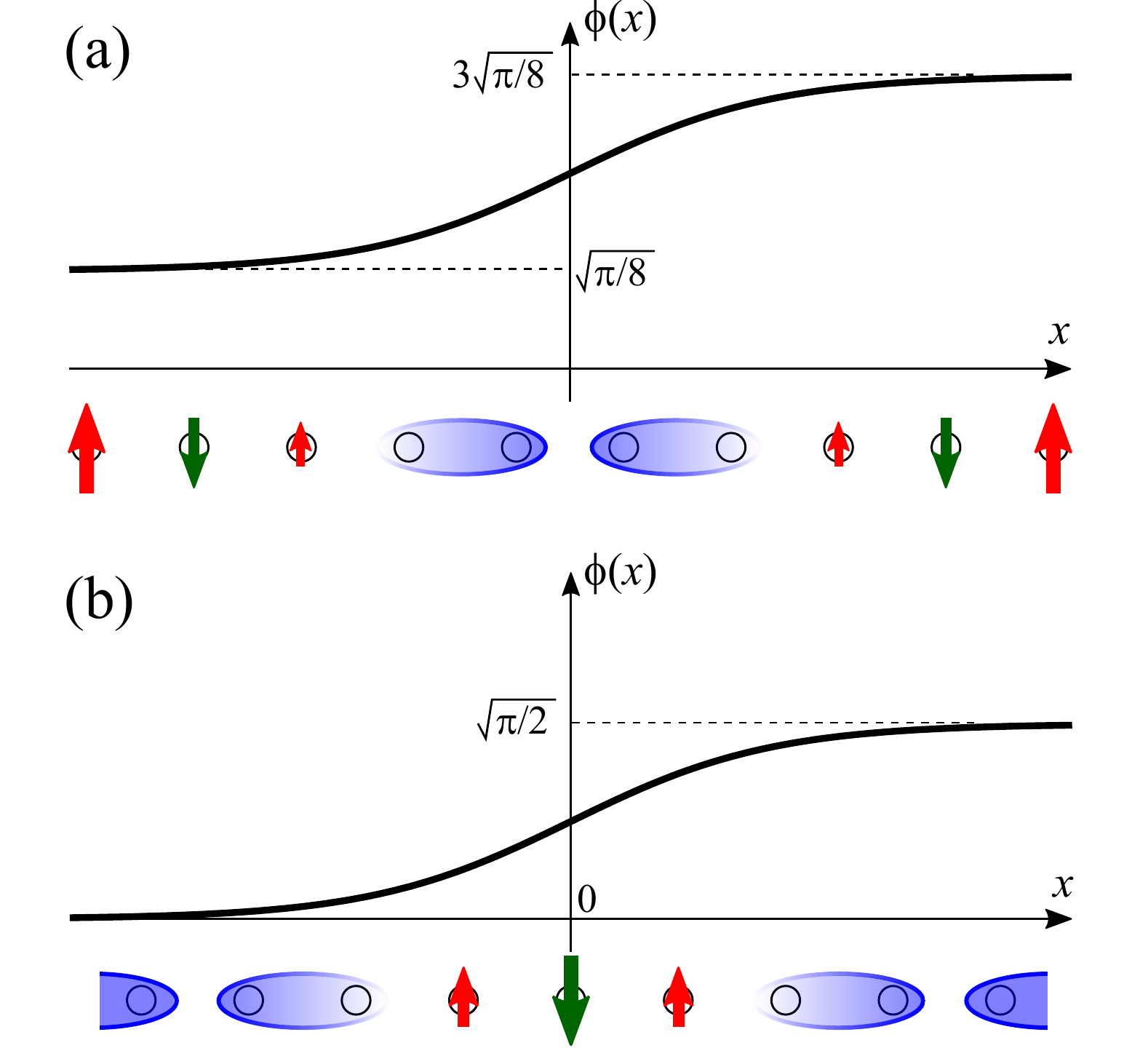}
\caption{(Color online)
Schematic picture of a domain wall in
(a) the Neel${}^{\,}_{z}$ phase and (b) the VBS (dimer) phase.
\label{fig: defects}
        }
\end{figure}

Next, we consider deriving from $\mathcal{H}^{\,}_{\mathrm{MF}}$ an effective
theory for the Neel${}^{\,}_{z}$-dimer phase transition by integrating
out the Dirac fermions $\psi^{\,}_{\mathrm{M}}$ with
$\mathrm{M}=\mathrm{L},\mathrm{R}$. To this end,
the two order-parameter fields $n^{\,}_{z}$ and $d$ are regarded as elements
of the vector field 
\begin{subequations}
\begin{equation}
\bm{n}\:=(d,n^{\,}_{z})
\end{equation}
on which we impose the non-linear constraint
\begin{equation}
\bm{n}^{2}=1.
\end{equation}
\end{subequations}
A by-product of this non-linear constraint is that
it regularizes the domain walls supported by either one of
the pair $n^{\,}_{z}$ and $d$
in such a way that $\bm{n}$ points locally towards $d$ when
$n^{\,}_{z}$ vanishes and conversely. In this way, the zero modes
from the previous paragraph are moved to a finite energy. Integration
over the Dirac fermions in the partition function is now safe.

Expanding the Dirac-fermion determinant in a gradient expansion of the
smooth fluctuations about a saddle point
corresponding to a mean-field solution $\bm{n}=\bm{n}^{\,}_{0}$,
we obtain,
after analytical continuation of time $t$ to imaginary time $\tau$,
the Euclidean effective action given by
\begin{equation}
S^{\,}_{0}=
\frac{1}{2g}
\int\mathrm{d}\tau\int\mathrm{d}x\,
\left[(\partial^{\,}_{\tau}\bm{n})^{2}+(\partial^{\,}_{x}\bm{n})^{2}\right].
\label{eq: integrating massive fermions} 
\end{equation}
Here, we have set the velocity to be unity for simplicity.
The stiffness $g$ is a positive dimensionless coupling constant.
The smooth unit vector field $\bm{n}$
may be parameterized by a smooth angle $\varphi$ through
\begin{equation}
\bm{n}=(d,n)=(\cos\varphi,\sin\varphi),
\end{equation}
in which case the action $S^{\,}_{0}$ is now represented by the Gaussian
action
\begin{equation}
S^{\,}_{0}=
\frac{1}{2g}
\int\mathrm{d}\tau\int\mathrm{d}x\,
[(\partial^{\,}_{\tau}\varphi)^{2}+(\partial^{\,}_{x}\varphi)^{2}].
\label{eq: integrating massive fermions bis}
\end{equation}
Very much as was the case
with Eqs.\ (\ref{eq: order parameter Neel z bis}) and
(\ref{eq: order parameter VBS bis}),
the long-range Neel${}^{\,}_{z}$ order corresponds to pinning $\varphi$
to the values
\begin{subequations}
\label{eq: pinning varphi field}
\begin{equation}
\varphi=\frac{\pi}{2},\frac{3\pi}{2} \hbox{ (mod $2\pi$)},
\label{eq: pinning varphi field a}
\end{equation}
while the long-range dimer order corresponds to pinning
$\varphi$
to the values
\begin{equation}
\varphi=0,\pi \hbox{ (mod $2\pi$)}.
\label{eq: pinning varphi field b}
\end{equation}
\end{subequations}
The four extrema
(\ref{eq: pinning varphi field a})
and
(\ref{eq: pinning varphi field b})
of
$\cos\varphi$
and
$\sin\varphi$
on the interval $0\leq\varphi<2\pi$ are the four local minima of
$-\lambda^{\,}_{4}\cos(4\varphi)$ on the same interval with $\lambda^{\,}_{4}>0$.
We thus add to $S^{\,}_{0}$ the potential $-\lambda^{\,}_{4}\cos(4\varphi)$ to
stabilize the Neel${}^{\,}_{z}$ and dimer phases.  Furthermore, we can
introduce another potential $\lambda^{\,}_{\varphi}\cos(2\varphi)$ that
selects either the Neel${}^{\,}_{z}$ or dimer order depending on the sign of
the coupling constant $\lambda^{\,}_{\varphi}$.  We note that the potential
$-\lambda^{\,}_{4}\cos(4\varphi)$ reduces the symmetry from U(1) in
$S^{\,}_{0}$ to
$\mathbb{Z}^{\,}_{4}$, and the potential $\lambda^{\,}_{\varphi}\cos(2\varphi)$
further reduces the symmetry to $\mathbb{Z}^{\,}_{2}\times\mathbb{Z}^{\,}_{2}$.

Similarly to the classical XY model in two-dimensional space,
the unit vector $\bm{n}$
need not be smooth as it may support point defects in the form of
vortices in (1+1)-dimensional spacetime. The vorticities of such point
defects are a topological attribute such as
the charge one vortex at the origin given by
\begin{equation}
\varphi^{\,}_{\mathrm{vtx}}(\tau,x)\:=
\mathrm{arctan}\left(\frac{x}{\tau}\right),  
\end{equation}
say. It turns out, however, that the relevant vortices to the
Neel${}^{\,}_{z}$-dimer transition are the charge $\pm2$ vortices, as
we will explain below. The presence of such charge $\pm2$ vortices
can be taken into account by
adding to the Lagrangian density in Eq.\
(\ref{eq: integrating massive fermions bis})
the cosine potential%
~\cite{Jose77,Wiegmann78,Lecheminant2002}
\begin{equation}
\mathcal{L}^{\,}_{\mathrm{vtx}}\:=
\lambda^{\,}_{\vartheta}\cos(4\pi\vartheta),
\end{equation}
where the coupling $\lambda^{\,}_{\vartheta}$ is dimensionful.
Here, the field $\vartheta$ is related to the field $\varphi$
by the Cauchy-Riemann conditions
\begin{subequations}
\begin{equation}
\partial^{\,}_{x}\varphi=
+\mathrm{i}g\,\partial^{\,}_{\tau}\vartheta,
\qquad
\partial^{\,}_{\tau}\varphi=
-\mathrm{i}g\,\partial^{\,}_{x}\vartheta,
\end{equation}
once the measure for $\varphi$ has been augmented to accommodate vortices.
Alternatively, in the operator formalism, we must demand the equal-time
commutation relation
\begin{equation}
\left[\varphi(x),\vartheta(y)\right]=\mathrm{i}\Theta(y-x)
\end{equation}
\end{subequations}
with the convention $\Theta(0)\equiv1/2$ for the Heaviside function.

We have therefore deduced the effective Lagrangian density,%
~\footnote{
The same action was obtained in
Appendix E of Ref.\~ \onlinecite{Jiang2018}
in the limit when $\Delta^{\,}_{y}=0$. 
          }
\begin{align}
\mathcal{L}^{\,}_{\mathbb{Z}^{\,}_{2}\times\mathbb{Z}^{\,}_{2}}\:=&\,
-
\mathrm{i}\partial^{\,}_{x}\vartheta\,\partial^{\,}_{\tau}\varphi
+
\frac{g}{2}(\partial^{\,}_{x}\vartheta)^{2}
+
\frac{1}{2g}(\partial^{\,}_{x}\varphi)^{2}
\nonumber\\
&\,
+
\lambda^{\,}_{\vartheta}\,\cos(4\pi\vartheta)
+
\lambda^{\,}_{\varphi}\,\cos(2\varphi)
\nonumber\\
&\,
-
\lambda^{\,}_{4}\,\cos(4\varphi),
\label{eq: integrating massive fermions bis bis}
\end{align}
where $\lambda^{\,}_{\vartheta}$, $\lambda^{\,}_{\varphi}$, $\lambda^{\,}_{4}$ are
dimensionful coupling constants, and $\lambda^{\,}_{4}>0$. 
The effective Lagrangian density
$\mathcal{L}^{\,}_{\mathbb{Z}^{\,}_{2}\times\mathbb{Z}^{\,}_{2}}$ 
is to be compared with the Hamiltonian $H^{\,}_{0}$
perturbed by the current-current interaction $H^{\,}_{JJ}$
through the identifying $(\varphi,\vartheta)$ with
$(\sqrt{2\pi}\phi,\theta/\sqrt{2\pi})$.
As we have discussed in Sec.~\ref{subsec: global phase diagram},
the critical theory at the
Neel${}^{\,}_{z}$-dimer phase transition is the Gaussian Hamiltonian
(\ref{eq: Gaussian fixed point}) with the continuous parameter $\eta>1$
and a U(1)$\times$U(1) symmetry.
At the Neel${}^{\,}_{z}$-dimer transition, the renormalized coupling constant
$\lambda^{\,}_{\varphi}$ vanishes,
while the interaction $\lambda^{\,}_{4}\cos(4\varphi)$ is irrelevant
and thus vanishes in the long-distance limit.
The parameter $\eta$ is related to $1/g$.
In the Hamiltonian picture, the role of the dual potential
$\lambda^{\,}_{\vartheta}\cos(4\pi\vartheta)$
is to create a $4\pi$ kink in the $\varphi$ field, as seen from the relation
\begin{equation}
e^{\mathrm{i}4\pi\vartheta(y)}\varphi(x)e^{-\mathrm{i}4\pi\vartheta(y)}=
\varphi(x)+4\pi\,\Theta(y-x).
\end{equation}
It is important to realize that the shift of $\varphi(x)$ at $x=y$ is $2\pi$,
the period of the $\varphi$ field.
The potential $\cos(2\pi\vartheta)$ corresponding
to charge one vortices would introduce
a $\pm\pi$ shift and therefore is not allowed in the effective action.
As we have seen in Sec.~\ref{sec: Delta_{y}=1}, the physical origin of the
$\cos(4\pi\vartheta)$ potential,
or the $\cos(\sqrt{8\pi}\theta)$ potential
invariant under (\ref{TPThetaR}), is
$S^{+}_{l}\,S^{+}_{l+1}+S^{-}_{l}\,S^{-}_{l+1}$.
It is also interesting to point out the analogy to the case of
$(2+1)$-dimensional spacetime
for which no monopoles with a charge less than four
appear in the effective theory of deconfined quantum criticality,%
\cite{Senthil04,Senthil04PRB,Senthil04Levin}
while no vortices with a vorticity less than two appear here.

\section{Dirac semimetallic phase in $d>1$-dimensional space
perturbed by a contact interaction}
\label{sec: Dirac semimetallic phase in d>1-dimensional space ...}

When the dimensionality $d$ of space is $d=1$,
we have shown in Sec.\ \ref{sec: J1-J2 XYZ model on a linear chain}
that the quantum spin-1/2 antiferromagnetic $J^{\,}_{1}$-$J^{\,}_{2}$ XYZ chain
supports a pair of gapped phases at zero temperature,
each of which breaks spontaneously an Ising symmetry,
that are separated by a continuous phase transition with an enlarged
U(1)$\times$U(1) continuous symmetry.
One phase is an Ising Neel phase.
The other phase is a valance bond solid (VBS or dimer) phase.
The driving mechanism for this transition is the proliferation
of a dual pair of domain walls. The duality means here that a domain
wall in the Ising Neel ordered phase nucleates locally the Ising VBS order,
while the converse also holds, i.e.,
a domain wall in the Ising VBS ordered phase nucleates
locally the Ising Neel order.
At the quantum critical point, both dual domain walls have
proliferated extensively.

Inspired by Sec.\ \ref{subsec: Revisiting the Neelz-VBS transition},
we are going to present a general framework to describe Neel-VBS
quantum phase transition beyond Landau-Ginzburg theory and related
phenomena by using a model of Dirac fermions in
three-dimensional space ($d=3$) with contact interactions.  This is a
generalization of the approach taken for $d=2$ by Tanaka and Hu
\cite{TanakaHuPRL2005} and by Senthil and
Fisher.%
~\cite{Senthil06Fisher}
In their work, it was shown that the
effective action for the Neel and VBS order parameter fields takes the
form of a nonlinear sigma model (NLSM) with a topological term.  In
particular, for a quantum phase transition between an easy-plane Neel
phase and a VBS phase, the O(4) nonlinear sigma model with a theta
term is obtained as an effective theory and its connection to the
noncompact CP$^{1}$ model was discussed.%
~\cite{Senthil06Fisher}

Our model for $d=3$ supports a pair of long-range ordered phases
at vanishing temperature that break spontaneously
the symmetries of the Hamiltonian, namely
Neel ordering that breaks the (internal) spin-1/2 SU(2) symmetry
and VBS (dimer) ordering that breaks the
translation and rotation symmetries of some underlying cubic lattice model.
We propose the two possibilities that these phases
are either separated by a continuous phase transition
with an enlarged continuous symmetry or by
a gapless spin-liquid phase that is extended
in parameter space.
Hereto, the driving mechanism for these two possibilities
is the proliferation of a dual pair of topological defects. 

We will first introduce a tight-binding model
on the cubic lattice and a Dirac Hamiltonian in the continuum limit.
The Neel and dimer order parameters are related to Dirac mass terms
in the Dirac Hamiltonian. Integrating out the Dirac fermions
gives a bosonic effective field theory for the order parameter fields,
which is a NLSM augmented by a Wess-Zumino term.
The RG flow for this NLSM will be used to conjecture the fate 
of the semi-metallic phase defined by the $\pi$-flux
phase on the cubic lattice at half-filling,
when perturbed by certain local quartic fermionic interactions
preserving an O(3)$\times$O(3) symmetry.

\subsection{The $\pi$ flux phase in three-dimensional space and
its instabilities}

In this section, we show that the $\pi$-flux phase on the cubic lattice
for spinless electrons accommodates an eight-dimensional representation
of the Dirac Hamiltonian at the corner $(\pi,\pi,\pi)$
of the $\pi$-flux phase Brillouin zone.
We show that there are 4 mass terms at the Dirac point
that anticommute pairwise and are compatible
with time-reversal symmetry and fermion-number conservation,
three of which are compatible with chiral symmetry,
one of which breaks the chiral symmetry.
If time-reversal symmetry and fermion-number conservation are both broken,
the corresponding sixteen-dimensional representation of the
Bogoliubov-de-Gennes Dirac Hamiltonian can be shown to accomodate
two additional massive channels that, together with the four
previous ones, anticommute pairwise.%
~\footnote{
C. Mudry, A. Furusaki, T.\ Morimoto, and T.\ Hikihara, unpublished.
          }
Alternatively, we show below that the $\pi$-flux phase
on the cubic lattice for spinful electrons also accommodates
a sixteen-dimensional representation of the Dirac Hamiltonian
with 6 mass terms at the Dirac point that
anticommute pairwise and are compatible
with fermion-number conservation.
Three of those masses are associated to valence-bond (dimer) ordering,
while the other three are associated to antiferromagnetic (Neel) order
that breaks the SU(2) spin symmetry down to a U(1) subgroup.

\begin{figure}
\includegraphics[width=0.8\linewidth]{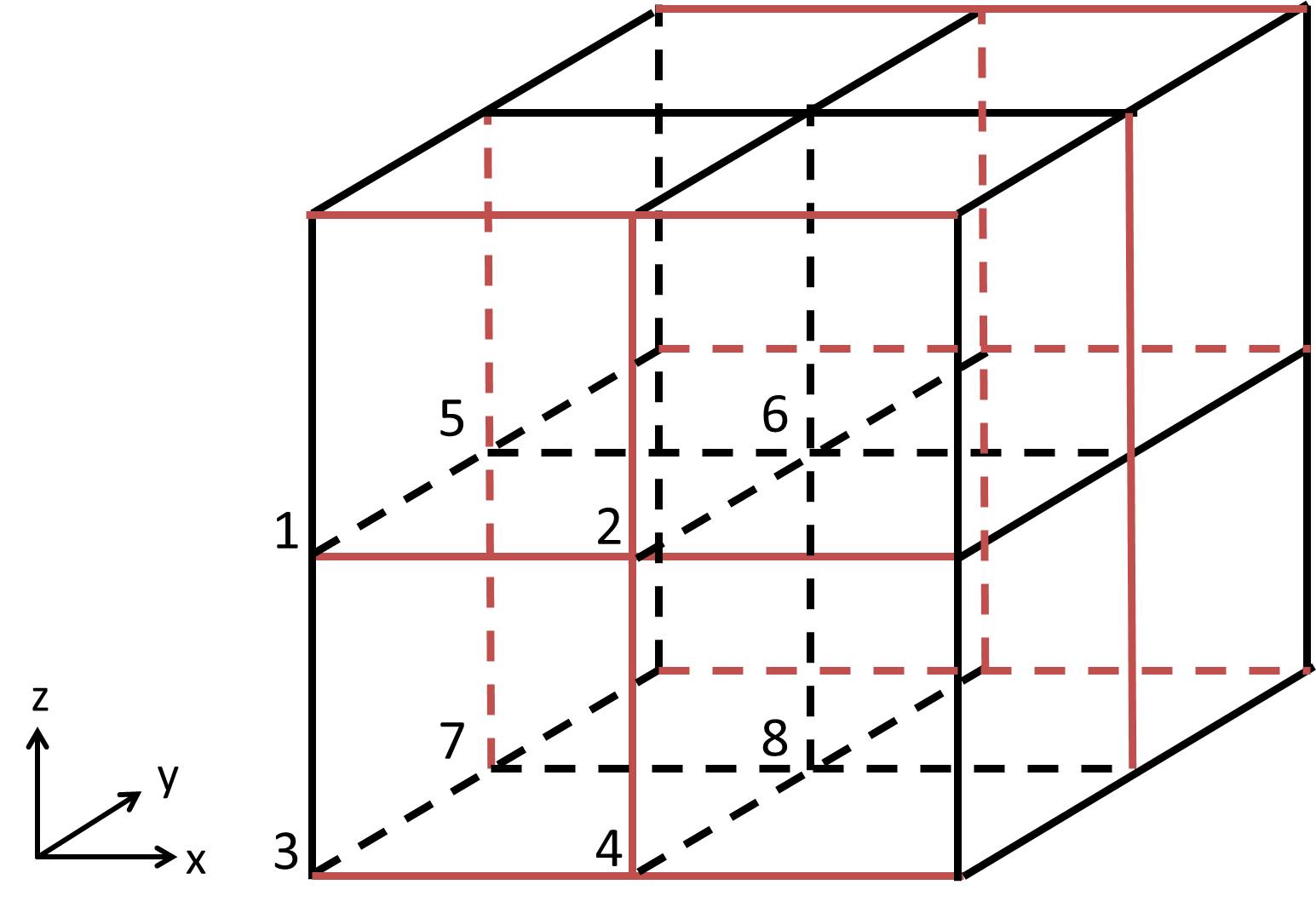}
\caption{(Color online)
We define the $\pi$-flux model on the three-dimensional cubic lattice
by the following rules.
Only nearest-neighbor hopping amplitudes are allowed.
All nearest-neighbor hopping amplitudes have the magnitude $t/2>0$
and are real valued.
The sign of the nearest-neighbor hopping amplitudes is $+1$ ($-1$)
when the nearest-neighbor is colored in black (red).
\label{fig: 3d pi flux model}
       }
\end{figure}

We consider a cubic lattice with the lattice spacing $\mathfrak{a}/2$,
which we partition into 8 sublattices with a cubic repeated unit cell
whose volume is $\mathfrak{a}^{3}$. This repeated unit cell
contains 8 non-equivalent sites, as is shown in Fig.\
\ref{fig: 3d pi flux model}.
The site 3 of this repeated unit cell is assigned the coordinate
$\bm{i}\in\mathbb{Z}^{3}$.
To each $\bm{i}$, we assign the eight-components wave function
\begin{equation}
\Psi^{\,}_{\bm{i}}\equiv
\begin{pmatrix}
\psi^{\,}_{\bm{i}1}&
\psi^{\,}_{\bm{i}2}&
\psi^{\,}_{\bm{i}3}&
\psi^{\,}_{\bm{i}4}&
\psi^{\,}_{\bm{i}5}&
\psi^{\,}_{\bm{i}6}&
\psi^{\,}_{\bm{i}7}&
\psi^{\,}_{\bm{i}8}
\end{pmatrix}^{\mathsf{T}}
\in\mathbb{C}^{8}.
\end{equation}
The local Hilbert space $\mathbb{C}^{8}$ can be represented by
the span of the 16 Hermitian matrices
\begin{subequations}
\begin{equation}
X^{\,}_{\mu^{\,}_{1}\mu^{\,}_{2}\mu^{\,}_{3}}\:=
\tau^{\,}_{\mu^{\,}_{1}}\otimes
\upsilon^{\,}_{\mu^{\,}_{2}}\otimes
\zeta^{\,}_{\mu^{\,}_{3}},
\qquad
\mu^{\,}_{1},\mu^{\,}_{2},\mu^{\,}_{3}=0,1,2,3,
\label{eq: def X mu1 mu2 mu3}
\end{equation}
where
$\tau^{\,}_{\mu}$,
$\upsilon^{\,}_{\mu}$,
and
$\zeta^{\,}_{\mu}$,
each denote a quadruplet of unit $2\times2$ matrix ($\mu=0$)
and Pauli matrices  ($\mu=1,2,3$) and the action of the
Pauli matrices
$\tau^{\,}_{3}$,
$\upsilon^{\,}_{3}$,
and
$\zeta^{\,}_{3}$
on the wave function $\Psi^{\,}_{i}$ is defined by the following rules.
The matrix
\begin{equation}
X^{\,}_{300}=\tau^{\,}_{3}\otimes\upsilon^{\,}_{0}\otimes\zeta^{\,}_{0}
\end{equation}
has the eigenvalue $+1$ and $-1$
when $\Psi^{\,}_{\bm{i}}$ is only non-vanishing on the face with the vertices
$(1,3,5,7)$ and $(2,4,6,8)$
in the $yz$ plane of the repeat unit cell, respectively.
The matrix
\begin{equation}
X^{\,}_{003}=\tau^{\,}_{0}\otimes\upsilon^{\,}_{0}\otimes\zeta^{\,}_{3}
\end{equation}
has the eigenvalue $+1$ and $-1$
when $\Psi^{\,}_{\bm{i}}$ is only non-vanishing on the face with the vertices
$(1,2,3,4)$ and $(5,6,7,8)$
in the $zx$ plane of the repeat unit cell, respectively.
The matrix
\begin{equation}
X^{\,}_{030}=\tau^{\,}_{0}\otimes\upsilon^{\,}_{3}\otimes\zeta^{\,}_{0}
\end{equation}
\end{subequations}
has the eigenvalue $+1$ and $-1$
when $\Psi^{\,}_{\bm{i}}$ is only non-vanishing on the face with the vertices
$(1,2,5,6)$ and $(3,4,7,8)$
in the $xy$ plane of the repeat unit cell, respectively.

We define a tight-binding model by the following rules.
Only nearest-neighbor hopping amplitudes are allowed.
All nearest-neighbor hopping amplitudes have the magnitude $t/2>0$
and are real valued.
The sign of the nearest-neighbor hopping amplitudes is $+1$ ($-1$)
when the nearest-neighbor is colored in black (red) in
Fig.\ \ref{fig: 3d pi flux model}.

Hopping along the positive $x$ direction in Fig.\ \ref{fig: 3d pi flux model}
takes wave functions localized on the face $(1,3,5,7)$
in Fig.\ \ref{fig: 3d pi flux model}
to wave functions localized on the face $(2,4,6,8)$
and conversely. This process is encoded by the product
\begin{subequations}
\begin{equation}
\Gamma^{\,}_{1}\equiv-X^{\,}_{103}
\label{Gamma_1}
\end{equation}
of the matrices
\begin{equation}
X^{\,}_{100}=\tau^{\,}_{1}\otimes\upsilon^{\,}_{0}\otimes\zeta^{\,}_{0},
\qquad
-X^{\,}_{003}=-\tau^{\,}_{0}\otimes\upsilon^{\,}_{0}\otimes\zeta^{\,}_{3},
\end{equation}
\end{subequations}
by inspection of Fig.\ \ref{fig: 3d pi flux model}.
Hopping along the positive $y$ direction in Fig.\ \ref{fig: 3d pi flux model}
takes wave functions localized on the face $(1,2,3,4)$
in Fig.\ \ref{fig: 3d pi flux model}
to wave functions localized on the face $(5,6,7,8)$
and conversely. This process is encoded by the product
\begin{subequations}
\begin{equation}
\Gamma^{\,}_{2}\equiv+X^{\,}_{001}
\label{Gamma_2}
\end{equation}
of the matrices
\begin{equation}
X^{\,}_{001}=\tau^{\,}_{0}\otimes\upsilon^{\,}_{0}\otimes\zeta^{\,}_{1},
\qquad
X^{\,}_{000}=\tau^{\,}_{0}\otimes\upsilon^{\,}_{0}\otimes\zeta^{\,}_{0},
\end{equation}
\end{subequations}
by inspection of Fig.\ \ref{fig: 3d pi flux model}.
Hopping along the positive $z$ direction in Fig.\ \ref{fig: 3d pi flux model}
takes wave functions localized on the face $(3,4,7,8)$
in Fig.\ \ref{fig: 3d pi flux model}
to wave functions localized on the face $(1,2,5,6)$
and conversely. This process is encoded by the product
\begin{subequations}
\begin{equation}
\Gamma^{\,}_{3}\equiv+X^{\,}_{313}
\label{Gamma_3}
\end{equation}
of the matrices
\begin{equation}
X^{\,}_{010}=\tau^{\,}_{0}\otimes\upsilon^{\,}_{1}\otimes\zeta^{\,}_{0},
\qquad
X^{\,}_{303}=\tau^{\,}_{3}\otimes\upsilon^{\,}_{0}\otimes\zeta^{\,}_{3},
\end{equation}
\end{subequations}
by inspection of Fig.\ \ref{fig: 3d pi flux model}.
If we choose units such that $\mathfrak{a}=t=1$,
the tight-binding Hamiltonian defined by
Fig.\ \ref{fig: 3d pi flux model}
is thus represented by
\begin{subequations}
\begin{equation}
\mathscr{H}^{\mathrm{kin}}_{\bm{k}}\:=
\cos k^{\,}_{x}\,
\Gamma^{\,}_{1}
+
\cos k^{\,}_{y}\,
\Gamma^{\,}_{2}
+
\cos k^{\,}_{z}\,
\Gamma^{\,}_{3}
\end{equation}
in the first Brillouin zone
\begin{equation} 
-\frac{\pi}{2}\le k^{\,}_{x}\le\frac{\pi}{2},
\quad
-\frac{\pi}{2}\le k^{\,}_{y}\le\frac{\pi}{2},
\quad
-\frac{\pi}{2}\le k^{\,}_{z}\le\frac{\pi}{2},
\end{equation}
\end{subequations}
associated with the repeated unit cell of unit volume,
where we have set $\mathfrak{a}=1$.

The eigenvalues of $\mathscr{H}^{\mathrm{kin}}_{\bm{k}}$
are four-fold degenerate and come in pairs of opposite signs
\begin{equation}
\varepsilon^{\,}_{\bm{k}}=
\pm
\sqrt{
\cos^{2}k^{\,}_{x}
+
\cos^{2}k^{\,}_{y}
+
\cos^{2}k^{\,}_{z}
     }.
\end{equation}
The upper four-fold degenerate band touches the
lower four-fold degenerate band at the eight corners
\begin{equation}
\bm{k}^{(\pm,\pm,\pm)}_{\mathrm{D}}\:=
\begin{pmatrix}
\pm\frac{1}{2},&
\pm\frac{1}{2},&
\pm\frac{1}{2}
\end{pmatrix}
\end{equation}
of the Brillouin zone.
All those corners are equivalent modulo a reciprocal wave vector.
We may then choose the Dirac point to be
\begin{equation}
\bm{k}^{\,}_{\mathrm{D}}\:=
\begin{pmatrix}
\frac{1}{2},&
\frac{1}{2},&
\frac{1}{2}
\end{pmatrix}
\label{eq: def dirac point}
\end{equation}
without loss of generality. Expanding to linear order around the
Dirac point $\mathscr{H}^{\mathrm{kin}}_{\bm{k}}$
delivers an eight-dimensional representation of the massless
Dirac Hamiltonian in three-dimensional space. This representation is
twice as large as the four-dimensional representation of the original
Dirac Hamiltonian. This is an example of fermion doubling.

So far, all hopping amplitudes from
Fig.\ \ref{fig: 3d pi flux model}
have the same magnitude $t/2$.
This assumption can be relaxed by demanding that
two consecutive nearest-neighbor bonds along the directions $x$, $y$, and $z$
within the repeat unit cell are changed by the substitutions
\begin{equation}
\frac{t}{2}\mapsto
\frac{t}{2}
\mp
\frac{d^{\,}_{x}}{2},
\qquad
\frac{t}{2}\mapsto
\frac{t}{2}
\pm
\frac{d^{\,}_{y}}{2},
\qquad
\frac{t}{2}\mapsto
\frac{t}{2}
\pm
\frac{d^{\,}_{z}}{2},
\end{equation}
with $d^{\,}_{x},d^{\,}_{y},d^{\,}_{z}\in\mathbb{R}$, respectively.
With this substitution,
\begin{subequations}
\begin{equation}
\mathscr{H}^{\mathrm{kin}}_{\bm{k}}\mapsto
\mathscr{H}^{\mathrm{kin}}_{\bm{k}}
+
\mathscr{V}^{\mathrm{VB}}_{\bm{k}}
\end{equation}
with
\begin{equation}
\mathscr{V}^{\mathrm{VB}}_{\bm{k}}=
d^{\,}_{x}\,
\sin k^{\,}_{x}\,
\Gamma^{\,}_{4}
+
d^{\,}_{y}\,
\sin k^{\,}_{y}\,
\Gamma^{\,}_{5}
+
d^{\,}_{z}\,
\sin k^{\,}_{z}\,
\Gamma^{\,}_{6},
\end{equation}
where
\begin{equation}
\Gamma^{\,}_{4}\equiv X^{\,}_{203},
\qquad
\Gamma^{\,}_{5}\equiv X^{\,}_{002},
\qquad
\Gamma^{\,}_{6}\equiv X^{\,}_{323},
\label{eq: def triplet dimer mass matrices}
\end{equation}
\end{subequations}
Here, $\mathscr{V}^{\mathrm{VB}}_{\bm{k}}$
follows from replacing in $\mathscr{H}^{\mathrm{kin}}_{\bm{k}}$
the Pauli matrices with index $1$ by the Pauli matrices
with index $2$ and the cosine by the sine function.
As it should be
\begin{equation}
\mathscr{V}^{\mathrm{VB}*}_{-\bm{k}}=
\mathscr{V}^{\mathrm{VB}}_{\bm{k}}.
\end{equation}
On the other hand, the cubic symmetry of
$\mathscr{H}^{\mathrm{kin}}_{\bm{k}}$
is reduced to an orthorhombic one
for generic values of $d^{\,}_{x}$, $d^{\,}_{y}$, and $d^{\,}_{z}$.
Hereto, each member of the triplet of (dimer) masses
$(\Gamma^{\,}_{4},\Gamma^{\,}_{5},\Gamma^{\,}_{6})$
anticommutes with each member of the triplet of Dirac matrices
$(\Gamma^{\,}_{1},\Gamma^{\,}_{2},\Gamma^{\,}_{3})$. Thus,
at the Dirac point (\ref{eq: def dirac point}),
this dimerization pattern opens up the gap
\begin{equation}
2|\bm{d}|\equiv  
2
\sqrt{
d^{2}_{x}
+
d^{2}_{y}
+
d^{2}_{z}
      }.
\end{equation}

Among all $8\times8$ Hermitian matrices, there is one more matrix
of the form  (\ref{eq: def X mu1 mu2 mu3})
that anticommutes with
the three Dirac matrices (\ref{Gamma_1})-(\ref{Gamma_3})
and the three dimerization mass matrices
(\ref{eq: def triplet dimer mass matrices}).
It is the diagonal matrix
\begin{equation}
\Gamma^{\,}_{7}\:= X^{\,}_{333}
\end{equation}
that represents
a staggered chemical potential
[a charge density wave with the momentum $(\pi,\pi,\pi)$].
We conclude that the most generic opening of a gap at the Dirac point
(\ref{eq: def dirac point}) is encoded by the Hamiltonian
\begin{subequations}
\begin{equation}
\mathscr{H}^{\,}_{\bm{k}}\:=
\mathscr{H}^{\mathrm{kin}}_{\bm{k}}
+
\mathscr{V}^{\mathrm{VB}}_{\bm{k}}
+
m\,X^{\,}_{333}
\end{equation}
with the gap
\begin{equation}
\Delta\equiv  
2
\sqrt{
d^{2}_{x}
+
d^{2}_{y}
+
d^{2}_{z}
+
m^{2}
      }
\end{equation}
\end{subequations}
that depends on four real-valued parameters.

So far, we have been considering spinless fermions and we have assumed that
the fermion number was a good quantum number.
We attach to each spinless fermion a spin-1/2 degree of freedom,
while preserving the conservation of the total fermion
quantum number. We thus introduce the three Pauli matrices
$\bm{\sigma}=(\sigma^{\,}_{1},\sigma^{\,}_{2},\sigma^{\,}_{3})$
and the $2\times2$ unit matrix $\sigma^{\,}_{0}$.
All four $2\times2$ matrices act of the spin-1/2 degrees of freedom.
We also introduce the basis
\begin{equation}
X^{\,}_{\mu^{\,}_{1}\mu^{\,}_{2}\mu^{\,}_{3}\mu^{\,}_{4}}\:=
\sigma^{\,}_{\mu^{\,}_{1}}
\otimes
\tau^{\,}_{\mu^{\,}_{2}}
\otimes
\upsilon^{\,}_{\mu^{\,}_{3}}
\otimes
\zeta^{\,}_{\mu^{\,}_{4}}
\end{equation}
with  $\mu^{\,}_{1},\mu^{\,}_{2},\mu^{\,}_{3},\mu^{\,}_{4}=0,1,2,3$ for all
$16\times16$ Hermitian matrices.

Define the $16\times16$ Hermitian matrices
\begin{subequations}
\begin{align}
&
\alpha^{\,}_{x}\equiv-X^{\,}_{0103},
\quad
\alpha^{\,}_{y}\equiv X^{\,}_{0001},
\quad
\alpha^{\,}_{z}\equiv X^{\,}_{0313},
\label{eq: def gamma matrices a}
\\
&
\beta^{\mathrm{VB}}_{x}\equiv X^{\,}_{0203},
\quad
\beta^{\mathrm{VB}}_{y}\equiv X^{\,}_{0002},
\quad
\beta^{\mathrm{VB}}_{z}\equiv X^{\,}_{0323},
\\
&
\beta^{\mathrm{AF}}_{x}\equiv X^{\,}_{1333},
\quad
\beta^{\mathrm{AF}}_{y}\equiv X^{\,}_{2333},
\quad
\beta^{\mathrm{AF}}_{z}\equiv X^{\,}_{3333}.
\end{align}
\end{subequations}
All nine matrices are Hermitian, anticommute pairwise, and square to the
identity $16\times16$ matrix. We then define the single-particle
tight-binding model
\begin{subequations}
\label{eq: def 16 by 16 gapped tight-binding model with charge conservation}
\begin{equation}
\mathcal{H}^{\,}_{\bm{k}}\:=
\mathcal{H}^{\mathrm{kin}}_{\bm{k}}
+
\mathcal{H}^{\mathrm{VB}}_{\bm{k}}
+
\mathcal{H}^{\mathrm{AF}}_{\bm{k}},
\end{equation}
where
\begin{align}
&
\mathcal{H}^{\mathrm{kin}}_{\bm{k}}\:=
\alpha^{\,}_{x}\,
\cos k^{\,}_{x}
+
\alpha^{\,}_{y}\,
\cos k^{\,}_{y}
+
\alpha^{\,}_{z}
\cos k^{\,}_{z},
\\
&
\mathcal{H}^{\mathrm{VB}}_{\bm{k}}\:=
\beta^{\mathrm{VB}}_{x}\,
d^{\,}_{x}\,
\sin k^{\,}_{x}
+
\beta^{\mathrm{VB}}_{y}\,
d^{\,}_{y}\,
\sin k^{\mathrm{VB}}_{y}
+
\beta^{\mathrm{VB}}_{z}\,
d^{\,}_{z}\,
\sin k^{\,}_{z},
\\
&
\mathcal{H}^{\mathrm{AF}}_{\bm{k}}\:=
\beta^{\mathrm{AF}}_{x}\,
n^{\,}_{x}
+
\beta^{\mathrm{AF}}_{y}\,
n^{\,}_{y}
+
\beta^{\mathrm{AF}}_{z}\,
n^{\,}_{z}.
\end{align}
Reversal of time is defined by conjugation with
\begin{equation}
\mathcal{T}\:=\mathrm{i}X^{\,}_{2000}\,\mathsf{K},
\end{equation}
\end{subequations}
where $\mathsf{K}$ represents complex conjugation.
Any non-vanishing value for any one of
$n^{\,}_{x}$, $n^{\,}_{y}$, and $n^{\,}_{z}$
breaks time-reversal symmetry.

There are 16 bands that form an 8-fold degenerate valence band and an 8-fold
degenerate conduction band. Conduction and valence bands are separated by the
direct gap
\begin{subequations}
\begin{equation}
\Delta\equiv2\sqrt{\bm{d}^{2}+\bm{n}^{2}}
\end{equation}
at the 8 corners of the Brillouin zone. The gap at the Dirac point
(\ref{eq: def dirac point}) thus depends on 6 real-valued parameters
that can be interpreted as the pair of three-component vectors
\begin{equation}
\bm{d}\:=
\begin{pmatrix}
d^{\,}_{x}&d^{\,}_{y}&d^{\,}_{z}
\end{pmatrix}
\end{equation}
and
\begin{equation}
\bm{n}\:=
\begin{pmatrix}
n^{\,}_{x}&n^{\,}_{y}&n^{\,}_{z}
\end{pmatrix}.
\end{equation}
\end{subequations}
The vector $\bm{d}$ realizes dimerization of the hopping amplitude
within the repeat unit cell of Fig.\ \ref{fig: 3d pi flux model}.
Using two different color codes to represent the sign of the
dimerized hopping amplitude realizes a valence-bond covering of the cubic
lattice by which each site is the end point of one and only one colored
nearest-neighbor bond.
The vector $\bm{n}$ realizes a colinear magnetic order with the
antiferromagnetic wave vector $(\pi,\pi,\pi)$
within the repeat unit cell of Fig.\ \ref{fig: 3d pi flux model}.

\subsection{Dualities between point defects}

We work with the single-particle tight-binding Hamiltonian
(\ref{eq: def 16 by 16 gapped tight-binding model with charge conservation})
that we linearize about the Dirac point (\ref{eq: def dirac point}).
The cubic lattice is thus replaced by Euclidean space $\mathbb{R}^{3}$,
whose points we denote with $\bm{r}=(r^{\,}_{x},r^{\,}_{y},r^{\,}_{z})$.
We consider static configurations of the vector fields
$\bm{d}(\bm{r})$ and $\bm{n}(\bm{r})$ that support a monopole at the origin of
$\mathbb{R}^{3}$.

For example, in the presence of one such defect, say in $\bm{d}$,
the single-particle tight-binding Hamiltonian
(\ref{eq: def 16 by 16 gapped tight-binding model with charge conservation}).
is approximated to linear order in a gradient expansion around the Dirac
point  (\ref{eq: def dirac point}) by
\begin{subequations}
\label{eq: index thm}
\begin{align}
\mathcal{H}\:=&\,
\mathrm{i}\alpha^{\,}_{1}\partial^{\,}_{x}
+
\mathrm{i}\alpha^{\,}_{2}\partial^{\,}_{y}
+
\mathrm{i}\alpha^{\,}_{3}\partial^{\,}_{z} 
\nonumber\\
&\,
+
f(\bm{r})
\left[
\hat{d}^{\,}_{x}(\bm{r})\,\beta^{\mathrm{VB}}_{x}
+
\hat{d}^{\,}_{y}(\bm{r})\,\beta^{\mathrm{VB}}_{y}
+
\hat{d}^{\,}_{z}(\bm{r})\,\beta^{\mathrm{VB}}_{z}
\right],
\label{eq: index thm a}
\end{align}
where $f(\bm{r})$ is a smooth monotonic function satisfying
\begin{equation}
f(\bm{0})=0,
\qquad
\lim_{|\bm{r}|\to\infty}f(\bm{r})=1,
\label{eq: index thm b}
\end{equation}
while the function $\hat{\bm{d}}$ is singular at the origin
\begin{equation}
\bm{\hat{d}}(\bm{r})\:=\frac{\bm{r}}{|\bm{r}|},
\label{eq: index thm bb}
\end{equation}
for $\bm{r}\in\mathbb{R}^{3}$. 
The singularity of $\bm{d}(\bm{r})$ has a topological character,
for the order parameter $\bm{d}$ has the integer-valued winding number
\begin{equation}
W\:=
\int\limits_{\mathbb{R}^{3}}
\frac{\mathrm{d}^{3}\bm{r}}{8\pi}
\frac{\partial}{\partial r^{\,}_{i}}
\left[
\epsilon^{\,}_{ijk}\,
\epsilon^{\,}_{abc}\,
\hat{d}^{\,}_{a}(\bm{r})
\left(
\frac{\partial \hat{d}^{\,}_{b}}{\partial r^{\,}_{j}}
\right)(\bm{r})
\left(
\frac{\partial \hat{d}^{\,}_{c}}{\partial r^{\,}_{k}}
\right)(\bm{r})
\right]
\label{eq: index thm c}
\end{equation}
of magnitude one around the origin.
The single-particle Hamiltonian
(\ref{eq: index thm a})
obeys the index theorem
(see Ref.\ \onlinecite{Jackiw76}
and, in a slightly more general context, Ref.\ \onlinecite{Nishida10})
\begin{equation}
\mathrm{Index}\,\mathcal{H}=
W\,\mathrm{tr}\,\sigma^{\,}_{0}=
2W,
\label{eq: index thm d}
\end{equation}
where the left-hand sides is the analytical index of $\mathcal{H}$
that counts the difference in the number of
zero modes of $\mathcal{H}$
with the chiral eigenvalues $\pm1$, respectively,
of a chiral operator that one may choose to be
\begin{equation}
\beta^{\mathrm{AF}}_{z}\:=
X^{\,}_{3333}
\label{eq: index thm e}
\end{equation}
\end{subequations}
without loss of generality.

If $\beta^{\mathrm{AF}}_{z}$ is used as a probe, i.e., as a small perturbation
to the  single-particle Hamiltonian
(\ref{eq: index thm a}),
it will lift the spin degeneracy of the zero modes through the Zeeman effect.
Which of the spin projections acquires a positive energy depends on the
eigenvalue of the chiral zero modes with respect to $\beta^{\mathrm{AF}}_{z}$.
In turn, the sign of this eigenvalue depends on which sublattice [even
versus odd sites as measured by $(-1)^{i^{\,}_{x}+i^{\,}_{y}+i^{\,}_{z}}$]
the chiral zero modes is non-vanishing, i.e.,
on the sign of the winding number. Hence, the core of the monopole in the VBS
order parameter $\bm{d}$ nucleates Neel order.
The same argument can be reversed to
infer that a monopole in the Neel order parameter
$\bm{n}$ nucleates dimer order at its core.

\subsection{Functional bosonization\\
Non-linear sigma model with a Wess-Zumino term}

We define the dimensionless vector field
$\bm{N}(\tau,\bm{r})\in\mathbb{R}^{6}$
comprised of the Neel, $\bm{n}(\tau,\bm{r})\in\mathbb{R}^{3}$,
and VBS (dimer), $\bm{d}(\tau,\bm{r})\in\mathbb{R}^{3}$,
order parameters through its components
\begin{subequations}
\label{eq: HS partition fct}
\begin{align}
&
\bm{N}(\tau,\bm{r})\equiv
\Big(
\bm{n}(\tau,\bm{r}),\bm{d}(\tau,\bm{r})
\Big),
\\
&
\bm{n}(\tau,\bm{r})\equiv
\Big(
n^{\,}_{x}(\tau,\bm{r}),
n^{\,}_{y}(\tau,\bm{r}),
n^{\,}_{z}(\tau,\bm{r})
\Big),
\\
&
\bm{d}(\tau,\bm{r})\equiv
\Big(
d^{\,}_{x}(\tau,\bm{r}),
d^{\,}_{y}(\tau,\bm{r}),
d^{\,}_{z}(\tau,\bm{r})
\Big).
\label{eq: HS partition fct a}
\end{align}
We define the single-particle Dirac Hamiltonian
\begin{equation}
\mathcal{H}\:=
\mathrm{i}\bm{\alpha}\cdot\bm{\partial}
+
m\,
\bm{N}(\tau,\bm{r})\cdot\bm{\beta},
\label{eq: HS partition fct b}
\end{equation}
where the constant $m$ has the dimension of inverse length
\begin{equation}
\bm{\alpha}\:=
\left(
\alpha^{\,}_{x},
\alpha^{\,}_{y},
\alpha^{\,}_{z}
\right)
\label{eq: HS partition fct c}
\end{equation}
and
\begin{equation}
\bm{\beta}\:=
\left(
\beta^{\mathrm{AF}}_{x},
\beta^{\mathrm{AF}}_{y},
\beta^{\mathrm{AF}}_{z},
\beta^{\mathrm{VB}}_{x},
\beta^{\mathrm{VB}}_{y},
\beta^{\mathrm{VB}}_{z}
\right).
\label{eq: HS partition fct d}
\end{equation}
We define the four Hermitian $16\times16$ matrices
\begin{equation}
\begin{split}
&
\gamma^{\,}_{0}\:=\beta^{\mathrm{AF}}_{x},
\\
&
\gamma^{\,}_{1}\:=\mathrm{i}\beta^{\mathrm{AF}}_{x}\,\alpha^{\,}_{x},
\\
&
\gamma^{\,}_{2}\:=\mathrm{i}\beta^{\mathrm{AF}}_{x}\,\alpha^{\,}_{y},
\\
&
\gamma^{\,}_{3}\:=\mathrm{i}\beta^{\mathrm{AF}}_{x}\,\alpha^{\,}_{z},
\end{split}
\label{eq: HS partition fct e}
\end{equation}
together with the six $16\times16$ matrices
\begin{equation}
\begin{split}
&
\Gamma^{\,}_{1}\:=
\beta^{\mathrm{AF}}_{x}\,\beta^{\mathrm{AF}}_{x},
\qquad
\Gamma^{\,}_{4}\:=
\beta^{\mathrm{AF}}_{x}\,\beta^{\mathrm{VB}}_{x},
\\
&
\Gamma^{\,}_{2}\:=
\beta^{\mathrm{AF}}_{x}\,\beta^{\mathrm{AF}}_{y},
\qquad
\Gamma^{\,}_{5}\:=
\beta^{\mathrm{AF}}_{x}\,\beta^{\mathrm{VB}}_{y},
\\
&
\Gamma^{\,}_{3}\:=
\beta^{\mathrm{AF}}_{x}\,\beta^{\mathrm{AF}}_{z},
\qquad
\Gamma^{\,}_{6}\:=
\beta^{\mathrm{AF}}_{x}\,\beta^{\mathrm{VB}}_{z}.
\end{split}
\label{eq: HS partition fct f}
\end{equation}
We define the partition function in $(3+1)$-dimensional Euclidean spacetime
to be
\begin{equation}
Z^{\,}_{\mathrm{HS}}\:=
\int\mathcal{D}[\bm{N}]\,
e^{-\int\mathrm{d}^{4}x\,\mathcal{L}^{\,}_{\mathrm{HS}}}\,
\int\mathcal{D}[\bar\psi,\psi]\,
e^{-\int\mathrm{d}^{4}x\,\mathcal{L}},
\label{eq: HS partition fct g}
\end{equation}
where
\begin{equation}
\mathcal{L}^{\,}_{\mathrm{HS}}\:=
\frac{1}{2}
\int\mathrm{d}^{4}x\,
\left[
\frac{\bm{n}^{2}(x)}{U^{\,}_{\mathrm{AF}}}
+
\frac{\bm{d}^{2}(x)}{U^{\,}_{\mathrm{VB}}}
\right]
\label{eq: HS partition fct h}
\end{equation}
and
\begin{equation}
\mathcal{L}\:=
\bar{\psi}(x)
\left[
\mathrm{i}\sum_{\mu=0}^{3}\gamma^{\,}_{\mu}\frac{\partial}{\partial x^{\,}_{\mu}}
+
\mathrm{i}m\,
\sum_{a=1}^{6}
N^{\,}_{a}(x)
\Gamma^{\,}_{a}
\right]
\psi(x).
\label{eq: HS partition fct i}
\end{equation}
\end{subequations}
Here, $U^{\,}_{\mathrm{AF}}\geq0$ and $U^{\,}_{\mathrm{VB}}\geq0$
are couplings with the dimension of
length raised to the power $(d+1)$.
Moreover, the sixteen components of
$\bar{\psi}(x)$ and the sixteen components of
$\psi(x)$ are Grassmann valued and independent.
Each component depends on the
position $x\equiv(x^{\,}_{\mu})\:=(\tau,\bm{r})\in\mathbb{R}^{4}$
in $(3+1)$-dimensional Euclidean spacetime. 
If we integrate over
the vector field (order parameter) $\bm{N}$ in the partition function,
there follows the quartic contact fermionic interaction density
\begin{subequations}
\label{eq: quartic interaction integrating neel and dimer order parameters}
\begin{equation}
\begin{split}
& 
\frac{m^{2}\,U^{\,}_{\mathrm{AF}}}{2}\,
\sum_{a=1}^{3}
(\bar{\psi}^{\,}\,\Gamma^{\,}_{a}\,\psi)^{2}
+
\frac{m^{2}\,U^{\,}_{\mathrm{VB}}}{2}\,
\sum_{a=4}^{6}
(\bar{\psi}^{\,}\,\Gamma^{\,}_{a}\,\psi)^{2}
\\
&
=
-
\frac{m^{2}\,U^{\,}_{\mathrm{AF}}}{2}\,
\sum_{a=1}^{3}
(\psi^{\dag}\beta^{\mathrm{AF}}_{a}\psi)^{2}
-
\frac{m^{2}\,U^{\,}_{\mathrm{VB}}}{2}\,
\sum_{a=1}^{3}
(\psi^{\dag}\beta^{\mathrm{VB}}_{a}\psi)^{2},
\label{eq: quartic interaction integrating neel and dimer order parameters a}
\end{split}
\end{equation}
where
\begin{equation}
\bar{\psi}\equiv\psi^{\dag}\,\mathrm{i}\beta^{\mathrm{AF}}_{x}.
\label{eq: quartic interaction integrating neel and dimer order parameters b}
\end{equation}
\end{subequations}
When the fermionic quartic interaction in the channel $a=1,\cdots,6$
is expressed in terms of
$(\psi^{\dag}\,\beta^{\,}_{a}\,\psi)^{2}$,
it may be interpreted as an attractive interaction, since 
$(\psi^{\dag}\,\beta^{\,}_{a}\,\psi)$, as an operator, is Hermitian
so that its square can only have positive or vanishing eigenvalues.

Alternatively,
we may also define the partition function
\begin{equation}
Z^{\,}_{\mathrm{NLSM}}\:=
\int\mathcal{D}[\bm{N}]\,
\delta\left(\bm{N}^{2}-1\right)\,
\int\mathcal{D}[\bar\psi,\psi]\,
e^{-\int\mathrm{d}^{4}x\,\mathcal{L}}.
\label{eq: Z NLSM}
\end{equation}
As any $m\neq0$ opens a spectral gap
in the single-particle Dirac spectrum, we can integrate approximately
the Grassmann fields within a gradient expansion.
There follows the approximate bosonic partition function
\begin{subequations}
\label{eq: bosonization}
\begin{equation}
Z^{\,}_{\mathrm{NLSM}}\approx
\int\mathcal{D}[\bm{N}]\,
\delta\left(\bm{N}^{2}-1\right)\,
e^{\mathrm{i}S^{\,}_{\mathrm{topo}}[\bm{N}]}\,
e^{-\int\mathrm{d}\tau\int\mathrm{d}^{3}\bm{x}\,\mathcal{L}^{\,}_{\mathrm{eff}}}.
\label{eq: bosonization a}
\end{equation}
The Lagrangian density
\begin{equation}
\mathcal{L}^{\,}_{\mathrm{eff}}\:=
\frac{1}{2g}
\sum_{\mu=0}^{3}
\left(\partial^{\,}_{\mu}\bm{N}\right)^{2},
\qquad
x\equiv(\tau,\bm{x})\equiv(x^{\,}_{\mu}),
\label{eq: bosonization b}
\end{equation}
in imaginary time
$x^{\,}_{0}\equiv\tau\:=\mathrm{i}t$
is that of the non-linear sigma model (NLSM) with the unit sphere
$\mathsf{S}^{6-1}=\mathsf{S}^{5}$ as the target manifold.
The bare coupling
\begin{equation}
g\propto m^{1-d}
\end{equation}
has the dimension
\begin{equation}
[g]=[\mathrm{length}]^{d-1}
\end{equation}
with $d=3$.
A necessary condition for the presence of the phase factor
$\exp(\mathrm{i}S^{\,}_{\mathrm{topo}}[\bm{n}])$
-- \textit{one that is compatible with locality} --
is that the homotopy group
\begin{equation}
\pi^{\,}_{n}\left(\mathsf{S}^{5}\right)\neq\emptyset
\label{eq: bosonization c}
\end{equation}
\end{subequations}
is not trivial for one of the integers
$n=1,2,\cdots,5$ (the upper bound $5=d+2$ with $d=3$ on $n$ follows from
demanding that the equations of motion for $\bm{N}$ are local).
This condition is only met for $n=5$.

Explicit computation
(see Ref.\ \onlinecite{Abanov00} and references therein)
yields the non-vanishing topological action given by
\begin{subequations}
\begin{equation}
S^{\,}_{\mathrm{topo}}[\bm{N}]=
2\pi\,
S^{\,}_{\mathrm{WZ}}[\bm{N}].
\end{equation}
Here, $S^{\,}_{\mathrm{WZ}}[\bm{N}]$ is the Wess-Zumino action,
an action that is non-local in $(3+1)$-dimensional Euclidean spacetime,
but delivers local equations of motions.
When Euclidean spacetime $\mathbb{R}^{3+1}$
is compactified to $\mathsf{S}^{3+1}$,
the Wess-Zumino action is given by
\begin{equation}
\begin{split}
&
S^{\,}_{\mathrm{WZ}}[\bm{N}]=
\frac{1}{(3+2)!\,\mathrm{Area}(\mathsf{S}^{3+2})}\,
\int\limits_{0}^{1}\mathrm{d}u\,
\int\limits_{\mathsf{S}^{3+1}}\mathrm{d}^{3+1}x
\\
&\,
\epsilon^{\,}_{\mu^{\,}_{1}\cdots\mu^{\,}_{3+2}}\,
\epsilon^{\,}_{aa^{\,}_{1}\cdots a^{\,}_{3+2}}\,
\bar{N}^{\,}_{a}(u,x)\,
\partial^{\,}_{\mu^{\,}_{1}}
\bar{N}^{\,}_{a^{\,}_{1}}(u,x)
\cdots
\\
&\,
\cdots
\partial^{\,}_{\mu^{\,}_{3+2}}
\bar{N}^{\,}_{a^{\,}_{3+2}}(u,x),
\end{split}
\label{eq: main result hierarchy III b}
\end{equation}
whereby the vector field
\begin{equation}
\bm{\bar{N}}(u,x)=
\left(
\bar{N}^{\,}_{1}(u,x),
\cdots,
\bar{N}^{\,}_{3+2}(u,x),
\bar{N}^{\,}_{3+3}(u,x)
\right)
\label{eq: main result hierarchy III c}
\end{equation}
smoothly interpolates between
\begin{equation}
\bm{\bar{N}}(0,x)\:=
\left(0,\cdots,0,N^{\,}_{3+3}(x)\right)
\label{eq: main result hierarchy III d}
\end{equation}
and
\begin{equation}
\bm{\bar{N}}(1,x)\:=
\left(N^{\,}_{1}(x),\cdots,N^{\,}_{3+2}(x),N^{\,}_{3+3}(x)\right)\in
\mathsf{S}^{3+2}
\label{eq: main result hierarchy III e}
\end{equation}
as a function of $0\leq u\leq1$.
\end{subequations}
The real-valued vector field $\bm{\bar{N}}(u,x)$ is therefore defined on a disk
$D\subset\mathbb{R}^{3+2}$ such that its boundary is the compactified spacetime
$\mathsf{S}^{3+1}$, i.e., $\partial D=\mathsf{S}^{3+1}$. The existence of
the smooth vector field
(\ref{eq: main result hierarchy III c})
obeying conditions
(\ref{eq: main result hierarchy III d})
and
(\ref{eq: main result hierarchy III e})
is guaranteed from the identity
$\pi^{\,}_{3+1}(\mathsf{S}^{3+2})=\emptyset$.

\subsection{Phase diagrams}

\begin{figure*}[t]
\includegraphics[width=0.8\linewidth]{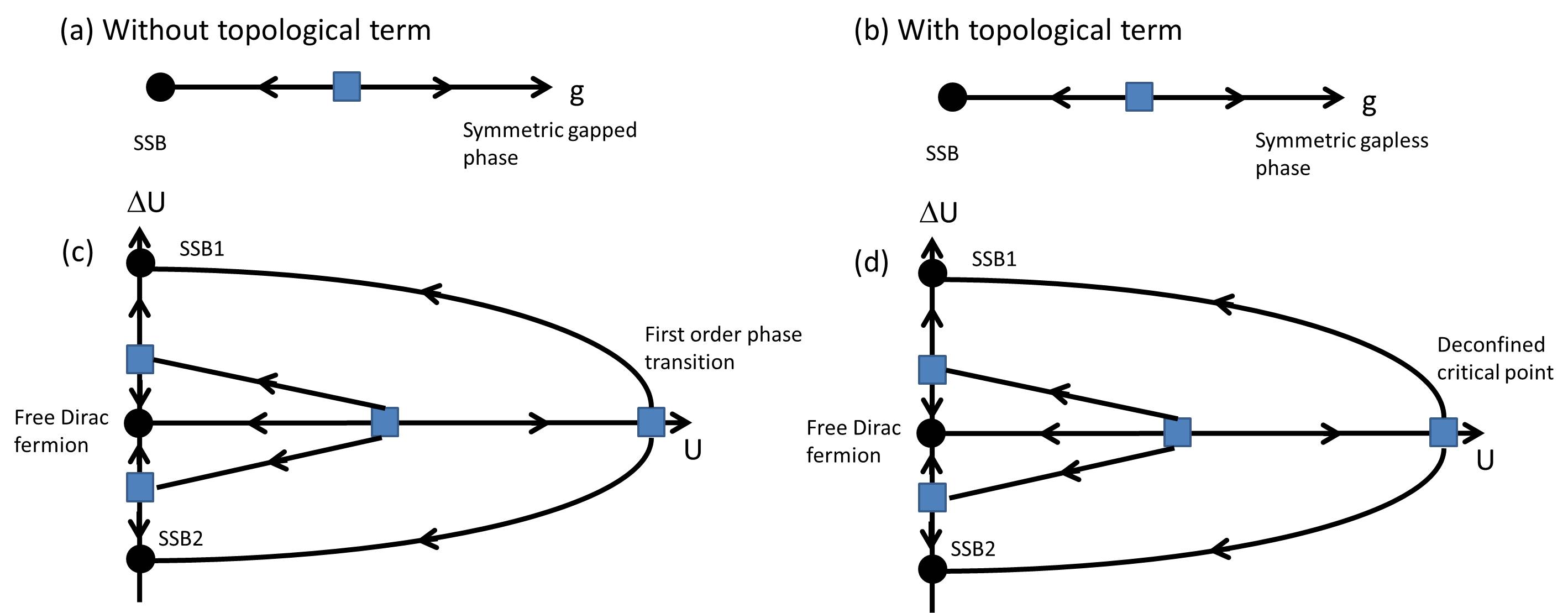}
\caption{(Color online)
Renormalization-group (RG) flows for the NLSM with O(6) symmetry
in $(3+1)$-dimensional spacetime
(a) without any topological term 
and (b) with a Wess-Zumino term.
Panels (c) and (d) show RG flows for Dirac fermions
perturbed by O(3)$\times$O(3) symmetric interactions
in $(3+1)$-dimensional spacetime,
assuming that bosonization
in their Mott insulating phases delivers
the RG flows in panel (a) and (b), respectively,
when the interactions are fine-tuned to an O(6) symmetric interaction.
\label{fig: phase diagrams}
        }
\end{figure*}

In this section, we discuss the phase diagrams of models
(\ref{eq: bosonization})
and
(\ref{eq: HS partition fct}),
in this order. We then discuss the phase diagram of
a cubic lattice model that realizes the $\pi$-flux phase in the
noninteracting limit and,
upon switching on local fermionic interactions with O(3)$\times$O(3) symmetry,
can be described by the effective field theory
(\ref{eq: HS partition fct}) in the low-energy limit.

With regard to the NLSM (\ref{eq: bosonization}), we need to review
the RG flow of NLSMs on Riemannian manifolds with
positive curvature at zero temperature when space has dimension $d\ge2$
(i.e., spacetime has dimension greater than or equal to 3).
In the absence of a topological term, there are two phases
as shown in Fig.~\ref{fig: phase diagrams}(a).
(1) There is a long-range-ordered phase with spontaneous breaking
of a continuous symmetry when $g<g^{\,}_{\mathrm{c}}$.
(2) There is a symmetric quantum disordered gapped phase when
$g>g^{\,}_{\mathrm{c}}$.
(3) The quantum critical point $g=g^{\,}_{\mathrm{c}}$ realizes
a continuous phase transition between these two phases of matter.
Adding a topological term does not modify the perturbative
RG flow when  $g\lesssim g^{\,}_{\mathrm{c}}$. However, it does change the nature
of the fixed point of the RG flow when $g>g^{\,}_{\mathrm{c}}$,
as this fixed point now describes a symmetric gapless phase,
as is indicated in Fig.~\ref{fig: phase diagrams}(b).

With regard to the model (\ref{eq: HS partition fct}),
we conjecture that its phase diagram at zero temperature
can be deduced from the phase diagram in
Figs.\  \ref{fig: phase diagrams}(a) and \ref{fig: phase diagrams}(b)
as follows. Let
\begin{subequations}
\begin{equation}
\Delta U\:=
\frac{
U^{\,}_{\mathrm{AF}}
-
U^{\,}_{\mathrm{VB}}
     }
     {
2
     }
\end{equation}
measure the anisotropy in the relative strength between
the coupling $U^{\,}_{\mathrm{AF}}\geq0$
of the interaction favoring Neel order and
the coupling $U^{\,}_{\mathrm{VB}}\geq0$
of the interaction favoring dimer order.
We denote with
\begin{equation}
U\:=
\frac{
U^{\,}_{\mathrm{AF}}
+
U^{\,}_{\mathrm{VB}}
     }
     {
2
     }
\end{equation}
the mean value of $U^{\,}_{\mathrm{AF}}\geq0$ and $U^{\,}_{\mathrm{VB}}\geq0$.
The isotropic case is defined by
\begin{equation}
\Delta U=0,
\qquad
U=
U^{\,}_{\mathrm{AF}}=
U^{\,}_{\mathrm{VB}}.
\label{eq: def isotropic case}
\end{equation}
\end{subequations}

We consider the isotropic case (\ref{eq: def isotropic case})
first, in which case the O(3)$\times$O(3) symmetry of
the anisotropic quartic contact fermionic interaction
(\ref{eq: quartic interaction integrating neel and dimer order parameters a})
becomes the O(6) symmetry of the isotropic contact fermionic interaction
\begin{equation}
\begin{split}
&
\frac{m^{2}\,U}{2}\,
\sum_{a=1}^{6}
(\bar{\psi}^{\,}\,\Gamma^{\,}_{a}\,\psi)^{2}=
\\
&\qquad
-
\frac{m^{2}\,U}{2}
\left[
\sum_{a=1}^{3}
(\psi^{\dag}\beta^{\mathrm{AF}}_{a}\psi)^{2}
+
\sum_{a=1}^{3}
(\psi^{\dag}\beta^{\mathrm{VB}}_{a}\psi)^{2}
\right].
\end{split}
\label{eq: quartic interaction Delta U=0}
\end{equation}
Since any point-contact interaction for Dirac fermions at half-filling
is irrelevant within perturbative RG when the dimensionality $d$
of space is larger than one ($d>1$),
the ground state for small $U$ is adiabatically connected to that
in the noninteracting limit $U=0$. Upon increasing
the coupling $U$ above some critical value $U^{\,}_{\mathrm{c}}$,
a single-particle gap opens, long-range order is established, and 
the mapping to an effective NLSM becomes a good approximation at low-energies.
Since the bare coupling $g$ in the NLSM and the bare fermionic coupling $U$
have the relationship $g\sim U$,
we expect a long-range ordered phase for intermediate values of $U$,
and a symmetric phase for larger values of $U$ that is either gapped in the
absence of a topological term
[Fig.~\ref{fig: phase diagrams}(c) with $\Delta U=0$]
or gapless in the presence of a topological term
[Fig.~\ref{fig: phase diagrams}(d) with $\Delta U=0$].

\begin{figure*}[t]
\includegraphics[width=0.7\linewidth]{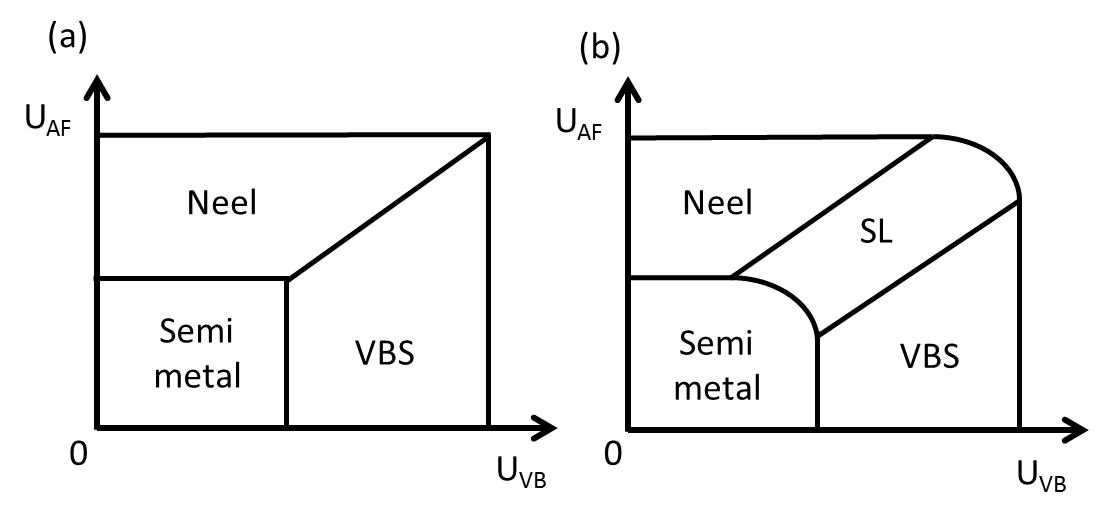}
\caption{
Two possible phase diagrams for interacting fermionic tight-binding models
at half-filling with an O(3)$\times$O(3) symmetry
that realize the $\pi$-flux phase on the cubic lattice
in the noninteracting limit.
(a) There exists a phase boundary between the
Neel and dimer (valence bond solid) phases
that is governed by a continuous quantum critical point
at which the dual point defects of either phases have
simultaneously proliferated.
(b) The Neel and dimer (valence bond solid) phases do not touch.
They are separated by a gapless spin liquid (SL) phase
characterized by the dual point defects of either phases being
simultaneously deconfined.
\label{fig: phase diagram 3D}
        }
\end{figure*}

If we break the O(6) symmetry of the quartic interaction
(\ref{eq: quartic interaction Delta U=0})
by assuming that $\Delta U\neq0$ in
\begin{equation}
U^{\,}_{\mathrm{AF}}=U+\Delta U,
\qquad
U^{\,}_{\mathrm{VB}}=U-\Delta U,
\label{eq: anisotropic U}
\end{equation}
we infer the phase diagrams
in Fig.~\ref{fig: phase diagrams}(c)
and
Fig.~\ref{fig: phase diagrams}(d)
depending on the absence or presence of dual topological point defects,
respectively.
Indeed, the choice $U\gtrsim U^{\,}_{\mathrm{c}}$ with $\Delta U>0$
selects the mean-field order parameter
$\bar{\bm{N}}=(\bar{\bm{n}},\bar{\bm{d}})$
aligned along the antiferromagnetic direction
$(\bar{\bm{n}},\bm{0})$.
The choice $U\gtrsim U^{\,}_{\mathrm{c}}$ with $\Delta U<0$
selects the mean-field order parameter
$\bar{\bm{N}}=(\bar{\bm{n}},\bar{\bm{d}})$
aligned along the direction of $(\bm{0},\bar{\bm{d}})$.
These two ordered phases are separated by a first-order
phase transition point in the NLSM perturbed by a symmetry breaking term
in the absence of a topological term as shown in
Fig.~\ref{fig: phase diagrams}(c)
[we have ignored the case of a phase with coexisting orders for simplicity].
However, in the presence of a topological term, the segment
$U^{\,}_{\mathrm{c}}<U<U^{\,}_{\star}$ and $\Delta U=0$
should instead be governed by a quantum critical
point (which is a $d=3$ analogue of the deconfined quantum
critical point proposed in
Refs.\ \onlinecite{Senthil04,Senthil04PRB,Senthil04Levin} when $d=2$)
at $U=U^{\,}_{\star}$ and $\Delta U=0$,
see Fig.~\ref{fig: phase diagrams}(d).

The phase diagram of a local fermionic lattice regularization of the model
(\ref{eq: HS partition fct})
would then look as follows, see Fig.\ \ref{fig: phase diagram 3D}.
When both $U^{\,}_{\mathrm{AF}}$ and $U^{\,}_{\mathrm{VB}}$ are small, fermionic 
interactions are irrelevant perturbations to
the semi-metallic phase (the noninteracting $\pi$-flux phase).
When $U^{\,}_{\mathrm{VB}}=0$ while increasing $U^{\,}_{\mathrm{AF}}>0$
across the critical  value $U^{\mathrm{c}}_{\mathrm{AF}}>0$,
the semi-metallic phase is unstable to
a Neel long-ranged-ordered antiferromagnetic (Mott) insulating phase. 
When $U^{\,}_{\mathrm{AF}}=0$ while increasing $U^{\,}_{\mathrm{VB}}$
across the critical  value $U^{\mathrm{c}}_{\mathrm{VB}}>0$,
the semi-metallic phase is unstable to
a dimer long-ranged-ordered (Mott) insulating phase. 
Now, the $\pi$-flux phase on the square lattice
perturbed by local fermionic interactions with
O(3)$\times$O(2) symmetry shows the direct phase transition between
Neel and dimer phases
for sufficiently large $U^{\,}_{\mathrm{AF}}=U^{\,}_{\mathrm{VB}}$.
This direct phase transition is governed by an unstable fixed point,
namely the fixed point of the NLSM with a topological term that describes
a gapless symmetric phase with O(5) symmetry.%
~\cite{TanakaHuPRL2005,Senthil06Fisher,Sato2017}
Similarly, one possible scenario for the $\pi$-flux on the cubic lattice
that is perturbed by local fermionic interactions with
O(3)$\times$O(3) symmetry 
is that the Neel and dimer phases are separated by a phase boundary
that is governed by a single unstable fixed point,
namely the fixed point of the 
NLSM with a topological term that describes a gapless
symmetric phase with O(6) symmetry, as is shown in
Fig.\ \ref{fig: phase diagram 3D}(a). However, working in $d=3$
allows for another scenario that is shown in
Fig.\ \ref{fig: phase diagram 3D}(b).
In three-dimensional space, the antiferromagnetic and dimer dual
point defects might be simultaneously
deconfined in an extended region of coupling space
instead of a single point in coupling space as is the case
in two-dimensional space.%
~\cite{Hosur10}
If so, a gapless spin-liquid (SL) phase, in which
some putative matter fields are coupled to Abelian gauge fields
in a Coulomb-like phase,
could separate the Neel phase from the dimer long-range ordered phases.

\section{Summary}
\label{sec: Summary}

The phase diagram at vanishing temperature
of the quantum spin-1/2 antiferromagnetic $J^{\,}_{1}$-$J^{\,}_{2}$} XYZ chain
was studied using both bosonization and numerical techniques. 
The symmetry group of the quantum spin-1/2 $J^{\,}_{1}$-$J^{\,}_{2}$ XYZ chain
obeying periodic boundary conditions is generated by
\begin{subequations}
\begin{equation}
\mathfrak{G}\:=
R^{\alpha}_{\pi}
\times
R^{\beta}_{\pi}
\times
T
\times
P
\times
\Theta.
\end{equation}
Here, $R^{\alpha}_{\pi}$ and $R^{\beta}_{\pi}$
denote any pair of distinct $\pi$-rotations around the $\alpha\neq\beta=x,y,z$
axis in spin space,
$T$ denotes a translation by one lattice spacing,
$P$ denotes a site inversion, and $\Theta$ denotes reversal of time.
It was shown that there are four gapped long-ranged ordered phases
consisting of three Neel phases and one dimer phase.
The corresponding patterns of spontaneous symmetry breaking (SSB) are
\begin{align}
\mathfrak{G}\to
\mathfrak{G}^{\,}_{\mathrm{N}^{\,}_{\alpha}}\:=
R^{\alpha}_{\pi}
\times
P
\times
\left(T\,\Theta\right)
\end{align}
for the Neel${}^{\,}_{\alpha}$ phase
with $\alpha=x,y,z$
and
\begin{align}
\mathfrak{G}\to
\mathfrak{G}^{\,}_{\mathrm{VBS}}\:=
R^{\alpha}_{\pi}
\times
R^{\beta}_{\pi}
\times
\Theta
\end{align}
\end{subequations}
for the dimer (VBS) phase. Because no pair of these residual symmetry groups
obeys an ordering relation through the inclusion, Landau's theory of
phase transitions precludes a direct continuous phase transition between
any pair of these long-range ordered phases. Instead, Landau's theory of
phase transitions predicts either coexistence or a direct first-order
phase transition. Contrary to this expectation, we have shown that
the three Neel phases and the dimer phase
are separated from each other by six planes of phase boundaries
realizing Gaussian criticality when $0\leq J^{\,}_{2}/J^{\,}_{1}<1/2$.
We also have shown that each long-range ordered phase harbors
topological point defects (domain walls)
that are dual to those across the phase boundary
in that a defect in one ordered phase locally binds
the other type of order around its core.
The Landau-forbidden continuous phase transitions are driven
by the simultaneous proliferation (deconfinement) of these dual
topological point-like defects.

We have also shown that a one-dimensional model of interacting
fermions with a suitable choice of interactions can undergo
a Landau-forbidden phase transition belonging to the same
Gaussian universality class as those in the
quantum spin-1/2 antiferromagnetic $J^{\,}_{1}$-$J^{\,}_{2}$ XYZ chain.
Moreover, the mechanism at play here is not tied to the dimensionality of
space. This observation led us
to consider a tight-binding model on the cubic lattice
that realizes a semi-metallic phase in the noninteracting limit
(the $\pi$-flux phase). Upon linearization of the
noninteracting spectrum about the
Fermi points (Dirac points) and the addition of interactions displaying an
O(3)$\times$O(3) symmetry in the continuum,
sufficiently strong interactions can stabilize two Mott phases.
One Mott phase supports colinear antiferromagnetic order.
The other Mott phase supports dimer long-range order on the lattice.
Both ordered phases were shown to support topological point defects,
hedgehogs, that are dual to each other
in that a defect in one ordered phase locally binds
the other type of order around its core.
When the bare interaction strengths are fine-tuned so as to
display the symmetry O(6) and assuming that the bare interactions
select a Mott insulating phase with the 
pattern of symmetry breaking
O(6)$\to$O(5),
functional bosonization yields a non-linear sigma model augmented
by a Wess-Zumino term. From this fact, we conjectured that
the Mott insulating phases are either separated by a
phase boundary governed by a quantum critical point 
or by a gapless spin liquid phase, both displaying
an O(6) symmetry and simultaneous proliferation of the dual hedgehogs.
A lattice regularization of the fermionic field theory could be amenable
to sign-free Monte-Carlo simulations, as was done in Refs.\
\onlinecite{Sato2017,Liu18assaad}
in two dimensional space.

\begin{acknowledgements}
  
A.F.\ was supported in part by JSPS KAKENHI Grant Number 15K05141.
T.H.\ was supported by JSPS KAKENHI
Grant Numbers 15K05198 and 17H02931.
T.M. was supported by the Gordon and Betty Moore Foundation's EPiQS
Initiative Theory Center Grant (to UC Berkeley), and the Quantum
Materials program at Lawrence Berkeley National Laboratory (LBNL)
funded by the US DOE under Contract DE-AC02-05CH11231.

\end{acknowledgements}

\appendix

\section{More on numerics}
\label{sec:Appendix}

\begin{figure}
\begin{center}
\includegraphics[width=0.4\textwidth]{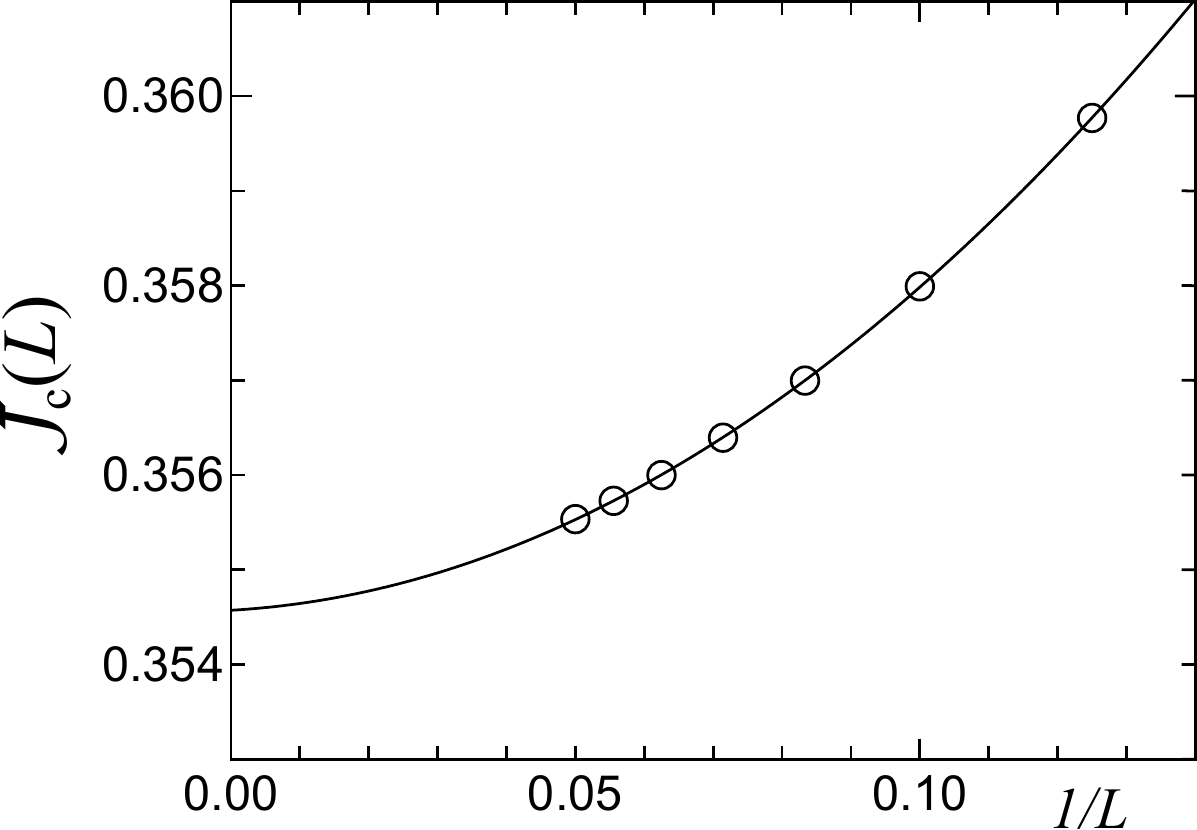}
\caption{
Extrapolation of the critical coupling $\mathcal{J}^{\,}_{\mathrm{c}}(L)$
to its limit $\mathcal{J}^{\,}_{\mathrm{c}}$ when $L\to\infty$
for $\Delta^{\,}_{z}=2.0$ and $\Delta^{\,}_{y}=0.5$.
}
\label{fig:extrpl_J2c_App}
\end{center}
\end{figure}

\subsection{The critical coupling $\mathcal{J}^{\,}_{\mathrm{c}}(L)$}

The finite-size critical couplings
$\mathcal{J}^{\,}_{\mathrm{c}}(L)$,
obtained from the position as a function of $0\leq\mathcal{J}<1/2$
of the cusp singularity of $\Delta E^{\,}_{0}(L)$
for given values of $L$, $\Delta^{\,}_{y}$, and $\Delta^{\,}_{z}$ entering
the quantum spin-1/2 antiferromagnetic $J^{\,}_{1}$-$J^{\,}_{2}$ XYZ Hamiltonian
(\ref{eq: def HXYZ a}),
is extrapolated to its thermodynamic limit
\begin{subequations}
\begin{equation}
\mathcal{J}^{\,}_{\mathrm{c}}\equiv\lim_{L\to\infty}\mathcal{J}^{\,}_{\mathrm{c}}(L)
\end{equation}
using the second-order polynomial in $1/L$ given by
\begin{equation}
\mathcal{J}^{\,}_{\mathrm{c}}(L)=
\mathcal{J}^{\,}_{\mathrm{c}}
+
\frac{\alpha^{\,}_{1}}{L}
+
\frac{\alpha^{\,}_{2}}{L^{2}}.
\end{equation}
\end{subequations}
The fitting was done using the values of $\mathcal{J}^{\,}_{\mathrm{c}}(L)$
for $L=8,10,...,20$ obtained from exact diagonalization as input
and taking $\mathcal{J}^{\,}_{\mathrm{c}}$,
$\alpha^{\,}_{1}$, and $\alpha^{\,}_{2}$ as free parameters.
Figure\ \ref{fig:extrpl_J2c_App} shows the results
for $\Delta^{\,}_{z}=2.0$ and $\Delta^{\,}_{y}=0.5$.
The errors in $\mathcal{J}^{\,}_{\mathrm{c}}$ are estimated from
the difference between the extrapolated value
and $\mathcal{J}^{\,}_{\mathrm{c}}(L=20)$.
They are less than 0.4 \% of $\mathcal{J}^{\,}_{\mathrm{c}}$.
Incidentally, the coefficient $\alpha^{\,}_{1}$ is
much smaller than $\mathcal{J}^{\,}_{\mathrm{c}}$ and $\alpha^{\,}_{2}$
for all the cases calculated:
$\alpha^{\,}_{1}$ is of the order $10^{-3}$
while $\mathcal{J}^{\,}_{\mathrm{c}}$ and $\alpha^{\,}_{2}$ are
of the order $10^{-1}$.
This suggests that $\alpha^{\,}_{1}=0$
as was already found for
the quantum spin-1/2 antiferromagnetic $J^{\,}_{1}$-$J^{\,}_{2}$ XXZ chain.%
~\cite{Nomura1994}

\subsection{DMRG}

We performed the DMRG calculations
for open chains hosting up to $L=192$ spins
($L$ is chosen a multiple of four).
The maximum number of the states that were kept was $\chi=160$.
We checked that the average of
the weight of discarded states at each step over the final DMRG sweep
was smaller than $6\times10^{-11}$.
We thereby confirmed that the DMRG data are accurate enough
for our analysis.

\subsection{Entanglement entropy}

We consider an open chain made hosting
$L$ spins with $L$ being a multiple of four.
Let ${{l}}\geq1$ be an integer smaller than $L$.
The entanglement entropy $\mathcal{S}({{l}})$ is defined by
\begin{equation}
\mathcal{S}({{l}})\:=
-
\sum_{j}
\rho^{\,}_{{{l}}}(j)\ln\rho^{\,}_{{{l}}}(j),
\label{eq:EEnt_definition}
\end{equation}
where $\rho^{\,}_{{{l}}}(j)$ is the $j$th eigenvalue of the sub-density matrix
for the left ${{l}}$-site block in the ground state of the full open chain.
As seen from Fig.\ \ref{fig: entanglement entropy}, $\mathcal{S}({{l}})$
contains a sizable oscillating component that arises 
from the use of open boundary conditions.
It was found numerically%
~\cite{LaflorencieSCA2006,AffleckLS2009}
that the oscillating contribution to the entanglement entropy
$\mathcal{S}({{l}})$
that originates from choosing open boundary conditions
is proportional to the oscillating component of 
the local bond energy expectation value,
\begin{subequations}
\label{eq: def Esc}
\begin{equation}
E^{\,}_{\mathrm{osc}}({{l}})\:=E^{\,}_{\mathrm{bond}}({{l}})
-E^{\,}_{\mathrm{uni}},
\label{eq: def Esc a}
\end{equation}
with
\begin{equation}
\begin{split}
E^{\,}_{\mathrm{bond}}({{l}})&\:=
J^{\,}_{1}
\left\langle
\left(
S^{x}_{{{l}}}\,S^{x}_{{{l}}+1}
+
\Delta^{\,}_{y}\,S^{y}_{{{l}}}\, S^{y}_{{{l}}+1}
+
\Delta^{\,}_{z}\,S^{z}_{{{l}}\,} S^{z}_{{{l}}+1}
\right)
\right\rangle^{\,}_{L}
\\
&
+
\frac{J^{\,}_{2}}{2}
\left\langle
\left(
S^{x}_{{{l}}-1}\,S^{x}_{{{l}}+1}
+
\Delta^{\,}_{y}\,S^{y}_{{{l}}-1}\,S^{y}_{{{l}}+1}
+
\Delta^{\,}_{z}\,S^{z}_{{{l}}-1}\,S^{z}_{{{l}}+1}
\right.
\right.
\\
&
\left.
\left.
~~~~~~~~
+
S^{x}_{{{l}}}\,S^{x}_{{{l}}+2}
+
\Delta^{\,}_{y}\,S^{y}_{{{l}}}\,S^{y}_{{{l}}+2}
+
\Delta^{\,}_{z}\,S^{z}_{{{l}}}\,S^{z}_{{{l}}+2}
\right)
\right\rangle^{\,}_{L}.
\end{split}
\label{eq: def Esc b}
\end{equation}
The oscillating component $E^{\,}_{\mathrm{osc}}({{l}})$
enters $\mathcal{S}({{l}})$ through
\begin{equation}
\mathcal{S}({{l}})=
\frac{c}{6}\,
\ln\!\left[f\!\left(l+\frac{1}{2}\right)\right]
+
\alpha^{\,}_{\mathrm{osc}}\,
E^{\,}_{\mathrm{osc}}({{l}})
+
\mathcal{S}^{\,}_{0}.
\label{eq: def Esc c}
\end{equation}
We have computed the local bond energy expectation value
$E^{\,}_{\mathrm{bond}}({{l}})$ as well as
the entanglement entropy $\mathcal{S}({{l}})$ using the DMRG method
and obtained $E^{\,}_{\mathrm{osc}}({{l}})$ by subtracting 
\begin{equation}
E^{\,}_{\mathrm{uni}}\:=
\frac{1}{2}
\left[
E^{\,}_{\mathrm{bond}}\!\left({{\frac{L}{2}}}\right)
+
E^{\,}_{\mathrm{bond}}\!\left({{\frac{L}{2}+1}}\right)
\right],
\label{eq: def Esc d}
\end{equation}
\end{subequations}
from $E^{\,}_{\mathrm{bond}}({{l}})$.
Then, we have performed the least-square fitting of the data
of $\mathcal{S}({{l}})$ and $E^{\,}_{\mathrm{osc}}({{l}})$
to Eq.\ (\ref{eq: def Esc c})
taking $c$, $\alpha^{\,}_{\mathrm{osc}}$,
and $\mathcal{S}^{\,}_{0}$ as fitting parameters.
The data around the center of an open chain
(for $3L/8\leq l\leq 5L/8$) were used in the fitting.
We thereby determine the central charge $c$.

\subsection{Long-range order from DMRG} 

\begin{figure}
\begin{center}
\includegraphics[width=0.4\textwidth]{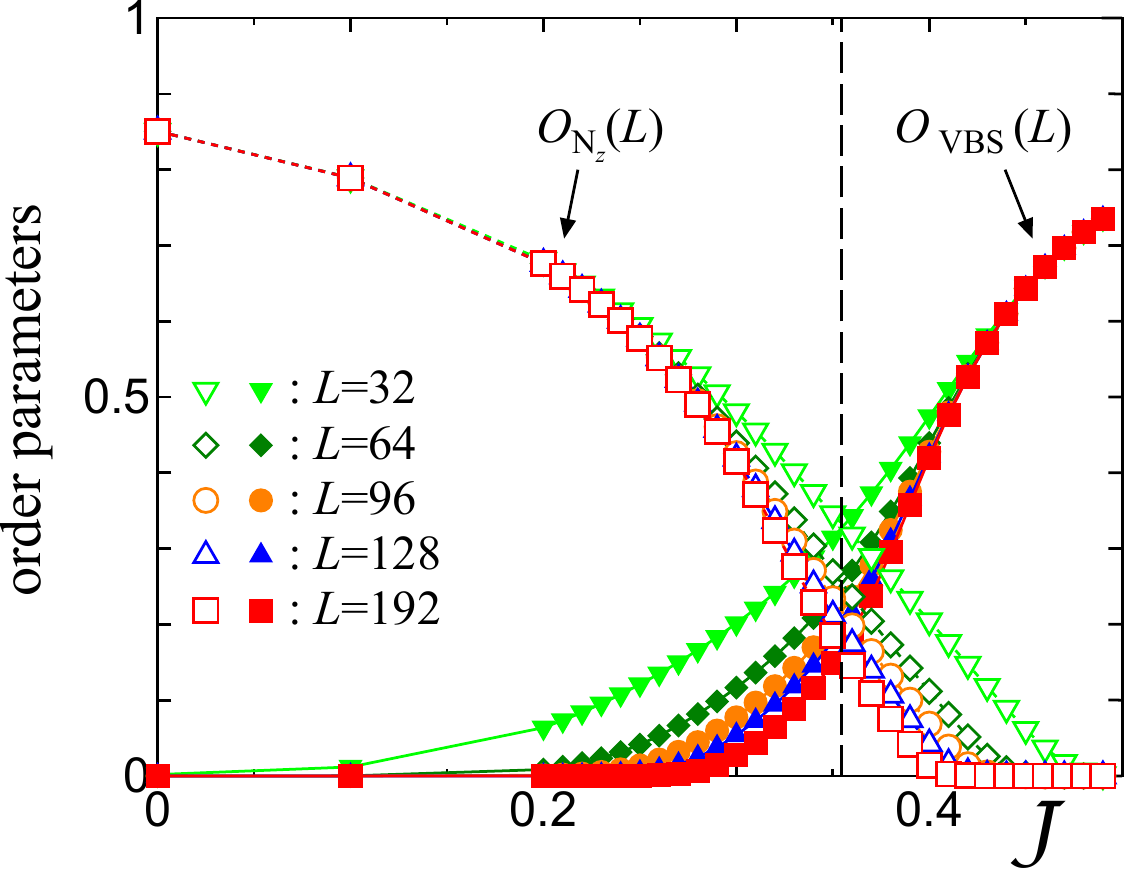}
\caption{(Color online)
Order parameters
$O^{\,}_{\mathrm{N}^{\,}_{z}}\!(L)$
(unfilled symbols)
and
$O^{\,}_{\mathrm{VBS}}(L)$
(filled symbols)
for $\Delta^{\,}_{z}=2.0$ and $\Delta^{\,}_{y}=0.5$
as functions of $\mathcal{J}$.
}
\label{fig:order-parameter_App}
\end{center}
\end{figure}

We have measured the dimer and Neel${}^{\,}_{z}$ order parameters
using the DMRG method on each sides of a continuous quantum critical point
reached by moving away from the value $\mathcal{J}^{\,}_{\mathrm{c}}$
holding the anisotropies $\Delta^{\,}_{y}$ and $\Delta^{\,}_{z}$ fixed.
Here, $\mathcal{J}^{\,}_{\mathrm{c}}$ was identified by using methods based on 
exact diagonalization. 

With regard to the dimer order,
we calculated the local order parameter
$O^{\,}_{\mathrm{VBS}}(L)$
defined in Eq.\ (\ref{eq:O_dimer_L}).
Whereas the two inequivalent dimer states are degenerate
for any finite chain obeying periodic boundary conditions,
this is not true anymore for any finite chain
obeying open boundary conditions, as translation symmetry is
broken by the two boundaries.

With regard to the Neel${}^{\,}_{z}$ order,
the choice of open boundary conditions for a chain hosting an even number of
spins is compatible with reversal of time,
the operation that exchanges the two inequivalent
classical Neel${}^{\,}_{z}$ states.
Hence, no spontaneous symmetry breaking of time-reversal symmetry
occurs for a chain hosting a finite and even number of spins. 
In order to detect numerically Neel${}^{\,}_{z}$ order,
we added to Hamiltonian (\ref{eq: def HXYZ a})
a symmetry-breaking term by coupling the first and last spins
to a staggered magnetic field, i.e.,
we added to Hamiltonian (\ref{eq: def HXYZ a}) the boundary energy cost
\begin{equation}
\mathcal{H}^{\,}_{h}\:=
-h\left[S^{z}_{1}-(-1)^{L}\,S^{z}_{L}\right]
\end{equation}
with $h=100\,J^{\,}_{1}$.
By design, $\mathcal{H}^{\,}_{h}$
lifts the degeneracy of the classical Neel${}^{\,}_{z}$ states.
We then calculated the local Neel${}^{\,}_{z}$-order parameter
$O^{\,}_{\mathrm{N}^{\,}_{z}}(L)$
defined by
\begin{equation}
\begin{split}
O^{\,}_{\mathrm{N}^{\,}_{z}}\!(L)\:=
\left\langle
\left(S^{z}_{\frac{L}{2}+1}-S^{z}_{\frac{L}{2}}\right)
\right\rangle^{\,}_{L}.
\end{split}
\end{equation}
We note that $\mathcal{H}^{\,}_{h}$ is added
only when we compute $O^{\,}_{\mathrm{N}^{\,}_{z}}(L)$.
We set $h=0$ for all other observables.

Figure\ \ref{fig:order-parameter_App}
shows the dependence on $\mathcal{J}$ of the 
Neel${}^{\,}_{z}$ order parameter $O^{\,}_{\mathrm{N}^{\,}_{z}}(L)$
and of the dimer order parameter $O^{\,}_{\mathrm{VBS}}(L)$
for $\Delta^{\,}_{z}=2.0$, $\Delta^{\,}_{y}=0.5$, and given $L$.
These data suggest that the model exhibits
the Neel${}^{\,}_{z}$ long-range order for
$\mathcal{J}<\mathcal{J}^{\,}_{\mathrm{c}}$
and the dimer long-range order for
$\mathcal{J} > \mathcal{J}^{\,}_{\mathrm{c}}$.
We have performed the same analysis along
all one-dimensional cuts 
with $\Delta^{\,}_{z}=2.0$ and
the values of $\Delta^{\,}_{y}$
given in Fig.\ \ref{fig: scaled gap and critical cal J}(b) 
for which a putative Neel${}^{\,}_{z}$ long-range ordered phase
is separated from a putative dimer long-range ordered phase
by a continuous quantum critical point
$\mathcal{J}^{\,}_{\mathrm{c}}$
as determined by exact diagonalization techniques.
For all cases, the long-range ordered phases are the
Neel${}^{\,}_{z}$ and dimer phases.

\subsection{The scaling exponent $\eta$ at quantum criticality}

In order to estimate the exponent $\eta$
at the Neel${}^{\,}_{z}$-dimer transition,
we make the scaling ansatz
\begin{eqnarray}
O^{\,}_{\mathrm{VBS}}(L)=
A\,
L^{-\frac{1}{2\eta}}
\label{eq:Odim-L-fit}
\end{eqnarray}
taking $\eta$ and $A$ as fitting parameters.
The estimate of $\eta$ was obtained from the fitting
using the data of $O^{\,}_{\mathrm{VBS}}(L)$ for $64\le L\le 192$
while the error in $\eta$ was determined by the difference
between the estimate and $\eta$ obtained using the data 
for $32 \le L \le 192$.

\begin{figure*}
\begin{center}

\includegraphics[width=0.4\textwidth]{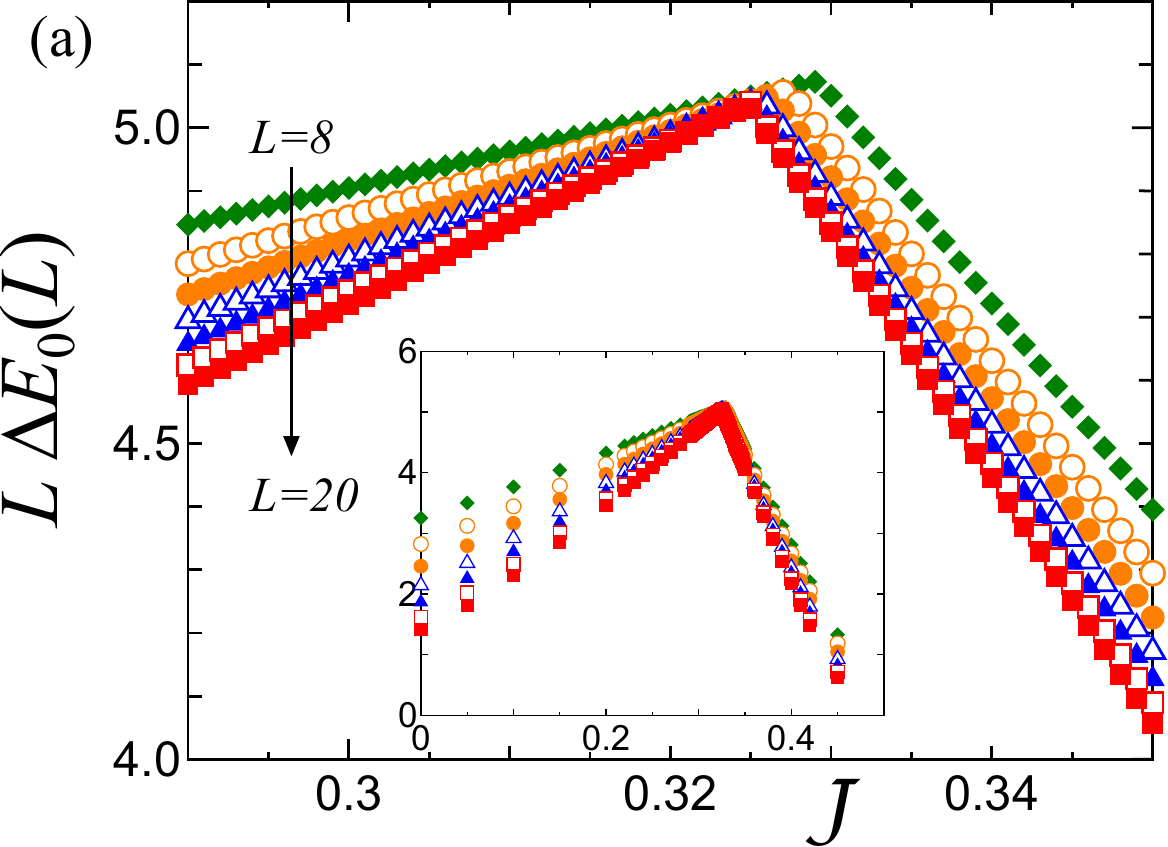}
~~~~~~~~~~~~~~~~~~~~~~~~
\includegraphics[width=0.39\textwidth]{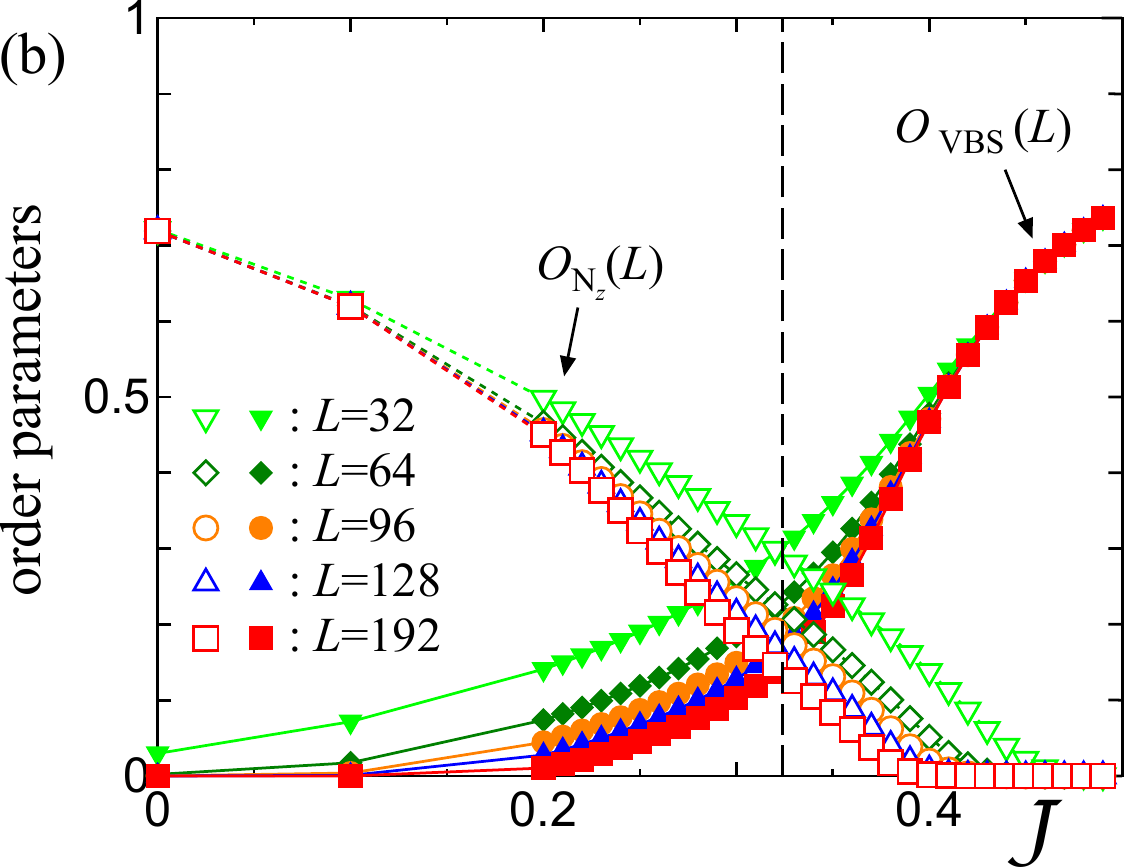}
\\
\includegraphics[width=0.4\textwidth]{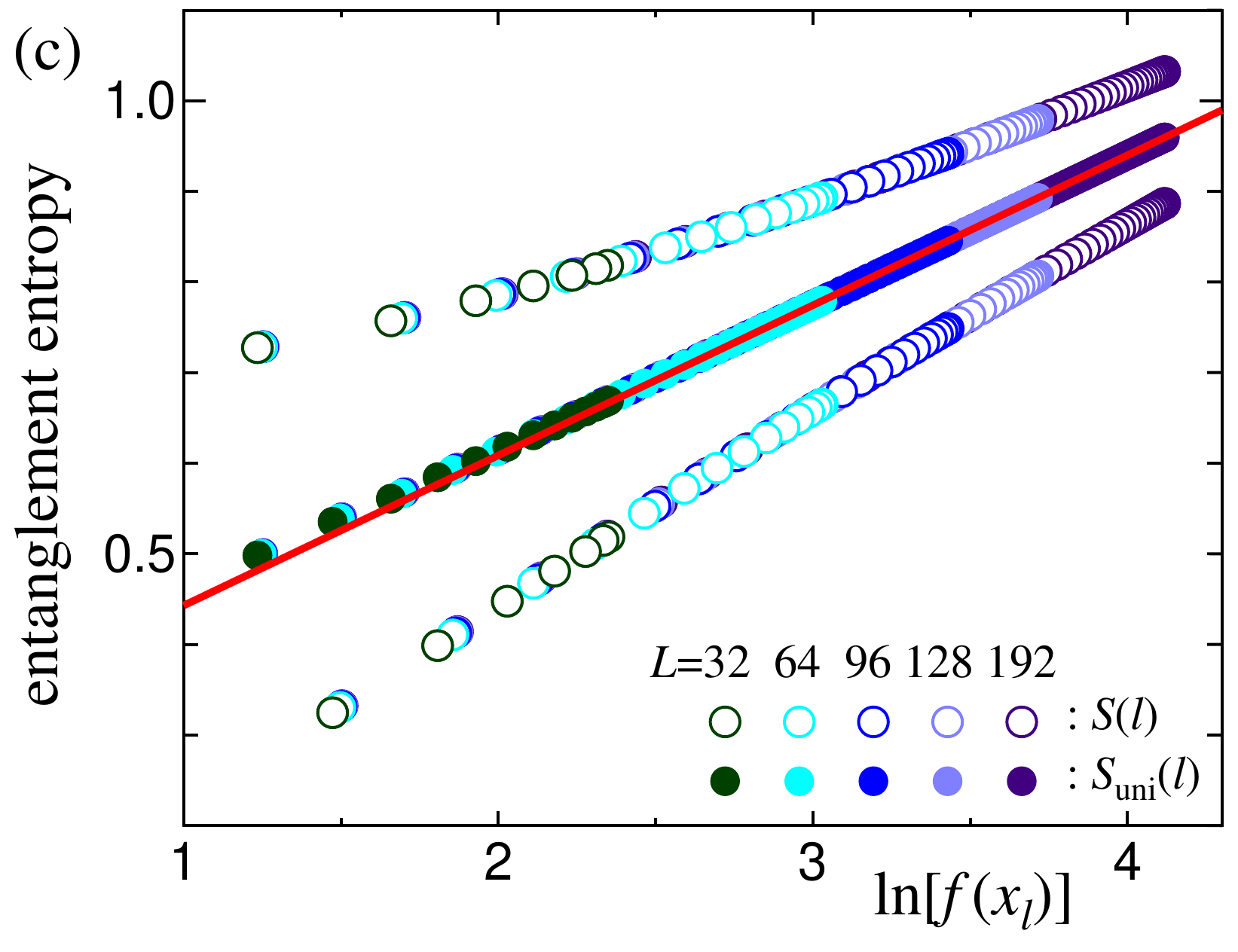}
~~~~~~~~~~~~~~~~~~~~~~~~
\includegraphics[width=0.4\textwidth]{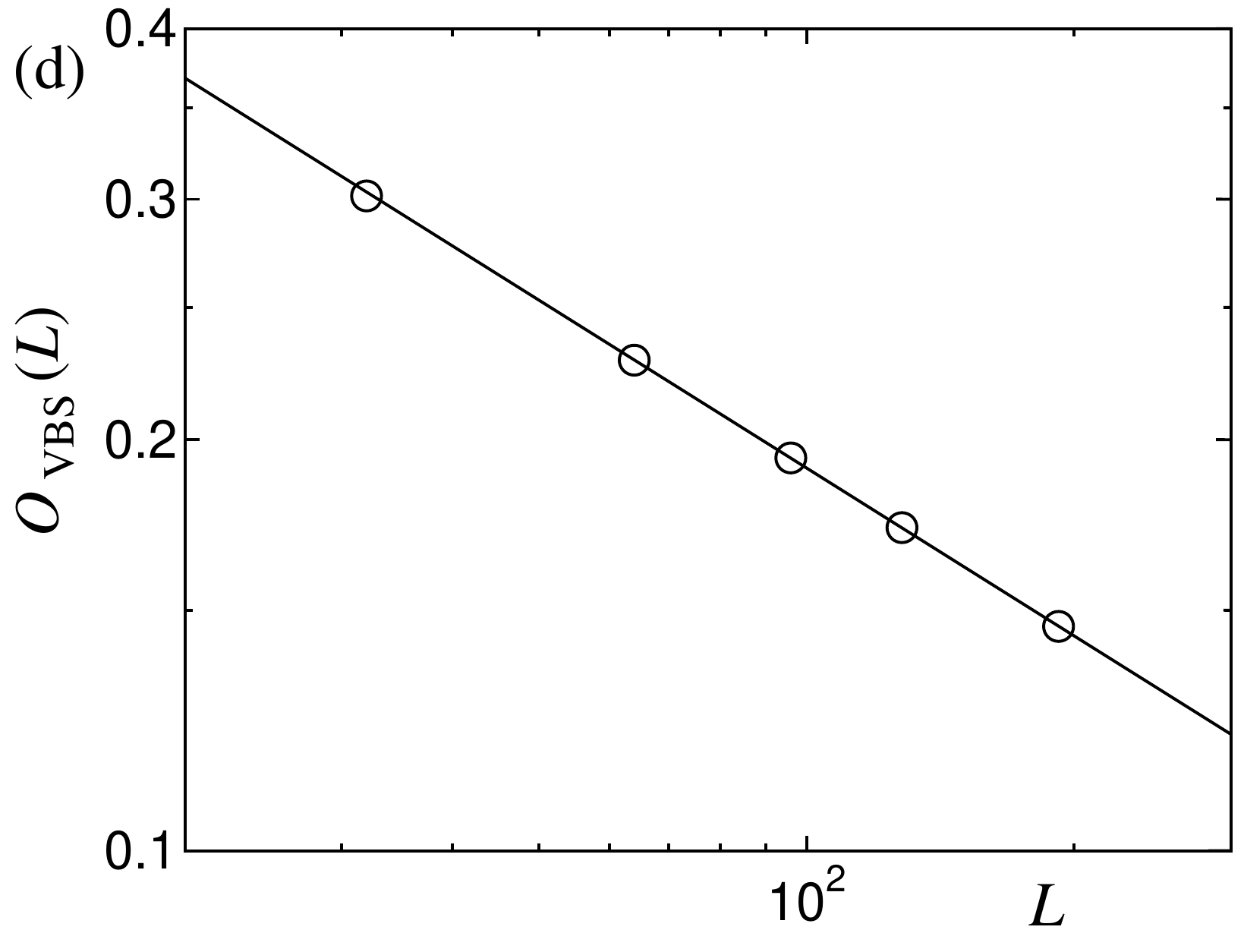}
\caption{(Color online)
Panels (a), (b), (c), and (d) are the
counterparts to Figs.\
\ref{fig: scaled gap and critical cal J}(a),
\ref{fig:order-parameter_App},
\ref{fig: entanglement entropy}(b),
and
\ref{fig: eta from DMRG}(a),
respectively,
when $\Delta^{\,}_{z}=3.0$ and $\Delta^{\,}_{y}=2.0$.
The central charge $c$ obtained in (c) (using the data for $L=192$)
is $c=0.994$, and the exponent $\eta$ obtained in (d) is $\eta=1.22$.
        }
\label{fig:data_for_Dz3}
\end{center}
\end{figure*}

\subsection{Complementary cuts}

We have repeated our numerical analysis 
along the one-dimensional cuts (\ref{eq: one-dimensional cut})
with $\Delta^{\,}_{y}$
given in Fig.\ \ref{fig: scaled gap and critical cal J}(b)
for the three one-dimensional cuts
\begin{subequations}
\label{eq: one-dimensional cuts}
\begin{align}
&
\Delta^{\,}_{z}=3.0,
\qquad
\Delta^{\,}_{y}=0.0,
\qquad
0\leq\mathcal{J}<0.5,
\label{eq: one-dimensional cuts a}
\\
&
\Delta^{\,}_{z}=3.0,
\qquad
\Delta^{\,}_{y}=0.5,
\qquad
0\leq\mathcal{J}<0.5,
\label{eq: one-dimensional cuts b}
\\
&
\Delta^{\,}_{z}=3.0,
\qquad
\Delta^{\,}_{y}=2.0,
\qquad
0\leq\mathcal{J}<0.5.
\label{eq: one-dimensional cuts c}
\end{align}
\end{subequations}
Figure\ \ref{fig:data_for_Dz3} 
shows the numerical data
along the one-dimensional cut (\ref{eq: one-dimensional cuts c}).
Hereto, the existence of a continuous quantum critical point 
with central charge $c=1$ separating
the Neel${}^{\,}_{z}$ phase from the dimer phase
is confirmed. The same is true for the
one-dimensional cuts (\ref{eq: one-dimensional cuts a})
and (\ref{eq: one-dimensional cuts b}).

\bibliography{references}
\end{document}